\documentclass[oneside,12pt]{book}

\usepackage[top=1.0in, bottom=2in, left=1.0in, right=1.0in]{geometry}

\newcommand{\beq}{\begin{eqnarray}}
\newcommand{\eeq}{\end{eqnarray}}

\usepackage{amsmath,amssymb,layout,setspace}

\usepackage{
color,
cite,graphicx}

\usepackage{fancyhdr}
\begin{document}

\topmargin 0pt
\headheight 0pt

\topskip 5mm

\hspace{4cm}

\thispagestyle{empty}

\vspace{20pt}

\begin{center}

{\Large {\bf 







YANG-MILLS THEORIES

\vspace{20pt}

AS DEFORMATIONS  

\vspace{20pt}

OF MASSIVE INTEGRABLE MODELS}}

\vspace{25pt}


\vspace{40pt}

{\large by

\vspace{40pt}

AXEL CORT\'ES CUBERO}

\end{center}




\vspace{40pt}

\noindent
A dissertation submitted to the Graduate Faculty in Physics in partial fulfillment of the requirements for the degree of Doctor of Philosophy, the City University of New York.

\vspace{40pt}

\begin{center}

2014

\end{center}

\pagenumbering{roman}

\newpage
\begin{center}

\[\]

\vspace{450pt}
\copyright 2014 \\
\vspace{5pt}
 Axel Cort\'es Cubero\\
\vspace{5pt}
All rights reserved

\end{center}
\newpage

\vspace{20pt}

\noindent
This manuscript has been read and accepted for the 
Graduate Faculty in Physics in satisfaction of the 
dissertation requirement for the degree of Doctor of Philosophy.

\vspace{20pt}
\hspace{0.2in}

\line(1,0){50}
\hspace{0.3in}
\line(1,0){70}

Date
\hspace{1.9in}
Chair of Examining Committee

\hspace{2.3in}
Peter Orland

\hspace{2.3in}
\vspace{30pt}

\line(1,0){50}
\hspace{0.3in}
\line(1,0){70}

Date
\hspace{1.9in}
Executive Officer

\hspace{2.3in}
Steven Greenbaum

\vspace{20pt}



\line(1,0){130}

Adrian Dumitru

\vspace{30pt}




\line(1,0){130}

Gerald  Dunne
\vspace{30pt}





\line(1,0){130}

Daniel Kabat

\vspace{30pt}

\line(1,0){130}

Alexios Polychronakos

\vspace{5pt}

Supervisory Committee

\begin{center}

CITY UNIVERSITY OF NEW YORK
\end{center}

\newpage
\doublespacing
\begin{center}

ABSTRACT

\vspace{20pt}

{\Large {\bf  YANG-MILLS THEORIES AS DEFORMATIONS OF MASSIVE INTEGRABLE MODELS}}
\vspace{20pt}

{\large by
\vspace{20pt}

Axel Cort\'es Cubero}
\end{center}

\vspace{20pt}

{\large{Advisor: Prof. Peter Orland}}\\

Yang Mills theory in 2+1 dimensions can be expressed as an array of coupled (1+1)-dimensional principal chiral sigma models. 
The $SU(N)\times SU(N)$ principal chiral sigma model in 1+1 dimensions is integrable, asymptotically free and has massive excitations. We calculate all the form factors and two-point correlation functions of the Noether current and energy-momentum tensor, in 
't~Hooft's large-$N$ limit (some form factors can be found even at finite $N$). We use these new form factors to calculate physical quantities in (2+1)-dimensional Yang-Mills theory, generalizing previous  $SU(2)$ results from references \cite{glueball}, \cite{horizontal}, \cite{vertical}, to $SU(N)$. The anisotropic gauge theory is related to standard isotropic one by a Wilsonian renormalization group with ellipsoidal cutoffs in momentum. We calculate quantum corrections to the effective action of QED and QCD, as the theory flows from isotropic to anisotropic. The exact principal chiral sigma model S-matrix is used to examine the spectrum of (1+1)-dimensional massive Yang Mills theory.

\newpage
\begin{center}
\[\]
\[\]
\[\]

\textit{To my cousin, Juanma.}

\textit{ I am still just trying to be more like you, when I grow up.}
\end{center}

\newpage

\begin{center}

ACKNOWLEDGMENTS

\end{center}

First of all I would like to thank my advisor, Prof. Peter Orland. From the first day of graduate school, when he stood in front of the incoming class and told us ``quantum field theory is hard, and it is not for everyone", I knew we would have a great working chemistry. He always has a new interesting idea when I walk into his office, and is always willing to spend hours talking about physics, life, the universe and everything.

I also owe very special thanks to Prof. Steven Greenbaum for that time I randomly met him in an empty hallway in Mayag\"{u}ez and he asked me ``Do you want to do a PhD in physics?". I answered ``Ok". He has given me endless support at every step of my degree since then.

I would like to thank professors Adrian Dumitru, Jamal Jalilian-Marian and Stefan Bathe for many interesting conversations. These have included topics like particle physics, proper espresso making and drinking techniques, German beer, and our complaints on teaching physics 1003.

I thank Prof. Justin Vazquez-Poritz for teaching me those initial quantum field theory and string theory classes, and for putting up with me and my questions inside and outside the classroom. I thank Prof. Gerald Dunne for being kind enough to be part of my committee twice, and for being such a hospitable  host when I visited Connecticut.  I also thank Professors Alexios Polychronakos, Daniel Kabat and V. Parameswaran Nair, for their very interesting and inspiring lectures, and for making me feel like I am part of their high energy theory community at the City College of New York. I also want to thank Mr. Daniel Moy, for keeping track of all the professors and students in the physics program and for always having time to talk to each of us.

I would not have been able to endure these past five years without the help of my friends. I thank John, Aline, Jason, Kay, Stephanie, Zach and Mark (yes, in that particular order!). We have too many memories together to discuss here, but to summarize, they made it easy and enjoyable to get my degree. I would do it again any day. 

I especially want to thank Elena Petreska. She has really been my one partner and ally all this time. We survived these five years, and I am sure we can take on anything else together. This is only the beginning. 

Finally I want to thank my family. I thank my brother Rey, with whom I have been able to share the New York experience. We have really grown a lot together, and gone from naive teenagers who dream of making it with their band, to naive old men who dream of making it with their band, and who now know how to dress for cold weather. I thank my mother, Brenda, who has done more for me and my brother than we could ever deserve or return. It is because of her effort and hard work that I am where I am today.  I also thank my sister, Isabella, and my grandma, aunts, and cousins for receiving me with so much warmth and love every time I go back home. Lastly I thank my father, Angel for his unconditional love and support, and for telling everyone he meets about ``his son, the doctor".


\tableofcontents

\chapter{Introduction}
\setcounter{equation}{0}
\renewcommand{\theequation}{1.\arabic{equation}}

\setcounter{page}{1}

\pagenumbering{arabic}

\section{Introduction}
\setcounter{equation}{0}
\renewcommand{\theequation}{1.1.\arabic{equation}}

\fancyhead[L]{}

Quantum chromodynamics (QCD) is the quantum field theory describing the interaction of quarks and gluons. The dynamics of hadrons and nuclei emerges from QCD. In QCD the interactions between particles are strong and the observed physical states are bound states of fundamental particles. Therefore to fully understand QCD, one needs methods beyond perturbation theory.
	
	The most successful non-perturbative tool in QCD is numerical lattice simulation. Numerical methods provide accurate results in scenarios where perturbation theory fails. Unfortunately, we lack a mathematical route to complement them. Numerical simulations are invaluable for finding bound state masses and string tensions, but they do not elucidate qualitatively the mechanism of how bound states form or how quarks are confined.

In this thesis, we develop analytical methods to examine field theories, with an eye towards QCD. We  discuss an anisotropic version of QCD, where the constants coupling the different vector components of the gluon field have different strengths. 
This is understood as a longitudinal rescaling of the coordinates. If  two of the four space-time coordinates in QCD (say one space direction, $x^3$,  and the time direction, $x^0$) are rescaled by a parameter $\lambda$ (replacing them by $\lambda x^3$ and $\lambda x^0$, respectively), we obtain an action with new couplings dependent on $\lambda$. We consider the limit where this parameter $\lambda$ goes to zero. 
We can perturb in powers of $\lambda$ instead of powers of the coupling constant. 

The anisotropic regime of QCD has important applications in the physics of heavy-ion collisions at RHIC and the LHC. Verlinde and Verlinde \cite{verlinde} have argued that the longitudinally rescaled action yields BFKL theory \cite{bfkl}. McLerran and Venugopalan used a similar idea to derive the Color-Glass Condensate picture \cite{colorglass}.

We view the anisotropic model as an array of two-dimensional field theories, coupled together to form a higher-dimensional theory. The strength of the coupling between these two-dimensional models depends on the rescaling parameter $\lambda$. The two-dimensional theory is the principal chiral sigma model (PCSM)\cite{twoplusone}. The PCSM is known to be integrable, and this property has been exploited to find  exact results. The main goal of our program is to use exact results from the PCSM to calculate physical quantities in anisotropic QCD, finding corrections for small $\lambda$.

This approach is especially interesting for $(2+1)$-dimensional QCD, since there are two different coupling constants for the gluon field, but they are both 
small compared to the cutoff. This makes our approach fundamentally different from other analytic studies of $(2+1)$-dimensional QCD (which are generally at large dimensionless coupling, far from the continuum limit) \cite{nair},\cite{greensite}.

In the remainder of this chapter, we give a brief introduction to QCD and Yang-Mills theory in 3+1, and 2+1 space-time dimensions, both in the continuum and on a 
lattice. 

In Chapter 2, we discuss the Hamiltonian of longitudinally-rescaled QCD. In 2+1 dimensions, we show how the anisotropic Hamiltonian is equivalent to an array of PCSM's.

In Chapter 3, we present a short introduction to the integrable bootstrap program in
1+1 dimensions. We discuss the general procedure of calculating exact 
S-matrices using elasticity and factorization. The S-matrix and some other assumptions can be used to calculate exact form factors and correlation functions. 

In Chapter 4, we apply the integrable bootstrap program to the PCSM. This is very difficult, in general, unless we take
't~Hooft's large-$N$ limit.
We calculate all the exact form factors of the Noether current and energy-momentum tensor operators. These form factors are then used to calculate the exact two-point function of these operators. For finite $N$, we are only able to find the first nontrivial form factor.

In Chapter 5, we study the ($1+1$)-dimensional massive Yang-Mills theory. This model can be understood as a gauged PCSM. Using the exact PCSM S-matrix, we calculate the mass spectrum.

In Chapter 6, we use the PCSM S-matrix and the exact finite-$N$ current form factor to compute some physical quantities in (2+1)-dimensional anisotropic Yang-Mills theory. We first compute the low-lying glueball spectrum (which is very similar to the massive spectrum from Chapter 5). We then compute the string tension for a static quark-antiquark pair. Because the theory is anisotropic, the string tension will be different if the quarks are separated in the $x^{1}-$ or the $x^{2}-$direction. These $SU(N)$ results generalize those obtained
by Orland for $N=2$.

In Chapter 7, we explore the quantum effects of longitudinal rescaling. Rescaling the coordinates changes the ultraviolet cutoffs of the theory. We therefore need a renormalization group to see how the couplings run in the quantum theory as we go from isotropic to anisotropic momentum cutoffs. To illustrate 
this idea, we first examine quantum electrodynamics in 3+1 dimensions, with massless electrons. We then apply the anisotropic-renormalization group to 
(3+1)-dimensional QCD, with massless quarks.

We summarize our results and discuss possible future projects in the last chapter.

\section{QCD and Yang-Mills theory}
\setcounter{equation}{0}
\renewcommand{\theequation}{1.2.\arabic{equation}}

In this section, we present the QCD action and Hamiltonian, as well as some notation and conventions that will be used throughout this thesis.

The Dirac action for $N$ free massless fermions in $3+1$ dimensions is
\beq
S_{\rm Dirac}=i\int d^4x\,\bar{\psi}^a\gamma^\mu \partial_\mu \psi^a,\label{diracaction}
\eeq
where $\gamma^\mu$ are the Dirac matrices, which satisfy $\{\gamma^\mu,\gamma^\nu\}=2\eta^{\mu\nu} 
\mathbf{1}
$, where $\eta^{\mu\nu}$ is the Minkowski metric with $\mu,\nu=0,1,2,3$, and $a=1,2,\dots,N$ (where $N=3$ in quantum chromodynamics). The action (\ref{diracaction}) is invariant under the global $SU(N)$ transformation, $\psi(x)\to V\psi(x)$, where $V\in SU(N)$. This symmetry is promoted to a local gauge symmetry by introducing an $SU(N)$-algebra-valued gauge field $A_\mu(x)^{ab}$ that transforms as $A_\mu(x)^{ab}\to V^\dag(x)^{ca} A_\mu(x)^{cd} V(x)^{db}-\frac{i}{g_0}V^\dag(x)^{ca} \delta^{cd}\partial_\mu V(x)^{db}$. Introducing the covariant derivative $D_\mu(x)^{ab}=\delta^{ab}\partial_\mu-ig_0A_\mu(x)^{ab}$, the gauged Dirac action is
\beq
S_{\rm Dirac}=i\int d^4x\,\bar{\psi}^a\gamma^\mu D_\mu^{ab} \psi^b.\label{diracactiongauged}
\eeq
The gauge field can be made dynamical by adding a Yang-Mills term to the action:
\beq
S_{\rm YM}=-\int d^4x \frac{1}{4}{\rm Tr} \,F_{\mu\nu} F^{\mu\nu},\label{yangmillsaction}
\eeq
where $F_{\mu\nu}=\frac{i}{g_0}[D_\mu,D_\nu]$. The gauge field can be expressed in terms of the generators of $SU(N)$, $A_{ab}=A^\alpha \,t^\alpha_{ab}$,  where $\alpha=1,2,\dots, N^2-1$, and the generators satisfy the lie algebra, $[t^\alpha,t^\beta]=if^{\alpha\beta\gamma} t^\gamma$, where $f^{\alpha\beta\gamma}$ are the $SU(N)$ structure constants. We normalize the generators by ${\rm Tr}\, t^\alpha\, t^\beta=\delta^{\alpha\beta}$. Henceforth we will rescale the gauge field as $A_\mu\to\frac{1}{g_0}A_\mu$, so that $S_{\rm YM}=-\int d^4x \frac{1}{4g_0^2}{\rm Tr}\, F_{\mu\nu} F^{\mu\nu}$

We now find the Hamiltonian that corresponds to the Yang-Mills action (\ref{yangmillsaction}). We will work in the temporal gauge, $A_0=0$. This is an incomplete gauge fixing, which still allows time-independent gauge transformations. In the next chapter we restrict this remaining
gauge freedom, by imposing the additional axial gauge condition, $A_{1}=0$ on the Hilbert space.

The Yang-Mills action in temporal gauge is
\beq
S_{\rm YM}=\int d^4x \left( \sum_{i=1}^3\frac{1}{2g_0^2}{\rm Tr}\,\partial_0A_i\partial_0 A^i-\sum_{i,j=1}^3\frac{1}{4g_0^2}F_{ij} F^{ij} \right)
\eeq
The conjugate-momentum field or electric field is $E_i=-i\delta/\delta A_i$. The Hamiltonian is
\beq
H_{\rm YM}=\int d^3x\sum_{i=1}^{3}\left(\ \frac{g_0^2}{2}E_i^2+\frac{1}{2g_0^2}B_i^2\right),\nonumber
\eeq
where $B_i=\frac{1}{2}\epsilon^{ijk}F_{jk}$ is the magnetic field. In the temporal gauge, we enforce the Gauss-law constraint on physical wave functionals $\Psi$:
\beq
\left(\sum_{i=1}^3D_iE^i-\rho\right)\Psi=0,\nonumber
\eeq
where $\rho$ is the charge density of the Dirac field. 

In 2+1 dimensions, there is only one component of the magnetic field, namely $B=\frac{1}{2}\epsilon^{ij}F_{ij}$, where $i,j=1,2$. The $(2+1)$-dimensional Yang-Mills Hamiltonian is
\beq
H_{\rm YM}^{(2+1)}=\int d^2 x\left(\sum_{i=1}^2\frac{g_0^2}{2}E_i^2+\frac{1}{2g_0^2}B^2\right).\nonumber
\eeq


\section{Lattice gauge theory}
\setcounter{equation}{0}
\renewcommand{\theequation}{1.3.\arabic{equation}}

In this section we present the action of the lattice Yang-Mills theory, first formulated by Wilson \cite{wilsonlattice}.  We find the Hamiltonian as the
$x^0$ direction becomes continuous, as first discussed by Kogut and Susskind \cite{kogutsusskind}.

Wilson's lattice gauge theory is formulated on a 4-dimensional, hypercubic lattice, in Euclidean space. The distance between two neighboring lattice sites is 
$a$, which we call the lattice spacing. We denote a unit vector in the $\mu$ direction by $\hat{\mu}$.
We label the link between the nearest-neighbor space-time points $x$ and $x+a\hat{\mu}$, by $(x,\mu)$.

We assign the $SU(N)$-group-valued gauge field on the link $(x,\mu)$, written as $U(x)_{\mu}\in SU(N)$. This field is related to the continuum $SU(N)$-algebra valued field by
\beq
U(x)_{\mu}=\mathcal{P}\exp i\int_0^a dt\,A_\mu(x+t\hat{\mu}),\nonumber
\eeq
where $\mathcal{P}$ denotes path ordering. Under a local gauge transformation, the gauge field transforms as $U_{x,\mu}\to V(x)U(x)_{\mu} V^\dag(x+a\hat{\mu})$, where $V(x)\in SU(N)$. The simplest gauge-invariant object we can construct out of the gauge fields is defined at a plaquette or elementary
square made of the four links $(x,\mu)$, $(x+a{\hat \mu},\nu)$, $(x+a{\hat \nu},\mu)$ and $(x,\nu)$. This object is ${\rm Tr}\,U^{\square}(x)_{\mu\nu}$, where
\beq
U^{\square}(x)_{\mu\nu}=U(x)_{\mu} U(x+a\hat{\mu})_{\nu}U^\dag(x+a\hat{\nu})_{\mu}U^\dag(x)_{\nu}.\nonumber
\eeq
The simplest gauge-invariant action one can construct is 
\beq
S_W=C\sum_{x,\mu,\nu}{\rm Tr}\,\left[ U^{\square}(x)_{\mu\nu}+U^{\square}(x)_{\mu\nu}^\dag\right],\label{wilsonaction}
\eeq
which is called the Wilson action. The Yang-Mills action is the continuum limit $a\to 0$, of the Wilson action, with $C=-a^2/g_0^2$.

A detailed introduction to Hamiltonian lattice gauge theory is given in References \cite{kogutreview},\cite{creutz}. The Kogut-Susskind lattice Hamiltonian is usually derived in the temporal gauge, $U(x)_{0}=1$. 

In the Hamiltonian formulation, there are  electric-field operators $l(x)_{j}^b$ in the adjoint representation of $SU(N)$, at every space link $(x,j)$, for $j=1,2,3$ and $b=1,2,\dots, N^2-1$. The  commutation relations are
\beq
\left[l(x)_{j}^b,l(y)_{k}^c\right]&=&i\delta_{x\,y}\delta_{j\,k} f^{dbc}l(x)_{j\,d},\nonumber\\
\left[l(x)_{ \,j}^b,\,U(y)_{k}\right]&=&-\delta_{x\,y}\delta_{j\,k}\,t^b U(x)_{j}.\nonumber
\eeq
The Hamiltonian is obtained by taking the continuum limit in the time direction. The Kogut-Susskind Hamiltonian, inside a box of size $(aL)^3$, is
\beq
H&=&\sum_{x^1,x^2,x^3=-\frac{L}{2}}^{\frac{L}{2}}\sum_{j=1}^{3}\sum_{b=1}^{N^2-1}\frac{g_0^2}{2a}\left[l(x)_{j}^b\right]^2\nonumber\\
&&-\sum_{x^1,x^2,x^3=-\frac{L}{2}}^{\frac{L}{2}}\sum_{b=1}^{3}\frac{1}{4g_0^2\,a}\,{\rm Tr}\,\left[U^{\square}(x)_{ jk}+\left(U^\square(x)_{jk}\right)^\dag\right],\nonumber
\eeq
where $L$ is an even integer. The quantities $U^\square(x)_{jk}$, assigned to space-like plaquettes, are associated with the magnetic-field in the continuum limit. They may be expanded in powers of the lattice spacing to yield 
\beq
U^{\square}(x)_{jk}=1+iaF_{jk}(x)-a^2F_{jk}(x)^2+\cdots\,.\nonumber
\eeq

In $2+1$ dimensions, the lattice Hamiltonian is
\beq
H&=&\sum_{x^1,x^2=-\frac{L}{2}}^{\frac{L}{2}}\sum_{j=1}^{2}\sum_{b=1}^{N^2-1}\frac{g_0^2}{2a}\left[l(x)_{j}^b\right]^2\nonumber\\
&&-\sum_{x^1,x^2=-\frac{L}{2}}^{\frac{L}{2}}\sum_{b=1}^{2}\frac{1}{4g_0^2\,a}\,{\rm Tr}\,\left[U^{\square}(x)_{ 12}+\left(U^\square(x)_{21}\right)^\dag\right].\nonumber
\eeq

In temporal gauge, physical states are those which satisfy Gauss's law:
\beq
\sum_{j=1}^{d-1} [\mathcal{D}_j l_j(x)]_b \Psi=0,\label{latticegauss}
\eeq
where
\beq
[\mathcal{D}_j l_j(x)]_b=l_j(x)-\mathcal{R}_j(x-\hat{j}a)_b^{\,\,c}\,\,l_j(x-\hat{j}a)_c,\nonumber
\eeq
where $\mathcal{R}_j(x)_b^{\,\,c}t_c$ is the adjoint representation of the gauge field,
\beq
\mathcal{R}_j(x)_b^{\,\,c}t_c=U_j(x)t_b U^\dag_j(x),\nonumber
\eeq
and $d$ is the number of space-time dimensions \cite{twoplusone}.


\chapter{Longitudinally rescaled Yang-Mills theory}
\setcounter{equation}{0}
\renewcommand{\theequation}{2.\arabic{equation}}

\section{Longitudinal rescaling of the Yang-Mills action}
\setcounter{equation}{0}
\renewcommand{\theequation}{2.1.\arabic{equation}}

We seek an effective theory of QCD at large center-of-mass energies. The particle beams in a collider are assumed to travel in the longitudinal 
$x^3$-direction. Thus we want (in our effective theory) four-momenta to have large components in the $x^0$- and $x^3$-directions, but not in the 
$x^1$- and $x^2$-directions. This effective theory is an anisotropic version of Yang-Mills theory, where two of the coordinates, $x^0\; {\rm and}\;x^3$, are rescaled to $ \lambda x^0 \;{\rm and} \;\lambda x^3$, respectively. Under this rescaling, the gauge fields transform as $A_{0,3}\to\lambda^{-1}A_{0,3}$. The remaining transverse space-time coordinates are left unchanged: $x_{1,2}\to x_{1,2},\,\,A_{1,2}\to A_{1,2}$.  Under this rescaling, the Yang-Mills action becomes
\beq
S\longrightarrow \frac{1}{2g_0^{2}}\int d^{4}x {\rm Tr}\, \left(F_{01}^{2}+F_{02}^{2}-F_{13}^{2}-F_{23}^{2}+\lambda^{-2}F_{03}^{2}-\lambda^{2}F_{12}^{2}\right).
\nonumber
\eeq

For much of this thesis, we will examine the simpler (2+1)-dimensional Yang-Mills theory. in this case we rescale the longitudinal coordinates as $x^{0,1}\to \lambda x^{0,1}$ and leave the transverse coordinate unchanged, $x^2\to x^2$. The rescaled (2+1)-dimensional action is
\beq
S\to \frac{1}{2g_0^2}\int d^4x {\rm Tr}\, (F_{02}^2-F_{12}^2+\lambda^{-2}F_{01}^2).\nonumber
\eeq

The longitudinally rescaled (2+1)-dimensional Hamiltonian, in temporal gauge $(A_0=0)$, is
\beq
H=H_0+\lambda^2 H_1,\label{rescaledhamiltonian}
\eeq
where
\beq
H_0=\int d^2x \left(\frac{g_0^2}{2}E_2^2+\frac{1}{2g_0^2}B^2\right),\nonumber
\eeq
and
\beq
H_1=\int d^2x \frac{g_0^2}{2}E_1^2,\nonumber
\eeq
where $E_i=-i\delta/\delta A_i$, $B=\epsilon^{jk}(\partial_j A_k+A_j\times A_k)$, where $i,j,k=1,2$ and $(A_j\times A_k)^a=f^a_{bc}A_j^b A_k^c$. Physical states $\Psi$ satisfy Gauss's law:
\beq
(D_1 E_1+D_2E_2)\Psi=0.\label{firstgauss}
\eeq

The Hamiltonian (\ref{rescaledhamiltonian}) is simplified in the highly rescaled limit, $\lambda\to 0$. In this case we can treat the term $\lambda^2 H_1$ as a small perturbation. We will show in the next sections that the unperturbed Hamiltonian $H_0$ is integrable. Our goal is to calculate exact quantities in the unperturbed theory, then calculate corrections in powers of the small parameter $\lambda^2$.

A similar perturbation theory is possible in 3+1 dimensions \cite{nearintegrability}. The longitudinally rescaled Hamiltonian in temporal gauge is
\beq
H=H_0+\lambda^2H_1 +\lambda^2H_2,\nonumber
\eeq
where
\beq
H_0=\int d^3x\left(\frac{g_0^2}{2} E_1^2+\frac{g_0^2}{2}E_2^2+\frac{1}{2g_0^2}B_1^2+\frac{1}{2g_0^2}B_2^2\right),\nonumber
\eeq
\beq
H_1=\int d^3x \frac{g_0^2}{2}E_3^2,\nonumber
\eeq
and
\beq
H_2=\int d^3x\frac{1}{2g_0^2}B_3^2,\nonumber
\eeq
where $B_i=\epsilon^{ijk}(\partial_jA_k+A_j\times A_k)$, with $i,j,k=1,2,3$.

In the next section we examine the (2+1)-dimensional Hamiltonian (\ref{rescaledhamiltonian}) in the axial gauge, $A_1=0$. Then we
consider this Hamiltonian on the lattice. As $\lambda\rightarrow 0$, the
lattice model is an array of integrable PCSM's.

\section{The Yang-Mills Hamiltonian in the axial gauge in 2+1 dimensions}
\setcounter{equation}{0}
\renewcommand{\theequation}{2.2.\arabic{equation}}

The Yang-Mills Hamiltonian in 2+1 dimensions, in $A_0=0$ gauge is 
\beq
H=\int d^2x \left( \frac{g_0^2}{2}E_1^2+\frac{g_0^2}{2}E_2^2+\frac{1}{2g_0^2}B^2\right).\nonumber
\eeq
Gauss's law (Eq. (\ref{firstgauss})) may be written more explicitly in terms of the gauge fields as
\beq
&&\left\{\left[\delta_{ac}\partial_1+g_o f_{abc}A_1(x^1,x^2)^b\right]E_1(x^1,x^2)^c\right.\nonumber\\
&&\left.\,\,\,\,\,\,\,\,\,\,+\left[\delta_{ac}\partial_2+g_o f_{abc}A_2(x^1,x^2)^b\right]E_2(x^1,x^2)^c\right\}\Psi=0.\label{secondgauss}
\eeq

We can solve (\ref{secondgauss}) to find $E_1$, in terms of $A_1,\,A_2,$ and $E_2$. With the boundary conditions, $E_1\to 0$ as $x^1\to\pm\infty$:
\beq
E_1(x^1,x^2)^a&=&\int_{-\infty}^{x^1}dy^1\left\{\mathcal{P}\exp\left[ ie\int_{-\infty}^{y^1} dz^1\mathcal{A}_1(x^1,x^2)\right]\right\}^{\,\,\,\,\,\,\,\,b}_{a}\nonumber\\
\,\nonumber\\
&&\times\left\{\left[\delta_{bd}\partial_2+g_0f_{bcd}A_{2}(y^1,x^2)^c\right]E_2(y^1,x^2)^d\right\}\nonumber
\eeq
where $\mathcal{A}_1(x^1,x^2)_{ab}=i f_{abc}A_1(x^1,x^2)^c$ is the gauge field in the adjoint representation. There remains a global invariance given by 
\beq
\!\!\!\!\left(\int_{-\infty}^{\infty}dy^1\! \left\{\mathcal{P}\exp\left[ ie\int_{-\infty}^{y^1} dz^1\mathcal{A}_1(x^1,x^2)\right]\right\}^{\,\,\,\,\,\,\,\,b}_{a}
\left[\left(\delta_{bd}\partial_2+g_0f_{bcd}A_{2}(y^1,x^2)^c\right)E_2(y^1,x^2)^d\right]\right)\!\!\Psi=\!0.\nonumber
\eeq

The temporal gauge condition, $A_0=0$ does not completely fix the gauge. Time-independent gauge transformations are consistent with this condition. We may now impose the axial gauge, $A_{1}(x^1,x^2)=0$. 

In the axial gauge, the electric field $E_1$ is 
\beq
E_1(x^1,x^2)^a&=&\int_{-\infty}^{x^1} dy^1\left[\delta_{ac}\partial_2+g_0 f_{abc} A_2(y^1, x^2)^b)E_2(y^1, x^2)^c\right]\nonumber\\
&=&\int_{-\infty}^{x^1} dy^1D_2(y^1,x^2)_{ac}E_2(y^1,x^2)^c.
\label{axialfixingcontinuum}
\eeq
After axial gauge fixing, there remains an invariance: 
\beq
\left\{\int_{-\infty}^{\infty} dy^1\left[(\delta_{ac}\partial_2+g_0 f_{abc} A_2(y^1, x^2)^b)E_2(y^1, x^2)^c\right]\right\}\Psi=0.\label{globalgaussaxial}
\eeq
The condition (\ref{globalgaussaxial}) is imposed on the wave functional for each value
of $x^{2}$.

Substituting Eq. (\ref{axialfixingcontinuum}) into the Hamiltonian:
\beq
&&H=\int d^2x\left\{\frac{g_0^2}{2}E_2(x^1,x^2)^2+\frac{1}{2}[\partial_1A_2(x^1,x^1)]^2\right\}\nonumber\\
&&\,\,\,\,\,\,\,-\int dx^1\int dy^1 \int dx^2\vert x^1-y^1\vert [D_2(x^1,x^2)E_2(x^1,x^2)]\,\nonumber\\
&&\,\,\,\,\,\,\,\,\,\,\,\,\,\times[D_2(y^1,x^2)E_2(y^1,x^2)].\label{axialhamiltoniancontinuum}
\eeq
The Hamiltonian (\ref{axialhamiltoniancontinuum}) depends only on the transverse degrees of freedom $A_2\; {\rm and}\;E_2$.  This Hamiltonian is nonlocal in the $x^1$-direction, which is an artifact of the axial gauge fixing.  Eq. (\ref{axialhamiltoniancontinuum}) can be made local, by reintroducing the $A_0$ component of the gauge field.

\section{The Kogut-Susskind lattice Hamiltonian in the axial gauge}
\setcounter{equation}{0}
\renewcommand{\theequation}{2.3.\arabic{equation}}


The Kogut-Susskind Hamiltonian in 2+1 dimensions, in temporal gauge ($A_0=0, U_0={\mathbf 1}$), is
\beq
H&=&\sum_{x^1=-\frac{L_1}{2}}^{\frac{L_1}{2}}\sum_{x^2=-\frac{L_2}{2}}^{\frac{L_2}{2}}\sum_{j=1}^{2}\sum_{b=1}^{N^2-1}\frac{g_0^2}{2a}[l_j(x)_b]^2\nonumber\\
&&-\sum_{x^1=-\frac{L_1}{2}}^{\frac{L_1}{2}}\sum_{x^2=-\frac{L_2}{2}}^{\frac{L_2}{2}}\frac{1}{4g_0^2 a}[{\rm Tr}\, U^\square_{12}(x)+{\rm Tr}\,U^\square_{21}(x)],\label{latticebeforeaxial}
\eeq
where
\beq
U^\square_{jk}(x)=U_j(x)U_k(x+\hat{j}a)U_j(x+\hat{k}a)^\dag U_k(x)^\dag.\nonumber
\eeq

We will impose axial gauge for the lattice model, just as we did in the previous section for the continuum gauge theory.
We find the electric field component $l_1$ by solving Gauss's law (\ref{latticegauss}):
\beq
l_1(x^1,x^2)_b=\sum_{y^1=-\frac{L_1}{2}}^{x^1}[\mathcal{D}_2l_2(y^1,x^2)]_b.\label{latticeelectric}
\eeq
There is a global invariance left after the axial gauge fixing $U_{1}(x)={\mathbf 1}$:
\beq
\sum_{x^1=-\frac{L_1}{2}}^{\frac{L_1}{2}}[\mathcal{D}_2l_2(x^1,x^2)]_b\Psi=0.\label{residualgausslattice}
\eeq

The lattice Hamiltonian in axial gauge is found by substituting the new nonlocal expression for the electric field (\ref{latticeelectric}) into (\ref{latticebeforeaxial}):
\beq
H&=&\sum_{x^1=-\frac{L_1}{2}}^{\frac{L_1}{2}}\sum_{x^2=-\frac{L_2}{2}}^{\frac{L_2}{2}}\frac{g_0^2}{2a}[l_2(x)]^2\nonumber\\
&&-\sum_{x^1=-\frac{L_1}{2}}^{\frac{L_1}{2}}\sum_{x^2=-\frac{L_2}{2}}^{\frac{L_2}{2}}\frac{1}{2g_0^2 a}[{\rm Tr}\, U_2(x^1,x^2)^\dag U_2(x^1+a,x^2)+c.c.]\nonumber\\
&&-\frac{(g_0^\prime)^2}{2a}\sum_{x^1=-\frac{L_1}{2}}^{\frac{L_1}{2}}\sum_{y^1=-\frac{L_1}{2}}^{\frac{L_1}{2}}\sum_{x^2=-\frac{L_2}{2}}^{\frac{L_2}{2}}\vert x^1-y^1\vert\nonumber\\
&&\times[l_2(x^1,x^2)-\mathcal{R}_2(x^1,x^2-a)l_2(x^1,x^2-a)]\nonumber\\
&&\times[l_2(y^1,x^2)-\mathcal{R}_2(y^1,x^2-a)l_2(y^1,x^2-a)].\label{latticeafteraxial}
\eeq
The Hamiltonian (\ref{latticeafteraxial}) is the discretized version of (\ref{axialhamiltoniancontinuum}). Like (\ref{axialhamiltoniancontinuum}), (\ref{latticeafteraxial}) is nonlocal in $x^1$, and depends only on the transverse degrees of freedom $U_2,\,l_2$. In the following section we explore the longitudinal rescaling of coordinates on the Hamiltonian (\ref{latticeafteraxial}).

\section{Anisotropic Yang-Mills as an array of sigma models}
\setcounter{equation}{0}
\renewcommand{\theequation}{2.4.\arabic{equation}}

Next we find the effect of longitudinal rescaling, $x^{0,1}\to\lambda x^{0,1}$, $x^2\to x^2$ to a lattice gauge theory. The longitudinally-rescaled lattice has spacing $\lambda a$ in the $x^{0,1}$ directions and spacing $a$ in the $x^2$ direction. In the $\lambda\to0$ limit, it is sensible to treat $x^0$ and $x^1$ as continuous directions, and $x^2$ discrete.

Longitudinally rescaling the lattice Hamiltonian ({\ref{latticeafteraxial}), gives $H=H_0+\lambda^2H_1$, where
\beq
H_0&=&\sum_{x^1=-\frac{L_1}{2}}^{\frac{L_1}{2}}\sum_{x^2=-\frac{L_2}{2}}^{\frac{L_2}{2}}\frac{g_0^2}{2a}[l_2(x)]^2\nonumber\\
&&-\sum_{x^1=-\frac{L_1}{2}}^{\frac{L_1}{2}}\sum_{x^2=-\frac{L_2}{2}}^{\frac{L_2}{2}}\frac{1}{2g_0^2 a}[{\rm Tr}\, U_2(x^1,x^2)^\dag U_2(x^1+a,x^2)+c.c.],\nonumber\\
H_1&=&-\frac{(g_0^\prime)^2}{2a}\sum_{x^1=-\frac{L_1}{2}}^{\frac{L_1}{2}}\sum_{y^1=-\frac{L_1}{2}}^{\frac{L_1}{2}}\sum_{x^2=-\frac{L_2}{2}}^{\frac{L_2}{2}}\vert x^1-y^1\vert\nonumber\\
&&\times[l_2(x^1,x^2)-\mathcal{R}_2(x^1,x^2-a)l_2(x^1,x^2-a)]\nonumber\\
&&\times[l_2(y^1,x^2)-\mathcal{R}_2(y^1,x^2-a)l_2(y^1,x^2-a)].\nonumber
\eeq
Henceforth we drop the Lorentz index $2$ from $U_2$, $l_2$.

We treat $H_1$ as a perturbation. In the interaction representation, $U$ satisfies the Heisenberg equation of motion, $\partial_0 U=i[H_0,U]$. The solution of this equation of motion is
\beq
l(x^1,x^2)_b&=&\frac{ia}{g_0^2}{\rm Tr}\, t_b\partial_0U(x^1,x^2)U(x^1,x^2)^\dag,\nonumber\\
\mathcal{R}(x^1,x^2)_b^{\,\,c}l(x^1,x^2)_c&=&\frac{ia}{g_0^2}{\rm Tr}\,t_b U(x^1,x^2)^\dag\partial_0 U(x^1,x^2).\label{heisenbergevolution}
\eeq
Substituting (\ref{heisenbergevolution}) into $H_0$, and taking the continuum limit in the $x^1$  direction, we find
\beq
H_0&=&\sum_{x^2}H_0(x^2)=\sum_{x^2} \int dx^1\frac{1}{2g_0^2}\left\{\left[j_0^L(x^1,x^2)_b\right]^2+\left[j_1^L(x^1,x^2)_b\right]^2\right\}\nonumber\\
&=&\sum_{x^2} \int dx^1\frac{1}{2g_0^2}\left\{\left[j_0^R(x^1,x^2)_b\right]^2+\left[j_1^R(x^1,x^2)_b\right]^2\right\},\nonumber
\eeq
where
\beq
j_\mu^L(x)_b=i{\rm Tr}\, t_b \partial_\mu U(x) U(x)^\dag,\,\,j_\mu^R(x)_b=i{\rm Tr}\, t_b U(x)^\dag\partial_\mu U(x),\label{leftandright}
\eeq
where $\mu=0,1$.

We now note that $H_0(x^2)$ is the Hamiltonian of a (1+1)-dimensional principal chiral sigma model located at $x^2$. The principal chiral sigma model has the action
\beq
\mathcal{L}_{\rm PCSM}=\int d^2x \frac{1}{2g_0^2}\eta^{\mu\nu}{\rm Tr}\,\partial_\mu U^\dag \partial_\nu U.\label{actionpcsmone}
\eeq
This model has a global $SU(N)\times SU(N)$ symmetry given by the transformation $U(x)\to V^L U(x) V^R$, where $V^{L,R}\in SU(N)$. The Noether currents corresponding to these global symmetries are $j^{L,R}$ given in (\ref{leftandright}). The Hamiltonian corresponding to the action (\ref{actionpcsmone}) of a single principal chiral sigma model at fixed $x^2$ is $H_0(x^2)$. The unperturbed Hamiltonian, $H_0$, is an array of principal chiral sigma models, one at each value of $x^2$,
\beq
H_0=\sum_{x^2}H_0(x^2)=\sum_{x^2}H_{\rm PCSM}(x^2).\nonumber
\eeq

The residual Gauss's law, (\ref{residualgausslattice}) becomes
\beq
\int dx^1\left[ j_0^L(x^1,x^2)_b-j_0^R(x^1,x^2-a)_b\right]\Psi=0,\nonumber
\eeq
for each value of $x^2$, when $x^1$ is continuous.

Using (\ref{heisenbergevolution}), we write the interaction Hamiltonian $H_1$ in the continuous $x^1$ limit:
\beq
H_1&=&\sum_{x^2}\int dx^1\int dy^1\frac{1}{4g_0^2 a}\vert x^1-y^1\vert\nonumber\\
&&\times\left[j_0^L(x^1,x^2)-j_0^R(x^1,x^2-a)\right]\nonumber\\
&&\times\left[j_0^L(y^1,x^2)-j_0^R(y^1,x^2-a)\right].\label{perturbationhamiltonian}
\eeq
The Hamiltonian (\ref{perturbationhamiltonian}) couples adjacent sigma models, which allows particles to propagate in the $x^2$ direction. The coupling is suppressed in the $\lambda\to 0$ limit.

To summarize, longitudinally-rescaled Yang-Mills theory in $2+1$ dimensions consists of an array of principal chiral sigma models. The principal chiral sigma model is integrable and many exact results can be found. 

The next chapter is a brief review of integrable quantum field theories in $1+1$ dimensions. In Chapter 4, we find exact form factors of the principal chiral sigma model, guided by its integrability. In Chapters 5 and 6 we apply these results to Yang-Mills theories.

\chapter{Integrable quantum field theory and form factor axioms}
\setcounter{equation}{0}
\renewcommand{\theequation}{3.\arabic{equation}}

\section{Integrability in quantum field theories}
\setcounter{equation}{0}
\renewcommand{\theequation}{3.1.\arabic{equation}}


An integrable field theory has an infinite set of nontrivial local conserved charges, which are Lorentz tensors of increasing rank. We will call $Q_n$ a conserved charge of rank $n$ (with $n$ Lorentz indices). 

The existence of this set conserved charges has dramatic consequences in quantum field theory. In an integrable quantum field theory, all scattering is elastic. This means that in any scattering event the total number of particles is conserved, and furthermore, the set of their momenta is conserved. All scattering in an integrable quantum field theory is factorizable. A many-particle S-matrix can be broken down into a product of two-particle scatterings. The different ways one can factorize an S-matrix must be equivalent. From this follows that the two-particle S-matrix satisfies the Yang-Baxter equation (which we will discuss in more detail in the next section).

These restrictions, along with other physical considerations (like unitarity of the S-matrix, crossing symmetry, and analyticity), allow us to find the two-particle S-matrix exactly in many integrable field theories. In the rest of this section we briefly explain how elasticity and factorization follow from the existence of the set of charges $Q_n$.


We now examine the scattering of $l$ particles into $k$ particles in a generic integrable quantum field theory. A state with one particle in 1+1 dimensions is characterized by its two-momentum, $p$, and some set of quantum numbers $c$ (these could include color, isospin, flavor, etc.). We define a state with $l$ incoming particles as
\beq
\vert l, \{p\}, \{c\}\rangle_{\rm in}=\vert p_1, c_1;p_2,c_2;\dots;p_l,c_l\rangle_{\rm in},\nonumber
\eeq
where we assume that the incoming particles are widely separated in the infinite past. We can similarly define the $k$-particle outgoing state
 \beq
 \vert k, \{p^\prime\},  \{c^\prime\}\rangle_{\rm out}=\vert p^\prime_1, c_1^\prime;p^\prime_2,c^\prime_2;\dots;p^\prime_l,c_l^\prime\rangle_{\rm out}.\nonumber
\eeq
 Acting with the conserved charge $Q_n$ in the incoming and outgoing states yields
\beq
Q_n\vert l, \{p\}, \{c\}\rangle_{\rm in}&=&\sum_{i=1}^l q_n(p_i)^{c^\prime_i}_{c_i}\vert l, \{p\}, \{c^\prime\}\rangle_{\rm in},\nonumber\\
Q_n\vert k, \{p^\prime\}, \{c\}\rangle_{\rm out}&=&\sum_{i=1}^kq_n(p^\prime_i)^{c^\prime_i}_{c_i}\vert k, \{p^\prime\}, \{c^\prime\}\rangle_{\rm out},\nonumber
\eeq
where $q_n(p_i)$ is some polynomial of rank $n$ of the energy and momentum of the $i$-th particle. The charges $Q_n$ are locally conserved, which means that we can add the individual contributions from each particle $q_n$ if the particles are widely separated. The fact that $Q_n$ is conserved implies
\beq
\sum_{i=1}^l q_n(p_i)^{c^\prime_i}_{c_i}=\sum_{i=1}^kq_n(p^\prime_i)^{c^\prime_i}_{c_i}.\label{polynomial}
\eeq
If there is an infinite number of charges $Q_n$, there is an infinite number of polynomial equations (\ref{polynomial}) of increasing rank that need to be satisfied by only $l+k$ variables. The only way way these infinite set of conditions can be satisfied is if $l=k$, and the set of momenta satisfies $\{p\}=\{p^\prime\}$. Thus all scattering events are elastic.

To verify the factorizability of the S-matrix, we examine the Fourier transform of the incoming state:
\beq
\vert l, \{x\}, \{c\}\rangle_{\rm in}=\left(\prod_{i=1}^{l} dp_i e^{ip_i(x_i-x_i^0)}\right)\vert l, \{p\}, \{c\}\rangle_{\rm in},\nonumber
\eeq
where the center of mass of the $i$-th particle is located at $x_i^0$. We can now operate on the state $\vert l, \{x\}, \{c\}\rangle_{\rm in}$ with the operator $e^{i\epsilon Q_n}$:
\beq
e^{i\epsilon Q_n}\vert l, \{x\}, \{c\}\rangle_{\rm in}=\left(\prod_{i=1}^l e^{i\epsilon  q_n(p_i)}\right)\vert l, \{x\}, \{c\}\rangle_{\rm in}.\label{centershifted}
\eeq
The operator, $e^{i\epsilon Q_n}$, commutes with the Hamiltonian, so it does not alter the S-matrix.
In the incoming state from Eq. (\ref{centershifted}), the $i$-th particle is now centered at $x_i^0-\epsilon(dq_n(p_i)/d p_i)$. The shift in the position of each particle depends on its momentum. By adjusting the parameter $\epsilon$, we can change the position of each particle independently. By applying the operator $e^{i\epsilon Q_n}$, we change the order in which particles interact. We can break any scattering event into multiple two-particle scatterings, and we can exchange the order of the two-particle scatterings without affecting the total S-matrix. The S-matrix is therefore factorizable.

In the next section, we show how the elasticity and factorizability, along with unitarity, crossing symmetry and analyticity can be used to find the exact S-matrix in integrable quantum field theories.

\section{Exact S-matrices}
\setcounter{equation}{0}
\renewcommand{\theequation}{3.2.\arabic{equation}}


As we discussed in the previous section, all scattering in an integrable field theory is elastic. The S-matrix is nonzero only if there is an equal number of incoming and outgoing particles, and the set of incoming and outgoing momenta is the same. The S-matrix with $l$ incoming and $l$ outgoing particles is called the $l$-particle S-matrix.

S-matrices are factorizable. For the three-particle S-matrix, this means
\beq
S_{PPP}(\theta_1,\theta_2,\theta_3)^{c_1^\prime\,c_2^\prime\,c_3^\prime}_{c_1\,c_2\,c_3}&=&S_{PP}(\theta_1, \theta_2)^{c_1^{\prime\prime}\,c_2^{\prime\prime}}_{c_1\,c_2} S_{PP}(\theta_1,\theta_3)^{c_1^\prime\,c_3^{\prime\prime}}_{c_1^{\prime\prime} \, c_3}S_{PP}(\theta_2,\theta_3)^{c_2^\prime \, c_3^\prime}_{c_2^{\prime\prime}\,c_3^{\prime\prime}}
\nonumber\\
&=&S_{PP}(\theta_2,\theta_3)^{c_2^{\prime\prime} \, c_3^{\prime\prime}}_{c_2\,c_3} S_{PP}(\theta_1,\theta_3)^{c_1^{\prime\prime}\,c_3^\prime}_{c_1 \, c_3^{\prime\prime}}S_{PP}(\theta_1, \theta_2)^{c_1^\prime\,c_2^\prime}_{c_1^{\prime\prime}\,c_2^{\prime\prime}},\nonumber
\\
\label{yangbaxterequation}
\eeq
Where $c_i$ is the set of quantum numbers of the $i$-th particle, and $\theta_i$ is the rapidity of the $i$-th particle, defined by its energy and momentum:  $E_i=m\cosh \theta_i$, $p_i=m\sinh\theta_i$. Equation (\ref{yangbaxterequation}) is called the Yang-Baxter equation.

 The S-matrix (and time evolution) is unitary. For the two-particle S-matrix, this implies
\beq
S_{PP}(\theta_2, \theta_1)^{c_2^{\prime\prime}\,c_1^{\prime\prime}}_{c_2\,c_1}S_{PP}(\theta_1, \theta_2)^{c_1^\prime\,c_2^\prime}_{c_1^{\prime\prime}\,c_2^{\prime\prime}}=\delta^{c_1^\prime}_{c_1}\delta^{c_2^\prime}_{c_2}.\label{unitaritycondition}
\eeq
The S-matrix should also be invariant under crossing symmetry. We can turn an incoming particle into an outgoing antiparticle by shifting is rapidity as $\theta\to\theta+\pi i$ (and an outgoing particle can be turned into an incoming antiparticle). The antiparticle-particle S-matrix, $S_{AP}$, can  be found from the particle-particle S-matrix by crossing. This is
\beq
S_{AP}(\theta)^{c_1^{\prime}\,c_2^{\prime}}_{c_1\,c_2}=S_{PP}(\pi i-\theta)^{c_1^{\prime}\,c_2^{\prime}}_{c_1\,c_2},\label{crossingsymmetry}
\eeq
where $\theta=\theta_1-\theta_2$.

The spectrum of a  quantum field theory might have bound states of elementary particles. The bound state mass, $m_B$, of a particle composed of two particles of mass $m_1$ and $m_2$ is given by
\beq
m_B=\sqrt{m_1^2+m_2^2+2m_1 m_2\cos \eta},\,\,\,(0<\eta<\pi),\label{fusionangle}
\eeq
where $\eta$ is the fusion angle. One can define the S-matrix for a bound state with an elementary particle, $S_{BP}$ from the two-particle S-matrix by
\beq
S_{BP}(\theta_B,\theta_1)^{d\,c_3^\prime}_{d^\prime\,c_3}\,\Gamma^{d^\prime}_{c_1\,c_2}=\Gamma^d_{c_1^\prime\,c_2^\prime}\,S_{PP}(\theta_{1},\theta_3)^{c_1^\prime\,c_3^\prime}_{c_1\,c_3^{\prime\prime}} S_{PP}(\theta_2,\theta_3)^{c_2^\prime c_3^{\prime\prime}}_{c_2\,c_3},\label{bootstrapequation}
\eeq
where $d$ are the quantum numbers of the bound state, and $\Gamma^d_{c_1\,c_2}$ is defined by
\beq
i{\rm Res}\vert_{\theta=i\eta}\,S_{PP}(\theta)^{c_1^\prime\,c_2^\prime}_{c_1\,c_2}=\Gamma^{c_1^\prime\,c_2^\prime}_d\,\Gamma^d_{c_1\,c_2}.\label{intertwiner}
\eeq

The two-particle S-matrix for can be found up to a CDD ambiguity \cite{cdd} by solving equations (\ref{yangbaxterequation}), (\ref{unitaritycondition}), (\ref{crossingsymmetry}) and (\ref{bootstrapequation}), for particles described by the quantum numbers $\{c\}$ and bound states described by the numbers $d$. The CDD ambiguity means that the two-particle S-matrix still satisfies the conditions (\ref{yangbaxterequation}), (\ref{unitaritycondition}), (\ref{crossingsymmetry}) and (\ref{bootstrapequation}) if we multiply with it with a factor 
\beq
\prod_k\frac{\sinh \theta+i\sin\alpha_k}{\sinh\theta-i\sin\alpha_k},\label{cddgeneral}
\eeq
for some set of numbers $\{\alpha\}$. This ambiguity can be fixed by requiring that the S-matrix be maximally analytic. The S-matrix should not have any poles in the physical strip, $0<{\rm Im} \theta<\pi$, that do not correspond to bound state fusion angles. One can multiply the S-matrix by factors of the form (\ref{cddgeneral}) to push any unnecessary poles outside the physical strip, obtaining the maximally analytic S-matrix.

This approach for finding S-matrices for integrable theories was first developed in References \cite{STW},\cite{KTTW},\cite{KT},\cite{BKW}, \cite{zamolodchikov}. The exact two-particle S-matrix of the principal chiral sigma model was found in Reference \cite{wiegmann}. In the next section we show how the exact S-matrix can be used to find form factors of local operators, and correlation functions.

\section{The form factor bootstrap program}
\setcounter{equation}{0}
\renewcommand{\theequation}{3.3.\arabic{equation}}

The $l$-excitation form factor of a local operator $\mathcal{O}(x)_{c_0}$ is defined as
\beq
F^{\mathcal{O}}(\theta_1,\dots,\theta_l)_{\{c\}}=\langle 0\vert \mathcal{O}(0)_{c_0}\vert \theta_1,c_1;\theta_2,c_2;\dots;\theta_l,c_l\rangle_{\rm in},\label{formfactorgeneral}
\eeq
where $c_0$ is the set of Lorentz, color, or flavor indices of the operator, and $\theta_1>\theta_2>\dots>\theta_l$. Other orderings of the rapidities can be defined by analytic continuation of the form factor.

The two-particle form factors were first studied in References \cite{vergeles}, \cite{karowski}. Generalizations for form factors with many particles were discussed in \cite{smirnovkirillov}. 
Form factors of integrable theories satisfy a set of conditions called the ``Smirnov axioms", \cite{smirnov},  which require knowledge of the exact two-excitation S-matrix. These axioms provide such strong restrictions, that they yield the exact form factors. Once all the form factors of a local operator are known, one can find correlation functions by inserting complete sets of intermediate states between these operators. This approach of finding the S-matrix, using it to calculate form factors, and using these form factors to calculate correlation functions is called the integrable bootstrap program.

In the rest of these section we list and explain briefly the Smirnov axioms. These are the scattering axiom, periodicity axiom, Lorentz invariance axiom, annihilation pole axiom, bound state axiom, and minimality axiom.

We will first discuss the so-called scattering axiom (also known as Watson's theorem). The scattering axiom states that exchanging two of the incoming excitations in (\ref{formfactorgeneral}) is equivalent to scattering them, explicitly,
\beq
&&F^{\mathcal{O}}(\theta_1,\dots,\theta_j,\theta_i,\dots, \theta_l)_{c_0 c_1\dots c_j c_i \dots c_l}\nonumber\\
&&\,\,\,\,\,\,\,\,\,\,\,\,\,\,\,\,\,\,\,\,\,\,\,\,=S(\theta_i-\theta_j)_{c_j c_i}^{c^\prime_i c^\prime_j}F^{\mathcal{O}}(\theta_1,\dots, \theta_i, \theta_j,\dots, \theta_l)_{c_0 c_1\dots c_i c_j \dots c_l}.\label{scatteringaxiomfirst}
\eeq
One can keep applying the scattering axiom on the form factor to exchange the order of any two particles. The form factor with any order of rapidities can be found by multiplying many times with the S-matrix.

Next we look at the periodicity axiom. This axiom follows from crossing symmetry in two dimensions. By crossing symmetry, one can turn the $l-th$ incoming particle (or antiparticle) in (\ref{formfactorgeneral}) into an outgoing antiparticle (or particle), by shifting $\theta_l\to \theta_l-\pi i$. This way, if we know the form factor (\ref{formfactorgeneral}), we can find the form factors of an operator with outgoing excitations. The periodicity axiom states that the outgoing antiparticle (or particle) can be turned into the first incoming excitation, by shifting its rapidity again by $-\pi i$. Explicitly, this axiom states:
\beq
F^{\mathcal{O}}(\theta_1,\dots,\theta_l)_{c_0 c_1\dots c_l}&=&F^{\mathcal{O}} (\theta_l-2\pi i,\theta_1,\dots, \theta_{l-1})_{c_l c_1 \, c_{l-1}}\nonumber\\
&=&F^{\mathcal{O}}(\theta_{l-1}-2\pi i,\theta_l-2\pi i,\theta_1,\dots,\theta_{l-2})_{c_0 c_{l-1} c_l c_1\dots c_{l-2}}\nonumber\\
&=&\dots\,\,\,.\label{periodicityaxiomfirst}
\eeq

The Lorentz invariance axiom restricts how the form factor (\ref{formfactorgeneral}) transforms under a Lorentz boost. A Lorentz boost in (\ref{formfactorgeneral}) means shifting all the rapidities by $\theta_i\to\theta_i +\alpha$, for some constant $\alpha$. If the Operator $\mathcal{O}$ has spin $s$, the form factor transforms as
\beq
F^{\mathcal{O}}(\theta_1+\alpha,\dots,\theta_l+\alpha)_{\{c\}}=e^{s\alpha} F^{\mathcal{O}}(\theta_1,\dots,\theta_l)_{\{c\}}.\label{lorentzaxiom}
\eeq
If the operator $\mathcal{O}$ is a Lorentz scalar, then the form factor is invariant under a boost. By Lorentz invariance, we can also write the $x$-dependent form factor:
\beq
\langle 0\vert \mathcal{O}(x)_{c_0}\vert \theta_1,c_1;\theta_2,c_2;\dots;\theta_l,c_l\rangle_{\rm in}=e^{ix\cdot(p_1+\cdots+p_l)}F^{\mathcal{O}}(\theta_1,\dots,\theta_l)_{\{c\}}.\nonumber
\eeq

There exists a possibility that a particle with rapidity $\theta_i$ and a neighboring antiparticle with rapidity $\theta_i+1$ in the incoming state in (\ref{formfactorgeneral}) annihilate before reaching the operator $\mathcal{O}$. The form factor must then have an kinematic pole at $\theta_i-\theta_j=-\pi i$.
The annihilation-pole axiom states that the residue of the $l$-particle form factor at this pole is proportional to the $(l-2)$-particle form factor. Precisely, this axiom states:
\beq
&&{\rm Res}\vert_{\theta_{l-1}-\theta_{l}=-\pi i}F^{\mathcal{O}}(\theta_1,\dots,\theta_l)_{\{c\}}=2 i F^{\mathcal{O}}(\theta_1,\dots,\theta_{l-2})_{c_0 c^\prime_1\dots c^\prime_{l-2}}\delta_{c_{l}c^\prime_{l-1}}\nonumber\\
&&\times\left(\delta_{c_1 c^\prime_1}\dots\delta_{c_{l-2}c^\prime_{l-2}}\delta_{c_{l-1} c_{l-1}^\prime}-S(\theta_1-\theta_{l-1})^{c^{\prime}_{l-1} c^\prime_1}_{c^{\prime\prime}_1 c_1} S(\theta_2-\theta_{l-1})^{c^{\prime\prime}_1 c^{\prime}_2}_{c^{\prime\prime}_2 c_2}\times\cdots\right.\nonumber\\
&&\left.\,\,\,\,\,\,\,\,\,\,\,\,\times S(\theta_i-\theta_{l-1})^{c^{\prime\prime}_{i-1} c^\prime_{i}}_{c^{\prime\prime}_i c_i}\times\cdots\times S(\theta_{l-2}-\theta_{l-1})^{c^{\prime\prime}_{l-3} c^\prime_{l-2}}_{c_{l-1} c_{l-2}}\right),\label{annihilationpolefirst}
\eeq
where the right hand of (\ref{annihilationpolefirst}) side includes the $(l-2)$-particle form factor. The product of S-matrices in the right hand side of (\ref{annihilationpolefirst}) accounts for the possibility that before the $(l-1)$-st particle (or antiparticle), annihilates with the $l$-th antiparticle (or particle), it can scatter with the rest of the $(l-2)$ excitations. The annihilation-pole axiom can be used as a recursion relation. Once the $l$-particle form factor is known, this axiom can be used to find information about the $(l+2)$-particle form factor, and so on.

As we discussed in the previous section, an integrable theory, may have bound states of elementary particles, with masses given by (\ref{fusionangle}). If the $i$-th and $(i+1)$-st excitations in Eq. (\ref{formfactorgeneral}) can form a bound state, the form factor must have a pole at $\theta_i-\theta_{i+1}=i\eta$, where $\eta$ is the fusion angle defined in (\ref{fusionangle}). One can use this to find information about form factors with incoming bound-state particles. The bound-state axiom states that 
\beq
&&{\rm Res}\vert_{\theta_1-\theta_2=i\eta}F^{\mathcal{O}}(\theta_1,\dots,\theta_l)_{c_0 c_1\dots c_l}\nonumber\\
&&\,\,\,\,\,\,\,\,\,\,\,\,\,=\sqrt{2}\Gamma^d_{c_1 c_2}F^{\mathcal{O}}(\theta^B_{(12)},\theta_3,\dots,\theta_l)_{c_0 d \,c_3\dots c_l},\label{boundstateaxiomfirst}
\eeq
where the right-hand side includes the form factor with an incoming bound state, $\theta^B_{(12)}$ is the rapidity of the bound state, and $\Gamma^d_{c_1c_2}$ is defined in Eq. (\ref{intertwiner}).

The axioms we have discussed in Equations (\ref{scatteringaxiomfirst}), (\ref{periodicityaxiomfirst}), (\ref{lorentzaxiom}), (\ref{annihilationpolefirst}), and (\ref{boundstateaxiomfirst}), are enough to determine the form factor up to an ambiguity, similar to the CDD ambiguity in the S-matrix. If we find a form factor, $F^{\mathcal{O}}_{\rm minimal}(\theta_1,\dots,\theta_l)_{\{c\}}$, which satisfies all the axioms we have mentioned, these axioms are also satisfied by 
\beq
F^{\mathcal{O}}(\theta_1,\dots,\theta_l)_{\{c\}}=\frac{P_l[\{\cosh(\theta_i-\theta_j)\}]}{Q_l[\{\cosh(\theta_i-\theta_j)\}]}F^{\mathcal{O}}_{\rm minimal}(\theta_1,\dots,\theta_l)_{\{c\}},\label{formambiguity}
\eeq
where $P_l$, and $Q_l$ are symmetric polynomials. The minimality axiom states that the form factor must be maximally analytic. Form factors should have no poles that are not  either kinematic annihilation poles, or bound state poles in the physical strip. If a form factor is found which has additional, unphysical poles, one can use the ambiguity (\ref{formambiguity}) to push these outside the physical strip. The physical form factor is then the one with the minimum number of poles.

Once all the exact form factors of an operator are known, they can be use to calculate non-time-ordered correlation functions. The simplest case is the two-point function of the operator $\mathcal{O}$, defined as
\beq
W^\mathcal{O}(x)_{c_0d_0}=\langle 0\vert \mathcal{O}(x)_{c_0} \mathcal{O}(0)_{d_0}\vert 0\rangle.\nonumber
\eeq
The function $W^{\mathcal{O}}(x)_{c_0d_0}$ can be computed by inserting a complete set of intermediate states between the two operators. The correlation function is then expressed as a sum over all the form factors:
\beq
W^\mathcal{O}(x)_{c_0d_0}&=&\sum_{l=1}^{\infty}\frac{1}{l!}\int\prod_{j=1}^{l}\frac{d \theta_j}{4\pi}e^{-ix\cdot(p_1+\cdots+p_l)}\nonumber\\
&&\times\langle 0\vert \mathcal{O}(0)_{c_0}\vert \theta_1, c_1,\dots,\theta_l, c_l\rangle_{\rm in}\langle 0 \vert\mathcal{O}(0)_{d_0}\vert \theta_1, c_1,\dots,\theta_l, c_l\rangle^*_{\rm in}\nonumber\\
&=&\sum_{l=1}^{\infty}\frac{1}{l!}\int\prod_{j=1}^{l}\frac{d \theta_j}{4\pi}e^{-ix\cdot(p_1+\cdots+p_l)}\nonumber\\
&&\times F^{\mathcal{O}}(\theta_1,\dots,\theta_l)_{c_0 c_1 \dots c_l}\times \left[ F^{\mathcal{O}}(\theta_1,\dots,\theta_l)_{d_0 c_1 \dots c_l}\right]^*.\nonumber
\eeq
Correlation functions of more than two operators can still be found by inserting a complete set of states between each pair of operators. One would then need the form factors with both incoming and outgoing particles, which can be found using crossing symmetry.

In the next chapter we use the integrable bootstrap program to find form factors and correlation functions of the principal chiral sigma model. We  show how to apply these new exact results to Yang-Mills theories.

\chapter{Exact form factors of Noether current and energy-momentum tensor of the principal chiral sigma model}

\section{The principal chiral sigma model}
\setcounter{equation}{0}
\renewcommand{\theequation}{4.1.\arabic{equation}}

This chapter contains material previously published in \cite{multi particle} and \cite{correlation}.

The principal chiral sigma model has the action
\beq
S_{\rm PCSM}=\int d^2x \frac{1}{2g_0^2} \eta^{\mu\nu} \,{\rm Tr}\,\partial_\mu U^\dag(x) \partial_\nu U(x),\label{actionpcsm}
\eeq
where the field $U(x)$ is in the fundamental representation of $SU(N), \,\mu,\nu=0,1,$ and $\eta^{00}=-\eta^{11}=1,\,\eta^{01}=\eta^{10}=0.$ The action (\ref{actionpcsm}) has a global $SU(N)\times SU(N)$ symmetry, given by the transformation $U(x)\to V_L U(x) V_R$, where $V_{L,R}\in SU(N)$. The Noether currents associated with these symmetries are 
\begin{eqnarray}
j_\mu^L(x)_a^c&=&\frac{-iN}{2g^2}\partial_\mu U_{ab}(x)U(x)^{\dag\,bc}(x),\nonumber\\
j_\mu^R(x)^d_b&=&\frac{-iN}{2g^2}U^{\dag\,da}(x)\partial_\mu U_{ab}(x),\label{noether}
\end{eqnarray}
respectively, where we have included the color indices, $a,b,c,d=1,\dots, N$, explicitly.

This model is asymptotically free, and has a mass gap, which we call $m$. It has been argued that this mass gap is generated by non-perturbative saddle points of the path integral \cite{unsal}. The quantum integrability of this model was shown in References \cite{integrablePCSM} and \cite{wiegmann2}.

The sigma model has elementary particles of mass $m$ which carry both left and right colors. These elementary particles form bound states that obey a sine formula \cite{schroer}
\beq
m_r=m\frac{\sin\left(\frac{\pi r}{N}\right)}{\sin\left(\frac{\pi}{N}\right)},\,\,\,\,\,\,r=1,\dots,N-1, \label{massspectrum}
\eeq
where $m_r$ is the mass of an $r$-particle bound state. In the large-$N$ limit, the mass of an $r$-particle bound state is $m_r=m\,r$, for finite $r$. This means that there are no bound states of a finite number of elementary particles in the planar limit, since the binding energy vanishes.This is the 't~Hooft large-$N$ limit, where the mass gap $m$ is fixed as $N$ goes to infinity. An alternative large-$N$ limit has been examined in References \cite{kazakov} where the mass $m_{N-1}$ is kept fixed and the mass gap goes to zero as $N$ goes to infinity.'t~Hooft

We introduce particle and antiparticle creation operators $\mathfrak{A}_P^{\dag}(\theta)_{ab}$ and $\mathfrak{A}_A^{\dag}(\theta)_{ba}$, respectively, where $\theta$ is the particle rapidity, defined in terms of the momentum vector by $p_0=m\cosh\,\theta, p_1=m\sinh\,\theta$, and $a,\,b=1,\dots,N$ are left and right color indices, respectively. A state with many particles is created by acting on the vacuum with a product of creation operators in order of increasing rapidity, from left to right,
\beq
\vert P, \theta_1,a_1,b_1; A,\theta_2,b_2,a_2;\dots\rangle_{\rm in}=\mathfrak{A}_P^{\dag}(\theta_1)_{a_1b_1}\mathfrak{A}_A^{\dag}(\theta_2)_{b_2a_2}\dots\vert\,0\rangle,\label{multiparticlestate}
\eeq
where $\theta_1>\theta_2>\dots\,.$

The S-matrix, $S_{PP}(\theta)_{a_1 b_1;a_2b_2}^{c_2d_2;c_1d_1}$ of two particles with incoming rapidities $\theta_1$ and $\theta_2$, and outgoing rapidities $\theta_1^\prime$ and $\theta_2^\prime$ is defined by
\beq
\,_{\rm out}\langle P, \theta_1^\prime,c_1,d_1; A,\theta_2^\prime,d_2,c_2\!\!\!\!\!&\vert&\!\!\!\!\! P, \theta_1,a_1,b_1; A,\theta_2,b_2,a_2\rangle_{\rm in}\nonumber\\
&=&S_{PP}(\theta)_{a_1 b_1;a_2b_2}^{c_2d_2;c_1d_1}\,4\pi\,\delta(\theta_1^\prime-\theta_1)\,4\pi\,\delta(\theta_2^\prime-\theta_2),\nonumber
\eeq
where $\theta=\theta_1-\theta_2$. This S-matrix was found exactly in Reference \cite{wiegmann} to be
\beq
S_{PP}(\theta)_{a_1 b_1;a_2b_2}^{c_2d_2;c_1d_1}=\chi(\theta) \,S_{\rm CGN}(\theta)_{a_1;a_2}^{c_2;c_1}\,S_{\rm CGN}(\theta)_{b_1;b_2}^{d_2d_1},\label{smatrix}
\eeq
where $S_{\rm CGN}$ is the S-matrix of two elementary excitations of the $SU(N)$ chiral Gross-Neveu model \cite{berg}, \cite{kurak}:
\beq
S_{\rm CGN}(\theta)_{a_1a;a_2}^{c_2c_1}=\frac{\Gamma(i\theta/2\pi+1)\Gamma(-i\theta/2\pi-1/N)}{\Gamma(i\theta/2\pi+1-1/N)\Gamma(-i\theta/2\pi)}\left(\delta_{a_1}^{c_1}\delta_{a_2}^{c_2}-\frac{2\pi i}{N\theta}\delta_{a_2}^{c_1}\delta_{a_1}^{c_2}\right),\nonumber
\eeq
and $\chi(\theta)$ is the CDD factor \cite{cdd}:
\beq
\chi(\theta)=\frac{\sinh\left(\frac{\theta}{2}-\frac{\pi i}{N}\right)}{\sinh\left(\frac{\theta}{2}+\frac{\pi i}{N}\right)}. \label{cdd}
\eeq

The particle-antiparticle S-matrix is related to the particle-particle S-matrix by crossing symmetry, i.e. $\theta\to\hat{\theta}=\pi i-\theta$. The S-matrix for a particle with incoming rapidity $\theta_1$ and outgoing rapidity $\theta_1^\prime$ and an antiparticle with incoming rapidity $\theta_2$ and outgoing rapidity $\theta_2^\prime$ is
\beq
S_{AP}(\theta)_{a_1b_1;b_2a_2}^{d_2c_2;c_1d_1}&=&S(\hat{\theta},N)\left[\delta_{a_1}^{c_1}\delta_{a_2}^{c_2}\delta_{b_1}^{d_1}\delta_{b_2}^{d_2}\right.\nonumber\\
&&\left.-\frac{2\pi i}{N\hat{\theta}}\left(\delta_{a_1a_2}\delta^{c_1c_2}\delta_{b_2}^{d_2}\delta_{b_1}^{d_1}+\delta_{a_2}^{c_2}\delta_{a_1}^{c_1}\delta_{b_1b_2}\delta^{d_1d_2}\right)\right.\nonumber\\
&&\left.-\frac{4\pi^2}{N^2\hat{\theta}^2}\delta_{a_1a_2}\delta^{c_1c_2}\delta_{b_1b_2}\delta^{d_1d_2}\right],\nonumber
\eeq
where
\beq
S(\theta,N)&=&\frac{\sinh\left(\frac{\theta}{2}-\frac{\pi i}{N}\right)}{\sinh\left(\frac{\theta}{2}+\frac{\pi i}{N}\right)}\left[\frac{\Gamma(i\theta/2\pi+1)\Gamma(-i\theta/2\pi-1/N)}{\Gamma(i\theta/2\pi+1-1/N)\Gamma(-i\theta/2\pi)}\right]^2\nonumber\\
&=&1+\mathcal{O}\left(\frac{1}{N^2}\right).\label{esfunction}
\eeq

The creation operators satisfy the Zamolodchikov algebra:
\beq
\mathfrak{A}_P^\dag(\theta_1)_{a_1b_1}\mathfrak{A}_P^\dag(\theta_2)_{a_2b_2}&=&S_{PP}(\theta)_{a_1b_1;a_2b_2}^{c_2d_2;c_1d_1}\mathfrak{A}_P^\dag(\theta_2)_{c_2d_2}\mathfrak{A}_P^\dag(\theta_1)_{c_1d_1},\nonumber\\
\mathfrak{A}_A^\dag(\theta_1)_{b_1a_1}\mathfrak{A}_A^\dag(\theta_2)_{b_2a_2}&=&S_{AA}(\theta)_{b_1a_1;b_2a_2}^{d_2c_2;d_1c_1}\mathfrak{A}_A^\dag(\theta_2)_{d_2c_2}\mathfrak{A}_A^\dag(\theta_1)_{d_1c_1},\nonumber\\
\mathfrak{A}_P^\dag(\theta_1)_{a_1b_1}\mathfrak{A}_A^\dag(\theta_2)_{b_2a_2}&=&S_{AP}(\theta)_{a_1b_1;b_2a_2}^{d_2c_2;c_1d_1}\mathfrak{A}_A^\dag(\theta_2)_{d_2c_2}\mathfrak{A}_P^\dag(\theta_1)_{c_1d_1}.\label{zamolodchikov}
\eeq

In the next section we will use the Smirnov form factor axioms to calculate the two-particle form factors of the Noether current operators (\ref{noether}) at finite and infinite $N$. We calculate explicitly the four-particle form factors at large $N$. We later find a general expression for the form factor with any number of particles. The two-point correlation function of two Noether currents is found by summing over all the form factors. The same procedure is later repeated for the energy-momentum tensor operator. This combination of the integrable bootstrap and the large-$N$ limit was also used by P. Orland in References \cite{renormalizedfield} to calculate all the form factors of the renormalized field operator.

\section{The two-particle form factor}
\setcounter{equation}{0}
\renewcommand{\theequation}{4.2.\arabic{equation}}

In this section, we calculate the first non vanishing form factor of the current operators. We will discuss only the left-handed current $j_\mu^L(x)_a^c$ in detail, since the same method yields the right-handed-current form factor.

Under a global $SU(N)\times SU(N)$ transformation, the current and the particle and antiparticle creation operators transform as 
\beq
j_\mu^L(x)\to V_Lj_\mu^L(x)V_L^\dag,\,\,\,&&
\mathfrak{A}_P^\dag(\theta)\to V_R^\dag\mathfrak{A}_P^\dag(\theta)V_L^\dag,\nonumber\\
\mathfrak{A}_A^\dag(\theta)&\to&V_R\mathfrak{A}_A^\dag(\theta)V_L^,\nonumber.
\eeq
Only form factors with an equal number of particles and antiparticles are invariant under such global transformations. 

The first nontrivial form factor is 
\beq
\langle 0\vert  j_\mu^L(x)_{a_0c_0}\!\!\!\!\!&\vert&\!\!\!\!\! A,\theta_1,b_1,a_1;P,\theta_2,a_2,b_2\rangle_{\rm in}=\langle 0\vert j_\mu^L(x)_{a_0c_0} \mathfrak{A}_A^\dag(\theta_1)_{b_1a_1}\mathfrak{A}_P^{\dag}(\theta_2)_{a_2b_2}\vert 0\rangle\nonumber\\
&=&e^{-ix\cdot(p_1+p_2)}F_\mu(\theta)_{a_0a_1a_2c_0;b_1b_2},\label{twoparticleform}
\eeq
for $\theta_1>\theta_2$.
The function $F_\mu(\theta)_{a_0a_1a_2c_0;b_1b_2}$ is restricted by the conservation and the tracelessness of the Noether current, such that
\beq
(p_1+p_2)^\mu F_\mu(\theta)_{a_0a_1a_2c_0;b_1b_2}=0,\label{conservation}
\eeq
and 
\beq
\delta^{a_0c_0}F_\mu(\theta)_{a_0a_1a_2c_0;b_1b_2}=0.\label{tracelessness}
\eeq
The condition (\ref{conservation}) means that we can write
\beq
F_\mu(\theta)_{a_0a_1a_2c_0;b_1b_2}&=&(p_1-p_2)_\mu F(\theta)_{a_0a_1a_2c_0;b_1b_2}\nonumber\\
&=&-\epsilon_{\mu\nu}(p_1+p_2)^\nu \tanh\left(\frac{\theta}{2}\right) F(\theta)_{a_0a_1a_2c_0;b_1b_2}.
\eeq
The condition (\ref{conservation}) is satisfied by choosing
\beq
F(\theta)_{a_0a_1a_2c_0;b_1b_2}=F(\theta)\left(\delta_{a_0a_2}\delta_{c_0a_1}\delta_{b_1b_2}-\frac{1}{N}\delta_{a_0c_0}\delta_{a_1a_2}\delta_{b_1b_2}\right).\nonumber
\eeq
For $\theta_2>\theta_1$, we define the form factor 
\beq
\langle 0\vert  j_\mu^L(x)_{a_0c_0}\!\!\!\!\!&\vert&\!\!\!\!\! P,\theta_2,a_2,b_2;A,\theta_1,b_1,a_1\rangle_{\rm in}\nonumber\\
&=&(p_1-p_2)_\mu e^{-ix\cdot(p_1+p_2)}F^\prime(\theta)\left(\delta_{a_0a_2}\delta_{c_0a_1}\delta_{b_1b_2}-\frac{1}{N}\delta_{a_0c_0}\delta_{a_1a_2}\delta_{b_1b_2}\right).\nonumber
\eeq

Next we further restrict the functions $F(\theta),\,F^\prime(\theta)$ by applying the Smirnov form factor axioms. First we apply the scattering axiom. The scattering axiom follows from the Zamolodchikov algebra (\ref{zamolodchikov}). This axiom implies
\beq
\langle 0\vert j_\mu^L(0)_{a_0c_0}\!\!\!\!\!&&\!\!\!\!\!\mathfrak{A}_{P}^\dag(\theta_2)_{a_2b_2}\mathfrak{A}_{A}^\dag(\theta_1)_{b_1a_1}\vert 0\rangle\nonumber\\
&&=S_{AP}(\theta)_{a_2b_2;b_1a_1}^{d_1c_1;c_2d_2}\langle 0\vert j_\mu^L(0)_{a_0c_0}\mathfrak{A}_A^\dag(\theta_1)_{d_1c_1}\mathfrak{A}_P^\dag(\theta_2)_{c_2d_2}\vert 0\rangle,\label{twoscattering}
\eeq
or
\beq
F^\prime(\theta)=S(\hat{\theta},N)\left(1-\frac{2\pi i}{\hat{\theta}}\right)F(\theta).\nonumber
\eeq

We next consider the Smirnov periodicity axiom. For the two-particle form factor, this axiom implies
\beq
\langle 0 \vert j_\mu^L(0)_{a_0c_0}\!\!\!\!\!&&\!\!\!\!\!\mathfrak{A}_P^\dag(\theta_2)_{a_2b_2}\mathfrak{A}_A^\dag(\theta_1)_{b_1a_1}\vert 0\rangle\nonumber\\
&&=\langle 0\vert j_\mu^L(0)_{a_0c_0}\mathfrak{A}_A^\dag(\theta_1-2\pi i)_{b_1a_1}\mathfrak{A}_P^\dag(\theta_2)_{a_2b_2}\vert 0\rangle,\nonumber
\eeq
or
\beq
F^\prime(\theta)=F(\theta-2\pi i).\label{twoperiodicity}
\eeq

Combining (\ref{twoscattering}) and (\ref{twoperiodicity}) gives
\beq
F(\theta-2\pi i)=\hat{S}(\theta,N)\left(\frac{\theta+\pi i}{\theta-\pi i}\right)F(\theta),\label{twocombination}
\eeq
where we have defined $\hat{S}(\theta,N)=S(\hat{\theta},N)$. 

Equation (\ref{twocombination}) can be easily solved at large $N$. For large $N$ we expand $\hat{S}(\theta,N)=1+\mathcal{O}\left(\frac{1}{N^2}\right)$, and $F(\theta)=F^0(\theta)+\frac{1}{N}F^1(\theta)+\frac{1}{N^2}F^2(\theta)+\dots$, and keep only terms up to order $\frac{1}{N^0}$, so that
\beq
F^0(\theta-2\pi i)=\left(\frac{\theta+\pi i}{\theta-\pi i}\right)F^0(\theta).\label{twoexpanded}
\eeq
The general solution to (\ref{twoexpanded}) is 
\beq
F^0(\theta)=\frac{g(\theta)}{\theta+\pi i},\nonumber
\eeq
where $g(\theta)$ satisfies the periodicity condition, $g(\theta-2\pi i)=g(\theta)$.
The minimal choice is taking $g(\theta)=g$ to be a constant.

Next, we determine the value of $g$. There is a conserved charge, $Q_{a_0c_0}^L$, associated with the current operator. This charge is
\beq
Q_{a_0c_0}^L=\int \,dx^1\,j_0^L(x)_{a_0c_0}.\nonumber
\eeq
We fix the value of $g$ by requiring that the charge is a generator of $SU(N)$, so it satisfies the Lie algebra
\beq
[Q_{a_1}^{L\,c_1},Q_{a_2}^{L\,c_2}]=-(\delta_{a_1}^{c_2}\delta_{a_2}^{a_3}\delta_{c_3}^{c_1}-\delta_{a_2}^{c_1}\delta_{a_1}^{a_3}\delta_{c_3}^{c_2})Q_{a_3}^{L\,c_3}.\label{liealgebra}
\eeq
We cross the incoming particle in Eq. (\ref{twoparticleform}) to an outgoing antiparticle, by shifting its rapidity, $\theta_2\to\theta_2-\pi i$:
\beq
\langle A,\theta_2,b_2,a_2\!\!\!\!\!&\vert&\!\!\!\!\! j_0^L(x)_{a_0c_0}\vert A,\theta_1,b_1,a_1\rangle\nonumber\\
&=&m\,(\cosh\,\theta_1+\cosh\,\theta_2)\exp\left\{ -im[x^0(\cosh\,\theta_1-\cosh\,\theta_2)\right.\nonumber\\
&&\left.-x^1(\sinh\,\theta_1-\sinh\,\theta_2)]\right\}F(\theta+\pi i)\nonumber\\&&\times\left(\delta_{a_0a_2}\delta_{b_1b_2}\delta_{c_0a_1}-\frac{1}{N}\delta_{a_0c_0}\delta_{b_1b_2}\delta_{a_1a_2}\right).\nonumber
\eeq
Integrating over $x^1$ gives the matrix element of the charge operator:
\beq
\langle A,\theta_2,b_2,a_2\vert Q_{a_0c_0}^L\vert A,\theta_1,b_1,a_1\rangle\!\!\!&=&\!\!\!(2\pi)^2 \,2(p_1)_0\delta(\theta_1-\theta_2)\nonumber\\
&\times&\left(\delta_{a_0a_2}\delta_{b_1b_2}\delta_{c_0a_1}-\frac{1}{N}\delta_{a_0c_0}\delta_{b_1b_2}\delta_{a_1a_2}\right)F(\pi i).\nonumber
\eeq
The matrix element of the commutator of two charges is found by inserting a complete set of intermediate states:
\beq
\langle A,\!\!\!\!\!\!\!&&\!\!\!\!\!\!\!\theta_2,b_2,a_2\vert [Q_{a_0c_0}^L,\,Q_{a_4c_4}^L]\vert A,\theta_1,b_1,a_1\rangle\nonumber\\
&=&\int \frac{d\theta_3}{4\pi}\langle A,\theta_2,b_2,a_2\vert Q_{a_0c_0}^L\vert A,\theta_3,b_3,a_3\rangle \langle A,\theta_3,b_3,a_3\vert Q_{a_4c_4}^L\vert A, \theta_1,b_1,a_1\rangle\nonumber\\
&&-\int \frac{d\theta_3}{4\pi}\langle A,\theta_2,b_2,a_2\vert Q_{a_4c_4}^L\vert A,\theta_3,b_3,a_3\rangle\nonumber\\
&&\times \langle A,\theta_3,b_3,a_3\vert Q_{a_0c_0}^L\vert A, \theta_1,b_1,a_1\rangle.\label{algebramatrixelement}
\eeq
With the choice $F(\pi i)=1$, Eq. (\ref{algebramatrixelement}) becomes
\beq
\langle A,\!\!\!\!\!\!\!&&\!\!\!\!\!\!\!\theta_2,b_2,a_2\vert [Q_{a_0c_0}^L,\,Q_{a_4c_4}^L]\vert A,\theta_1,b_1,a_1\rangle\nonumber\\
&=&-(\delta_{a_1}^{c_2}\delta_{a_2}^{a_3}\delta_{c_3}^{c_1}-\delta_{a_2}^{c_1}\delta_{a_1}^{a_3}\delta_{c_3}^{c_2})\langle A,\theta_2,b_2,a_2\vert Q_{a_5}^{L\,c_5}\vert A,\theta_1,b_1,a_1\rangle,\nonumber
\eeq
which is equivalent to Eq. (\ref{liealgebra}). This fixes the constant $g=2\pi i$.

To find the form factor for general, finite $N$, we need to solve the full Equation (\ref{twocombination}) instead of the simpler (\ref{twoexpanded}). We start with the ansatz
\beq
F(\theta)=\frac{g(\theta)}{\theta+\pi i}.\label{equationg}
\eeq
In terms of $g(\theta)$, Equation (\ref{twocombination}) becomes
\beq
g(\theta-2\pi i)=\hat{S}(\theta,N)g(\theta).\label{gequation}
\eeq

We solve Equation (\ref{gequation}) by a contour-integration method first used in Ref. \cite{karowski}. We define a contour $C$ to be that from $-\infty$ to $\infty$ and from $\infty+2\pi i$ to $-\infty+2\pi i$, bounding the strip in which the form factor is holomorphic. Then
\beq
\ln g(\theta)=\int_C\frac{dz}{4\pi i}\coth \frac{z-\theta}{2}\ln g(z)=\int_{-\infty}^\infty \frac{dz}{4\pi i}\coth \frac{z-\theta}{2}\ln\frac{g(z)}{g(z+2\pi i)}.\nonumber
\eeq
We differentiate both sides with repeat to $\theta$, and use Eq. (\ref{gequation}) to write
\beq
\frac{d}{d\theta}\left[\ln g(\theta)\right]=\frac{1}{8\pi i}\int_{-\infty}^\infty\frac{dz}{\sinh^2\frac{1}{2}(z-\theta)}\ln \hat{S}(z,N).\label{contourintegral}
\eeq
 The solution to (\ref{contourintegral}) is
\beq
g(\theta)=g\exp \int_0^\infty dx \,A(x,N)\frac{\sin^2[x(\pi i-\theta)/2\pi]}{\sinh x},\label{minimalform}
\eeq
where the function $A(x,N)$ is defined by
\beq
\hat{S}(\theta,N)=\exp \int_0^\infty dx \,A(x,N) \sinh \left(\frac{x\theta}{\pi i}\right),\label{exponentialexpression}
\eeq
and $g$ is a constant. Note that expanding the $S$ matrix in powers of $1/N$ yields $A(x,N)=\frac{1}{N^2}B(x)+\mathcal{O}(\frac{1}{N^3})$.

To express the function $\hat{S}(\theta,N)$, presented in (\ref{esfunction}), in the form 
(\ref{exponentialexpression}), we use the integral formula of the gamma function 
 \cite{karowski}, \cite{whitakerwatson},
\beq
\Gamma(z)=\exp\int_0^\infty\frac{dx}{x}\left[\frac{e^{-x z}-e^{-x}}{1-e^{-x}}+(z-1)e^{-x}\right],\,\,\,{\rm for}\,{\rm Re}\,z>0.\nonumber
\eeq
Then
\beq
&&\left[\frac{\Gamma\left(\frac{i\hat{\theta}}{2\pi}+1\right)\Gamma\left(\frac{-i\hat{\theta}}{2\pi}-\frac{1}{N}\right)}{\Gamma\left(\frac{i\hat{\theta}}{2\pi}+1-\frac{1}{N}\right)\Gamma\left(\frac{-i\hat{\theta}}{2\pi}\right)}\right]^2\nonumber\\
&&\,\,\,\,\,\,\,\,\,\,\,\,\,\,\,\,\,\,\,\,\,\,=\exp\int_0^\infty\frac{dx}{x}\frac{4e^{-x}\left(e^{2x/N}-1\right)}{1-e^{-2x}}\sinh\left(\frac{x\theta}{\pi i}\right),\label{gamma}
\eeq
for $N>2$. We use the formula \cite{babujian}
\beq
\frac{\sin\frac{\pi}{2}(z+a)}{\sin\frac{\pi}{2}(z-a)}=\exp 2\int_0^\infty\frac{dx}{x}\frac{\sinh x(1-z)}{\sinh x}\sinh(xa),\,\,\,{\rm for}\,\,0<z<1,\nonumber
\eeq
to write the CDD factor as
\beq
\frac{\sinh\left(\frac{\hat{\theta}}{2}-\frac{\pi i}{N}\right)}{\sinh\left(\frac{\hat{\theta}}{2}+\frac{\pi i}{N}\right)}&=&\frac{\sin\frac{\pi}{2}\left(\left(1-\frac{2}{N}\right)-\frac{\theta}{\pi i}\right)}{\sin\frac{\pi}{2}\left(\left(1-\frac{2}{N}\right)+\frac{\theta}{\pi i}\right)}\nonumber\\
&&=\exp\int_0^\infty\frac{dx}{x}\frac{-2\sinh(2x/N)}{\sinh x}\sinh \left(\frac{x\theta}{\pi i}\right),\label{sines}
\eeq
for $N>2$. Combining (\ref{gamma}) and (\ref{sines}) gives
\beq
\hat{S}(\theta,N)
&=&\exp\int_0^\infty \frac{dx}{x}\left[\frac{-2\sinh(2x/N)}{\sinh x}\right.\nonumber\\
&&\,\,\,\,\,\,\,\,\,\,\,\,\,\,\left.+\frac{4e^{-x}\left(e^{2x/N}-1\right)}{1-e^{-2x}}\right]\sinh\left(\frac{x\theta}{\pi i}\right).\label{afunction}
\eeq
From (\ref{equationg}) and (\ref{minimalform}), the form factor is
\beq
F_1(\theta)&=&\frac{g}{(\theta+\pi i)}\exp\int_0^\infty \frac{dx}{x}\left[\frac{-2\sinh\left(\frac{2x}{N}\right)}{\sinh x}\right.\nonumber\\
&&\,\,\,\,\,\left.+\frac{4e^{-x}\left(e^{2x/N}-1\right)}{1-e^{-2x}}\right]\frac{\sin^2[x(\pi i-\theta)/2\pi]}{\sinh x}.\label{finitenform}
\eeq
The condition $F_1(\pi i)=1$ implies $g=2\pi i$.

\section{Four-particle form factors}
\setcounter{equation}{0}
\renewcommand{\theequation}{4.3.\arabic{equation}}

In this section we find the four-excitation form factor of the Noether current operator, in the large-$N$ limit. Only the form factor with two particles and two antiparticles is nonzero, because of the global color symmetry. The most general Lorentz- and $SU(N)\times SU(N)$-invariant  four-particle form factor, respecting the tracelessness of the current operator is
\beq
\langle 0|j_\mu^L(0)_{a_0c_0}\!\!\!\!\!&|&\!\!\!\!\! A,\theta_1,b_1,a_1;A,\theta_2,b_2,a_2;P,\theta_3,a_3,b_3;P,\theta_4,a_4,b_4\rangle\nonumber\\
&=&\langle 0| j_\mu^L(0)_{a_0c_0} \mathfrak{A}_{A}^\dag(\theta_1)_{b_1a_1}\mathfrak{A}_A^\dag(\theta_2)_{b_2a_2}\mathfrak{A}_P^\dag(\theta_3)_{a_3b_3}\mathfrak{A}_P^\dag(\theta_4)_{a_4b_4}|0\rangle\nonumber\\
&=&-\epsilon_{\mu\nu}(p_1+p_2+p_3+p_4)^\nu\frac{1}{N}\nonumber\\
&&\times\vec{F}(\theta_1,\theta_2,\theta_3,\theta_4)\cdot \vec{D}_{a_0c_0a_1a_2a_3a_4;b_1b_2b_3b_4},\label{fourparticleone}
\eeq
for $\theta_1>\theta_2>\theta_3>\theta_4$,
\beq
\langle 0|j_\mu^L(0)_{a_0c_0}\!\!\!\!\!&|&\!\!\!\!\! A,\theta_1,b_1,a_1;P,\theta_2,a_2,b_2;A,\theta_3,b_3,a_3;P,\theta_4,a_4,b_4\rangle\nonumber\\
&=&\langle 0|j_\mu^L(0)_{a_0c_0}\mathfrak{A}_{A}^\dag(\theta_1)_{b_1a_1}\mathfrak{A}_P^\dag(\theta_3)_{a_3b_3}\mathfrak{A}_A^\dag(\theta_2)_{b_2a_2}\mathfrak{A}_P^\dag(\theta_4)_{a_4b_4}|0\rangle\nonumber\\
&=&-\epsilon_{\mu\nu}(p_1+p_2+p_3+p_4)^\nu\frac{1}{N}\nonumber\\
&&\times\vec{G}(\theta_1,\theta_2,\theta_3,\theta_4)\cdot \vec{D}_{a_0c_0a_1a_2a_3a_4;b_1b_2b_3b_4},\label{fourparticletwo}
\eeq
for $\theta_1>\theta_3>\theta_2>\theta_4$,
\beq
\langle 0|j_\mu^L(0)_{a_0c_0}\!\!\!\!\!&|&\!\!\!\!\! A,\theta_1,b_1,a_1;P,\theta_2,a_2,b_2;P,\theta_3,a_3,b_3;A,\theta_4,b_4,a_4\rangle\nonumber\\
&=&\langle 0|j_\mu^L(0)_{a_0c_0}\mathfrak{A}_{A}^\dag(\theta_1)_{b_1a_1}\mathfrak{A}_P^\dag(\theta_3)_{a_3b_3}\mathfrak{A}_P^\dag(\theta_4)_{a_4b_4}\mathfrak{A}_A^\dag(\theta_2)_{b_2a_2}|0\rangle\nonumber\\
&=&-\epsilon_{\mu\nu}(p_1+p_2+p_3+p_4)^\nu\frac{1}{N}\nonumber\\
&&\times\vec{H}(\theta_1,\theta_2,\theta_3,\theta_4)\cdot \vec{D}_{a_0c_0a_1a_2a_3a_4;b_1b_2b_3b_4},\label{fourparticlethree}
\eeq
for $\theta_1>\theta_3>\theta_4>\theta_2$,
\beq
\langle 0|j_\mu^L(0)_{a_0c_0}\!\!\!\!\!&|&\!\!\!\!\! P,\theta_1,a_1,b_1;A,\theta_2,b_2,a_2;P,\theta_3,a_3,b_3;A,\theta_4,b_4,a_4\rangle\nonumber\\
&=&\langle 0|j_\mu^L(0)_{a_0c_0}\mathfrak{A}_{P}^\dag(\theta_3)_{a_3b_3}\mathfrak{A}_A^\dag(\theta_1)_{b_1a_1}\mathfrak{A}_P^\dag(\theta_4)_{a_4b_4}\mathfrak{A}_A^\dag(\theta_2)_{b_2a_2}|0\rangle\nonumber\\
&=&-\epsilon_{\mu\nu}(p_1+p_2+p_3+p_4)^\nu\frac{1}{N}\nonumber\\
&&\times\vec{K}(\theta_1,\theta_2,\theta_3,\theta_4)\cdot \vec{D}_{a_0c_0a_1a_2a_3a_4;b_1b_2b_3b_4},\label{fourparticlefour}
\eeq
for $\theta_3>\theta_1>\theta_4>\theta_2$,
\beq
\langle 0|j_\mu^L(0)_{a_0c_0}\!\!\!\!\!&|&\!\!\!\!\! P,\theta_1,a_1,b_1;P,\theta_2,a_2,b_2;A,\theta_3,b_3,a_3;A,\theta_4,b_4,a_4\rangle\nonumber\\
&=&\langle 0|j_\mu^L(0)_{a_0c_0}\mathfrak{A}_{P}^\dag(\theta_3)_{a_3b_3}\mathfrak{A}_P^\dag(\theta_4)_{a_4b_4}\mathfrak{A}_A^\dag(\theta_1)_{b_1a_1}\mathfrak{A}_A^\dag(\theta_2)_{b_2a_2}|0\rangle\nonumber\\
&=&-\epsilon_{\mu\nu}(p_1+p_2+p_3+p_4)^\nu\frac{1}{N}\nonumber\\
&&\times\vec{L}(\theta_1,\theta_2,\theta_3,\theta_4)\cdot \vec{D}_{a_0c_0a_1a_2a_3a_4;b_1b_2b_3b_4},\label{fourparticlefive}
\eeq
for $\theta_3>\theta_4>\theta_1>\theta_2$,
\beq
\langle 0|j_\mu^L(0)_{a_0c_0}\!\!\!\!\!&|&\!\!\!\!\! P,\theta_1,a_1,b_1;A,\theta_2,b_2,a_2;A,\theta_3,b_3,a_3;P,\theta_4,a_4,b_4\rangle\nonumber\\
&=&\langle 0|j_\mu^L(0)_{a_0c_0}\mathfrak{A}_{P}^\dag(\theta_3)_{a_3a_3}\mathfrak{A}_A^\dag(\theta_1)_{b_1a_1}\mathfrak{A}_A^\dag(\theta_2)_{b_2a_2}\mathfrak{A}_P^\dag(\theta_4)_{a_4b_4}|0\rangle\nonumber\\
&=&-\epsilon_{\mu\nu}(p_1+p_2+p_3+p_4)^\nu\frac{1}{N}\nonumber\\
&&\times\vec{Q}(\theta_1,\theta_2,\theta_3,\theta_4)\cdot \vec{D}_{a_0c_0a_1a_2a_3a_4;b_1b_2b_3b_4},\label{fourparticlesix}
\eeq
for $\theta_3>\theta_1>\theta_2>\theta_4$,
\beq
\langle 0|j_\mu^L(0)_{a_0c_0}\!\!\!\!\!&|&\!\!\!\!\! A,\theta_2,b_2,a_2;A,\theta_1,b_1,a_1;P,\theta_3,a_3,b_3;P,\theta_4,a_4,b_4\rangle\nonumber\\
&=&\langle 0| j_\mu^L(0)_{a_0c_0} \mathfrak{A}_{A}^\dag(\theta_2)_{b_2a_2}\mathfrak{A}_A^\dag(\theta_1)_{b_1a_1}\mathfrak{A}_P^\dag(\theta_3)_{a_3b_3}\mathfrak{A}_P^\dag(\theta_4)_{a_4b_4}|0\rangle\nonumber\\
&=&-\epsilon_{\mu\nu}(p_1+p_2+p_3+p_4)^\nu\frac{1}{N}\nonumber\\
&&\times\vec{F}(\theta_2,\theta_1,\theta_3,\theta_4)\cdot \vec{D}_{a_0c_0a_1a_2a_3a_4;b_1b_2b_3b_4},\label{fourparticleseven}
\eeq
for $\theta_2>\theta_1>\theta_3,>\theta_4$, and
\beq
\langle 0|j_\mu^L(0)_{a_0c_0}\!\!\!\!\!&|&\!\!\!\!\! A,\theta_1,b_1,a_1;A,\theta_2,b_2,a_2;P,\theta_4,a_4,b_4;P,\theta_3,a_3,b_3\rangle\nonumber\\
&=&\langle 0| j_\mu^L(0)_{a_0c_0} \mathfrak{A}_{A}^\dag(\theta_1)_{b_1a_1}\mathfrak{A}_A^\dag(\theta_2)_{b_2a_2}\mathfrak{A}_P^\dag(\theta_4)_{a_4b_4}\mathfrak{A}_P^\dag(\theta_3)_{a_3b_3}|0\rangle\nonumber\\
&=&-\epsilon_{\mu\nu}(p_1+p_2+p_3+p_4)^\nu\frac{1}{N}\nonumber\\
&&\times\vec{F}(\theta_1,\theta_2,\theta_4,\theta_3)\cdot \vec{D}_{a_0c_0a_1a_2a_3a_4;b_1b_2b_3b_4},\label{fourparticleeight}
\eeq
for $\theta_1>\theta_2>\theta_4>\theta_3$,
where we define the eight-component vectors
\beq
[\vec{D}_{a_0c_0a_1a_2a_3a_4;b_1b_2b_3b_4}]=\left(\begin{array}{c}\delta_{a_0a_3}\delta_{a_1c_0}\delta_{a_2a_4}\delta_{b_1b_3}\delta_{b_2b_4}-\frac{1}{N}\delta_{a_0c_0}\delta_{a_1a_3}\delta_{a_2a_4}\delta_{b_1b_3}\delta_{b_2b_4}\\\,\\\delta_{a_0a_3}\delta_{a_1c_0}\delta_{a_2a_4}\delta_{b_1b_4}\delta_{b_2b_3}-\frac{1}{N}\delta_{a_0c_0}\delta_{a_1a_3}\delta_{a_2a_4}\delta_{b_1b_4}\delta_{b_2b_3}\\\,\\\delta_{a_0a_4}\delta_{a_1c_0}\delta_{a_2a_3}\delta_{b_1b_3}\delta_{b_2b_4}-\frac{1}{N}\delta_{a_0c_0}\delta_{a_1a_4}\delta_{a_2a_3}\delta_{b_1b_3}\delta_{b_2b_4}\\\,\\\delta_{a_0a_4}\delta_{a_1c_0}\delta_{a_2a_3}\delta_{b_1b_4}\delta_{b_2b_3}-\frac{1}{N}\delta_{a_0c_0}\delta_{a_1a_4}\delta_{a_2a_3}\delta_{b_1b_4}\delta_{b_2b_3}\\\,\\\delta_{a_0a_3}\delta_{a_1a_4}\delta_{a_2c_0}\delta_{b_1b_4}\delta_{b_2b_3}-\frac{1}{N}\delta_{a_0c_0}\delta_{a_2a_3}\delta_{a_1a_4}\delta_{b_1b_4}\delta_{b_2b_3}\\\,\\\delta_{a_0a_3}\delta_{a_1a_4}\delta_{a_2c_0}\delta_{b_1b_3}\delta_{b_2b_4}-\frac{1}{N}\delta_{a_0c_0}\delta_{a_2a_3}\delta_{a_1a_4}\delta_{b_1b_3}\delta_{b_2b_4}\\\,\\\delta_{a_0a_4}\delta_{a_1a_3}\delta_{a_2c_0}\delta_{b_1b_3}\delta_{b_2b_4}-\frac{1}{N}\delta_{a_0c_0}\delta_{a_2a_4}\delta_{a_1a_3}\delta_{b_1b_3}\delta_{b_2b_4}\\\,\\\delta_{a_0a_4}\delta_{a_1a_3}\delta_{a_2a_0}\delta_{b_1b_4}\delta_{b_2b_3}-\frac{1}{N}\delta_{a_0c_0}\delta_{a_2a_4}\delta_{a_1a_3}\delta_{b_1b_4}\delta_{b_2b_3}\end{array}\right),\nonumber
\eeq
\beq
[\vec{F}(\theta_1,\theta_2,\theta_3,\theta_4)]=\left(\begin{array}{c}F_1(\theta_1,\theta_2,\theta_3,\theta_4)\\\,\\F_2(\theta_1,\theta_2,\theta_3,\theta_4)\\\,\\F_3(\theta_1,\theta_2,\theta_3,\theta_4)\\\,\\F_4(\theta_1,\theta_2,\theta_3,\theta_4)\\\,\\F_5(\theta_1,\theta_2,\theta_3,\theta_4)\\\,\\F_6(\theta_1,\theta_2,\theta_3,\theta_4)\\\,\\F_7(\theta_1,\theta_2,\theta_3,\theta_4)\\\,\\F_8(\theta_1,\theta_2,\theta_3,\theta_4)\end{array}\right),\nonumber
\eeq
and similarly for $\vec{G},\,\vec{H},\,\vec{K},\,\vec{L}$ and $\vec{Q}$.

The scattering axiom relates the form factors with different ordering of rapidities, 
yielding
\beq
\langle 0|j_\mu^L(0\!\!\!\!\!&)&\!\!\!\!\! _{a_0c_0}\mathfrak{A}_A^\dag(\theta_1)_{b_1a_1}\mathfrak{A}_P^\dag(\theta_3)_{a_3b_3}\mathfrak{A}_A^\dag(\theta_2)_{b_2a_2}\mathfrak{A}_P^\dag(\theta_4)_{a_4b_4}|0\rangle\nonumber\\
&=&S_{AP}(\theta_{23})_{a_3b_3;b_2a_2}^{d_2c_2;c_3d_3}\nonumber\\
&&\times\langle 0|j_\mu^L(0)_{a_0c_0}\mathfrak{A}_A^\dag(\theta_1)_{b_1a_1}\mathfrak{A}_A^\dag(\theta_2)_{d_2c_2}\mathfrak{A}_P^\dag(\theta_3)_{c_3d_3}\mathfrak{A}_P^\dag(\theta_4)_{a_4b_4}|0\rangle,\nonumber
\eeq
\beq
\langle 0|j_\mu^L(0\!\!\!\!\!&)&\!\!\!\!\!_{a_0c_0}\mathfrak{A}_A^\dag(\theta_1)_{b_1a_1}\mathfrak{A}_P^\dag(\theta_3)_{a_3b_3}\mathfrak{A}_P^\dag(\theta_4)_{a_4b_4}\mathfrak{A}_A^\dag(\theta_2)_{b_2a_2}|0\rangle\nonumber\\
&=&S_{AP}(\theta_{24})_{a_4b_4;b_2a_2}^{d_2c_2;c_4d_4}\nonumber\\
&&\times\langle 0|j_\mu^L(0)_{a_0c_0} \mathfrak{A}_A^\dag(\theta_1)_{b_1a_1}\mathfrak{A}_P^\dag(\theta_3)_{a_3b_3}\mathfrak{A}_A^\dag(\theta_2)_{d_2c_2}\mathfrak{A}_P^\dag(\theta_4)_{c_4d_4}|0\rangle\nonumber,
\eeq
\beq
\langle 0| j_\mu^L(0\!\!\!\!\!&)&\!\!\!\!\!_{a_0c_0}\mathfrak{A}_P^\dag(\theta_3)_{a_3b_3}\mathfrak{A}_A^\dag(\theta_1)_{b_1a_1}\mathfrak{A}_P^\dag(\theta_4)_{a_4b_4}\mathfrak{A}_A^\dag(\theta_2)_{b_2a_2}|0\rangle\nonumber\\
&=&S_{AP}(\theta_{13})^{d_1c_1;c_3d_3}_{a_3b_3;b_1a_1}\nonumber\\
&&\times\langle 0|j_\mu^L(0)_{a_0c_0}\mathfrak{A}_A^\dag(\theta_1)_{d_1c_1}\mathfrak{A}_P^\dag(\theta_3)_{c_3d_3}\mathfrak{A}_P^\dag(\theta_4)_{a_4b_4}\mathfrak{A}_A^\dag(\theta_2)_{b_2a_2}|0\rangle,\nonumber
\eeq
\beq
\langle0|j_\mu^L(0\!\!\!\!\!&)&\!\!\!\!\!_{a_0c_0}\mathfrak{A}_P^\dag(\theta_3)_{a_3b_3}\mathfrak{A}_P^\dag(\theta_4)_{a_4b_4}\mathfrak{A}_A^\dag(\theta_1)_{b_1a_1}\mathfrak{A}_A^\dag(\theta_2)_{b_2a_1}|0\rangle\nonumber\\
&=&S_{AP}(\theta_{14})^{d_1c_1;c_4d_4}_{a_4b_4;b_1a_1}\nonumber\\
&&\times\langle0|j_\mu^L(0)_{a_0c_0}\mathfrak{A}_P^\dag(\theta_3)_{a_3b_3}\mathfrak{A}_A^\dag(\theta_1)_{d_1c_1}\mathfrak{A}_P^\dag(\theta_4)_{c_4d_4}\mathfrak{A}_A^\dag(\theta_2)_{b_2a_2}|0\rangle,\nonumber
\eeq
\beq
\langle 0|j_\mu^L(0\!\!\!\!\!&)&\!\!\!\!\!_{a_0c_0}\mathfrak{A}_P^\dag(\theta_3)_{a_3b_3}\mathfrak{A}_A^\dag(\theta_1)_{b_1a_1}\mathfrak{A}_A^\dag(\theta_2)_{b_2a_2}\mathfrak{A}_P^\dag(\theta_4)_{a_4b_4}|0\rangle\nonumber\\
&=&S_{AP}(\theta_{13})^{d_1c_1;c_3d_3}_{a_3b_3;b_1a_1}\nonumber\\
&&\times\langle 0|j_\mu^L(0)_{a_0c_0}\mathfrak{A}_A^\dag(\theta_1)_{d_1c_1}\mathfrak{A}_P^\dag(\theta_3)_{c_3d_3}\mathfrak{A}_A^\dag(\theta_2)_{b_2a_2}\mathfrak{A}_P^\dag(\theta_4)_{a_4b_4}|0\rangle,\nonumber
\eeq
\beq
\langle 0|j_\mu^L(0\!\!\!\!\!&)&\!\!\!\!\!_{a_0c_0}\mathfrak{A}_A^\dag(\theta_1)_{b_1a_1}\mathfrak{A}_A^\dag(\theta_2)_{b_2a_2}\mathfrak{A}_P^\dag(\theta_3)_{a_3b_3}\mathfrak{A}_P(\theta_4)_{a_4b_4}|0\rangle\nonumber\\
&=&S_{AA}(\theta_{12})^{d_2c_2;d_1c_1}_{b_1a_1;b_2a_2}\nonumber\\
&&\times\langle0|j_\mu^L(0)_{a_0c_0}\mathfrak{A}_A^\dag(\theta_2)_{d_2c_2}\mathfrak{A}_A^\dag(\theta_1)_{d_1c_1}\mathfrak{A}_P^\dag(\theta_3)_{a_3b_3}\mathfrak{A}_P^\dag(\theta_4)_{a_4b_4}|0\rangle,\nonumber
\eeq
\beq
\langle 0|j_\mu^L(0\!\!\!\!\!&)&\!\!\!\!\!_{a_0c_0}\mathfrak{A}_A^\dag(\theta_1)_{b_1a_1}\mathfrak{A}_A^\dag(\theta_2)_{b_2a_2}\mathfrak{A}_P^\dag(\theta_3)_{a_3b_3}\mathfrak{A}_P(\theta_4)_{a_4b_4}|0\rangle\nonumber\\
&=&S_{PP}(\theta_{34})^{c_4d_4;c_3d_3}_{a_3b_3;a_4b_4}\nonumber\\
&&\times\langle0|j_\mu^L(0)_{a_0c_0}\mathfrak{A}_A^\dag(\theta_1)_{b_1a_1}\mathfrak{A}_A^\dag(\theta_2)_{b_2a_2}\mathfrak{A}_P^\dag(\theta_4)_{c_4d_4}\mathfrak{A}_P^\dag(\theta_3)_{c_3d_3}|0\rangle,\nonumber
\eeq
where $\theta_{jk}=\theta_j-\theta_k$.
These imply, respectively,
\beq
\vec{G}(\theta_1,\theta_2,\!\!\!&\theta_3&\!\!\!,\theta_4)\nonumber\\
&=&\left(\begin{array}{ccc}
1&0&0\\
\frac{-2\pi i}{N\hat{\theta}_{23}}&\left(1-\frac{2\pi i}{\hat{\theta}_{23}}\right)&0\\
\frac{-2\pi i}{N\hat{\theta}_{23}}&0&\left(1-\frac{2\pi i}{\hat{\theta}_{23}}\right)\\
0&\frac{-1}{N}\left(\frac{2\pi i}{\hat{\theta}_{23}}+\frac{4\pi^2}{\hat{\theta}_{23}^2}\right)&\frac{-1}{N}\left(\frac{2\pi i}{\hat{\theta}_{23}}+\frac{4\pi^2}{\hat{\theta}_{23}^2}\right)\\
0&0&0\\
0&0&0\\
0&0&0\\
0&0&0
\end{array}\right\vert \nonumber\\
&&\left\vert\begin{array}{ccccc}
0&0&0&0&0\\
0&0&0&0&0\\
0&0&0&\frac{-2\pi i}{N\hat{\theta}_{23}}&0\\
\left(1-\frac{4\pi i}{\hat{\theta}_{23}}-\frac{4\pi^2}{\hat{\theta}_{23}^2}\right)&0&0&0&\frac{-1}{N}\left(\frac{2\pi i}{\hat{\theta}_{23}}+\frac{4\pi^2}{\hat{\theta}_{23}^2}\right)\\
0&\left(1-\frac{2\pi i}{\hat{\theta}_{23}}\right)&\frac{-2\pi i}{N\hat{\theta}_{23}}&0&0\\
0&0&1&0&0\\
0&0&0&1&0\\
0&0&0&\frac{-2\pi i}{N\hat{\theta}_{23}}&\left(1-\frac{2\pi i}{\hat{\theta}_{23}}\right)
\end{array}\right)\nonumber\\
&&\times\vec{F}(\theta_1,\theta_2,\theta_3,\theta_4) +\mathcal{O}\left(\frac{1}{N^2}\right)\nonumber\\
&\equiv& \overleftrightarrow{M}_1(\theta_2,\theta_3)\vec{F}(\theta_1\theta_2,\theta_3,\theta_4)+\mathcal{O}\left(\frac{1}{N^2}\right),\label{watsonone}
\eeq
\beq
\vec{H}(\theta_1,\theta_2,\!\!\!&\theta_3&\!\!\!,\theta_4)\nonumber\\
&=&\left(\begin{array}{ccc}
\left(1-\frac{4\pi i}{\hat{\theta}_{24}}-\frac{4\pi^2}{\hat{\theta}^2_{24}}\right)&\frac{-1}{N}\left(\frac{2\pi i}{\hat{\theta}_{24}}+\frac{4\pi^2}{\hat{\theta}^2_{24}}\right)&\frac{-1}{N}\left(\frac{2\pi i}{\hat{\theta}_{24}}+\frac{4\pi^2}{\hat{\theta}^2_{24}}\right)\\
0&\left(1-\frac{2\pi i}{\hat{\theta}_{24}}\right)&0\\
0&0&\left(1-\frac{2\pi i}{\hat{\theta}_{24}}\right)\\
0&0&0\\
0&0&0\\
0&0&0\\
0&0&0\\
0&0&0
\end{array}\right\vert\nonumber\\
&&\left\vert\begin{array}{ccccc}
0&0&\frac{-1}{N}\left(\frac{2\pi i}{\hat{\theta}_{24}}-\frac{4\pi^2}{\hat{\theta}_{24}^2}\right)&0&0\\
\frac{-2\pi i}{N\hat{\theta}_{24}}&\frac{-2\pi i}{N\hat{\theta}_{24}^2}&0&0&0\\
\frac{-2\pi i}{N\hat{\theta}_{24}}&0&0&0&0\\
1&0&0&0&0\\
0&1&0&0&0\\
0&\frac{-2\pi i}{N\hat{\theta}_{24}}&\left(1-\frac{2\pi i}{\hat{\theta}_{24}}\right)&0&0\\
0&0&0&\left(1-\frac{2\pi i}{\hat{\theta}_{24}}\right)&\frac{-2\pi i}{N\hat{\theta}_{24}}\\
0&0&0&0&1
\end{array}\right)\nonumber\\
&&\times \vec{G}(\theta_1,\theta_2,\theta_3,\theta_4)+\mathcal{O}\left(\frac{1}{N^2}\right)\nonumber\\
&\equiv&\overleftrightarrow{M}_2(\theta_2,\theta_4)\vec{G}(\theta_1,\theta_2,\theta_3,\theta_4)+\mathcal{O}\left(\frac{1}{N^2}\right),\label{watsontwo}
\eeq
\beq
\vec{K}(\theta_1,\theta_2,\!\!\!&\theta_3&\!\!\!,\theta_4)\nonumber\\
&=&\left(\begin{array}{ccc}
\left(1-\frac{2\pi i}{\hat{\theta}_{13}}\right)&\frac{-2\pi i}{N\hat{\theta}_{13}}&0\\
0&1&0\\
0&0&\left(1-\frac{2\pi i}{\hat{\theta}_{13}}\right)\\
0&0&0\\
0&0&0\\
0&0&0\\
0&0&0\\
0&0&0
\end{array}\right\vert\nonumber\\
&&\left\vert\begin{array}{ccccc}
0&0&0&0&0\\
0&0&0&0&0\\
\frac{-2\pi i}{N\hat{\theta}_{13}}&0&0&0&0\\
1&0&0&0&0\\
0&1&0&0&0\\
0&\frac{-2\pi i}{N\hat{\theta}_{13}}&\left(1-\frac{2\pi i}{\hat{\theta}_{13}}\right)&0&0\\
0&0&\frac{-1}{N}\left(\frac{2\pi i}{\hat{\theta}_{13}}+\frac{4\pi^2}{\hat{\theta}_{13}^2}\right)&\left(1-\frac{4\pi i}{\hat{\theta}_{13}}-\frac{4\pi^2}{\hat{\theta}_{13}^2}\right)&\frac{-1}{N}\left(\frac{2\pi i}{\hat{\theta}_{13}}+\frac{4\pi^2}{\hat{\theta}_{13}^2}\right)\\
\frac{-2\pi i}{N\hat{\theta}_{13}}&\frac{-2\pi i}{N\hat{\theta}_{13}}&0&0&\left(1-\frac{2\pi i}{\hat{\theta}_{13}}\right)
\end{array}\right)\nonumber\\
&&\times\vec{H}(\theta_1,\theta_2,\theta_3,\theta_4)+\mathcal{O}\left(\frac{1}{N^2}\right)\nonumber\\
&\equiv& \overleftrightarrow{M}_3(\theta_1,\theta_3)\vec{H}(\theta_1,\theta_2,\theta_3,\theta_4)+\mathcal{O}\left(\frac{1}{N^2}\right),\label{watsonthree}
\eeq
\beq
\vec{L}(\theta_1,\theta_2,\!\!\!&\theta_3&\!\!\!,\theta_4)\nonumber\\
&=&\left(\begin{array}{ccc}
1&0&0\\
\frac{-2\pi i}{N\hat{\theta}_{14}}&\left(1-\frac{2\pi i}{\hat{\theta}_{14}}\right)&0\\
0&0&1\\
0&0&\frac{-2\pi i}{N\hat{\theta}_{14}}\\
0&0&0\\
\frac{-2\pi i}{N\hat{\theta}_{14}}&0&0\\
0&0&0\\
0&0&0
\end{array}\right\vert\nonumber\\
&&\left\vert\begin{array}{ccccc}
0&0&0&0&0\\
0&0&0&0&0\\
0&0&0&0&0\\
\left(1-\frac{2\pi i}{\hat{\theta}_{14}}\right)&0&0&0&0\\
0&\left(1-\frac{4\pi i}{\hat{\theta}_{14}}-\frac{4\pi^2}{\hat{\theta}_{14}^2}\right)&\frac{-1}{N}\left(\frac{2\pi i}{\hat{\theta}_{14}}+\frac{4\pi^2}{\hat{\theta}_{14}^2}\right)&0&\frac{-1}{N}\left(\frac{2\pi i}{\hat{\theta}_{14}}+\frac{4\pi^2}{\hat{\theta}_{14}^2}\right)\\
0&0&\left(1-\frac{2\pi i}{\hat{\theta}_{14}}\right)&\frac{-2\pi i}{N\hat{\theta}_{14}}&0\\
0&0&0&1&0\\
0&0&0&\frac{-2\pi i}{N\hat{\theta}_{14}}&\left(1-\frac{2\pi i}{\hat{\theta}_{14}}\right)
\end{array}\right)\nonumber\\
&&\times\vec{K}(\theta_1,\theta_2,\theta_3,\theta_4)+\mathcal{O}\left(\frac{1}{N^2}\right)\nonumber\\
&\equiv&\overleftrightarrow{M}_4(\theta_1,\theta_4)\vec{K}(\theta_1,\theta_2,\theta_3,\theta_4)+\mathcal{O}\left(\frac{1}{N^2}\right),\label{watsonfour}
\eeq
\beq
\vec{Q}(\theta_1,\theta_2,\theta_3,\theta_4)=\overleftrightarrow{M}_3(\theta_1,\theta_3)\vec{G}(\theta_1,\theta_2,\theta_3,\theta_4)+\mathcal{O}\left(\frac{1}{N^2}\right),\label{watsonfive}
\eeq
\beq
\vec{F}(\theta_1,\theta_2,\theta_3,\theta_4)&=&\left(\begin{array}{cccccccc}
0&\frac{-2\pi i}{N{\theta}_{12}}&0&0&1&\frac{-2\pi i}{N\theta_{12}}&0&0\\
\frac{-2\pi i}{N\theta_{12}}&0&0&0&\frac{-2\pi i}{N\theta_{12}}&1&0&0\\
0&0&0&\frac{-2\pi i}{N\theta_{12}}&0&0&\frac{-2\pi i}{N\theta_{12}}&1\\
0&0&\frac{-2\pi i}{N\theta_{12}}&0&0&0&1&\frac{-2\pi i}{N\theta_{12}}\\
1&\frac{-2\pi i}{N\theta_{12}}&0&0&0&\frac{-2\pi i}{N\theta_{12}}&0&0\\
\frac{-2\pi i}{N\theta_{12}}&1&0&0&\frac{-2\pi i}{N\theta_{12}}&0&0&0\\
0&0&\frac{-2\pi i}{N\theta_{12}}&1&0&0&0&\frac{-2\pi i}{N\theta_{12}}\\
0&0&1&\frac{-2\pi i}{N\theta_{12}}&0&0&\frac{-2\pi i}{N\theta_{12}}&0
\end{array}\right)\nonumber\\
&&\times\vec{F}(\theta_2,\theta_1,\theta_3,\theta_4)+\mathcal{O}\left(\frac{1}{N^2}\right)\nonumber\\
&\equiv&\overleftrightarrow{I}_1(\theta_1,\theta_2)\vec{F}(\theta_2,\theta_1,\theta_3,\theta_4)+\mathcal{O}\left(\frac{1}{N^2}\right),\label{watsonsix}
\eeq
\beq
\vec{F}(\theta_1,\theta_2,\theta_3,\theta_4)&=&\left(\begin{array}{cccccccc}
0&\frac{-2\pi i}{N\theta_{34}}&\frac{-2\pi i}{N\theta_{34}}&1&0&0&0&0\\
\frac{-2\pi i}{N\theta_{34}}&0&1&\frac{-2\pi i}{N\theta_{34}}&0&0&0&0\\
\frac{-2\pi i}{N\theta_{34}}&1&0&\frac{-2\pi i}{N\theta_{34}}&0&0&0&0\\
1&\frac{-2\pi i}{N\theta_{34}}&\frac{-2\pi i}{N\theta_{34}}&0&0&0&0&0\\
0&0&0&0&0&\frac{-2\pi i}{N\theta_{34}}&1&\frac{-2\pi i}{N\theta_{34}}\\
0&0&0&0&\frac{-2\pi i}{N\theta_{34}}&0&\frac{-2\pi i}{N\theta_{34}}&1\\
0&0&0&0&1&\frac{-2\pi i}{N\theta_{34}}&0&\frac{-2\pi i}{N\theta_{34}}\\
0&0&0&0&\frac{-2\pi i}{N\theta_{34}}&1&\frac{-2\pi i}{N\theta_{34}}&0
\end{array}\right)\nonumber\\
&&\times\vec{F}(\theta_1,\theta_2,\theta_4,\theta_3)+\mathcal{O}\left(\frac{1}{N^2}\right)\nonumber\\
&\equiv&\overleftrightarrow{I}_2(\theta_3,\theta_4)\vec{F}(\theta_1,\theta_2,\theta_4,\theta_3)+\mathcal{O}\left(\frac{1}{N^2}\right).\label{watsonseven}
\eeq

Next we apply the Smirnov periodicity axiom (\ref{periodicityaxiom}):
\beq
\langle 0|j_\mu^L(0\!\!\!\!\!&)&\!\!\!\!\!_{a_0c_0}\mathfrak{A}_A^\dag(\theta_1-2\pi i)_{b_1a_1}\mathfrak{A}_A^\dag(\theta_2)_{b_2a_2}\mathfrak{A}_P^\dag(\theta_3)_{a_3b_3}\mathfrak{A}_P^\dag(\theta_4)_{a_4b_4}|0\rangle\nonumber\\
&=&\langle0|j_\mu^L(0)_{a_0c_0}\mathfrak{A}_A^\dag(\theta_2)_{b_2a_2}\mathfrak{A}_P^\dag(\theta_3)_{a_3b_3}\mathfrak{A}_P^\dag(\theta_4)_{a_4b_4}\mathfrak{A}_A^\dag(\theta_1)_{b_1a_1}|0\rangle,\nonumber\\
&&\,\nonumber\\
\langle 0|j_\mu^L(0\!\!\!\!\!&)&\!\!\!\!\!_{a_0c_0}\mathfrak{A}_A^\dag(\theta_2-2\pi i)_{b_2a_2}\mathfrak{A}_P^\dag(\theta_3)_{a_3b_3}\mathfrak{A}_P^\dag(\theta_4)_{a_4b_4}\mathfrak{A}_A^\dag(\theta_1)_{b_1a_1}|0\rangle\nonumber\\
&=&\langle0|j_\mu^L(0)_{a_0c_0}\mathfrak{A}_P^\dag(\theta_3)_{a_3b_3}\mathfrak{A}_P^\dag(\theta_4)_{a_4b_4}\mathfrak{A}_A^\dag(\theta_1)_{b_1a_1}\mathfrak{A}_A^\dag(\theta_2)_{b_2a_2}|0\rangle,\nonumber\\
&&\,\nonumber\\
\langle 0|j_\mu^L(0\!\!\!\!\!&)&\!\!\!\!\!_{a_0c_0}\mathfrak{A}_P^\dag(\theta_3-2\pi i)_{a_3b_3}\mathfrak{A}_P^\dag(\theta_4)_{a_4b_4}\mathfrak{A}_A^\dag(\theta_1)_{b_1a_1}\mathfrak{A}_A^\dag(\theta_2)_{b_2a_2}|0\rangle\nonumber\\
&=&\langle0|j_\mu^L(0)_{a_0c_0}\mathfrak{A}_P^\dag(\theta_4)_{a_4b_4}\mathfrak{A}_A^\dag(\theta_1)_{b_1a_1}\mathfrak{A}_A^\dag(\theta_2)_{b_2a_2}\mathfrak{A}_P^\dag(\theta_3)_{a_3b_3}|0\rangle,\nonumber\\
&&\,\nonumber\\
\langle 0|j_\mu^L(0\!\!\!\!\!&)&\!\!\!\!\!_{a_0c_0}\mathfrak{A}_P^\dag(\theta_4-2\pi i)_{a_4b_4}\mathfrak{A}_A^\dag(\theta_1)_{b_1a_1}\mathfrak{A}_A^\dag(\theta_2)_{b_2a_2}\mathfrak{A}_P^\dag(\theta_3)_{a_3b_3}|0\rangle\nonumber\\
&=&\langle0|j_\mu^L(0)_{a_0c_0}\mathfrak{A}_A^\dag(\theta_1)_{b_1a_1}\mathfrak{A}_A^\dag(\theta_2)_{b_2a_2}\mathfrak{A}_P^\dag(\theta_3)_{a_3b_3}\mathfrak{A}_P^\dag(\theta_4)_{a_4b_4}|0\rangle,\nonumber
\eeq
which imply, respectively,
\beq
\vec{F}(\theta_1-2\pi i,\theta_2,\theta_3,\theta_4)&=&\vec{H}(\theta_2,\theta_1,\theta_3,\theta_4),\label{smirnovone}\\
\vec{H}(\theta_2-2\pi i,\theta_1,\theta_3,\theta_4)&=&\vec{L}(\theta_1,\theta_2,\theta_3,\theta_4),\label{smirnovtwo}\\
\vec{L}(\theta_1,\theta_2,\theta_3-2\pi i,\theta_4)&=&\vec{Q}(\theta_1,\theta_2,\theta_4,\theta_3),\label{smirnovthree}\\
\vec{Q}(\theta_1,\theta_2,\theta_4-2\pi i,\theta_3)&=&\vec{F}(\theta_1,\theta_2,\theta_3,\theta_4).\label{smirnovfour}
\eeq

We combine Watson's theorem with the periodicity axiom, to express Equations (\ref{smirnovone}),
 (\ref{smirnovtwo}),  (\ref{smirnovthree}) and (\ref{smirnovfour})
in terms of only 
$\vec{F}(\theta_1,\theta_2,\theta_3,\theta_4)$. We combine (\ref{smirnovone}) with (\ref{watsonfour}), (\ref{watsonthree}) and (\ref{watsonsix}), and find 
\beq
\vec{F}(\!\!\!\!\!&&\!\!\!\!\!\theta_1-2\pi i,\theta_2,\theta_3,\theta_4)\nonumber\\
&&=\overleftrightarrow{M}_4(\theta_1,\theta_4)\overleftrightarrow{M}_3(\theta_1,\theta_3)\left[\overleftrightarrow{I}_1(\theta_1,\theta_2)\right]^{-1}\vec{F}(\theta_1,\theta_2,\theta_3,\theta_4).\label{watsonandsmirnovone}
\eeq
Combining (\ref{smirnovtwo}) with (\ref{watsontwo}), (\ref{watsonone}) and (\ref{watsonsix}) gives
\beq
\left[\overleftrightarrow{I}_1(\theta_1,\theta_2-2\pi i)\right]^{-1}\!\!\!\!\!&&\!\!\!\!\!\vec{F}(\theta_1,\theta_2-2\pi i,\theta_3,\theta_4)\nonumber\\
=\!\!\!\!\!&&\!\!\!\!\!\overleftrightarrow{M}_2(\theta_2,\theta_4)\overleftrightarrow{M}_1(\theta_2,\theta_4)\vec{F}(\theta_1,\theta_2,\theta_3,\theta_4).\label{watsonandsmirnovtwo}
\eeq
Combining (\ref{smirnovthree}) with (\ref{watsonthree}), (\ref{watsonone}) and (\ref{watsonseven}) gives
\beq
\overleftrightarrow{M}_3(\theta_1,\theta_3-2\pi i)\!\!\!\!\!&&\!\!\!\!\!\overleftrightarrow{M}_1(\theta_2,\theta_3-2\pi i)\vec{F}(\theta_1,\theta_2,\theta_3,\theta_4)\nonumber\\
=\!\!\!\!\!&&\!\!\!\!\!\left[\overleftrightarrow{I}_2(\theta_3,\theta_4)\right]^{-1}\vec{F}(\theta_1,\theta_2,\theta_3,\theta_4).\label{watsonandsmirnovthree}
\eeq
Finally, we combine (\ref{smirnovfour}) with (\ref{watsonfour}), (\ref{watsontwo}) and (\ref{watsonseven}) to find
\beq
\overleftrightarrow{M}_4(\theta_1,\theta_4-2\pi i)\!\!\!\!\!&&\!\!\!\!\!\overleftrightarrow{M}_2(\theta_2,\theta_4-2\pi i)\left[\overleftrightarrow{I}_2(\theta_3,\theta_4-2\pi i)\right]^{-1}\vec{F}(\theta_1,\theta_2,\theta_3,\theta_4-2\pi i)\nonumber\\
=\!\!\!\!\!&&\!\!\!\!\!\vec{F}(\theta_1,\theta_2,\theta_3,\theta_4).\label{watsonandsmirnovfour}
\eeq
The set of equations (\ref{watsonandsmirnovone}), (\ref{watsonandsmirnovtwo}), (\ref{watsonandsmirnovthree}) and (\ref{watsonandsmirnovfour}) are difficult to solve, for finite $N$. In the large-$N$ limit, the matrices $\overleftrightarrow{M}_{1,2,3,4}$ become diagonal and mutually commute, and the matrices $\overleftrightarrow{I}_{1,2}$ become their own inverses. This greatly simplifies the problem, allowing us to find the form factors. We expand the form factors in powers of $1/N$ as $\vec{F}(\theta_1,\theta_2,\theta_3,\theta_4)=\vec{F}^0(\theta_1,\theta_2,\theta_3,\theta_4)+\frac{1}{N}\vec{F}^1(\theta_1,\theta_2,\theta_3,\theta_4)+\dots$, simplifying the periodicity conditions for $\vec{F}^0(\theta_1,\theta_2,\theta_3,\theta_4)$. We combine (\ref{watsonandsmirnovone}) and (\ref{watsonandsmirnovtwo}) to get
\beq
\vec{F}^0\!\!\!\!\!&&\!\!\!\!\!(\theta_1-2\pi i,\theta_2-2\pi i,\theta_3,\theta_4)\nonumber\\
=\!\!\!\!\!&&\!\!\!\!\!\overleftrightarrow{M}_4(\theta_1,\theta_4)\overleftrightarrow{M}_3(\theta_1,\theta_3)\overleftrightarrow{M}_2(\theta_2,\theta_4)\overleftrightarrow{M}_1(\theta_2,\theta_3)\vec{F}^0(\theta_1,\theta_2,\theta_3,\theta_4),\label{watsonandsmirnovoneandtwo}
\eeq
or explicitly, in terms of the components of $\vec{F}^0(\theta_1,\theta_2,\theta_3,\theta_4)$,
\beq
F_1^0(\theta_1-2\pi i,\theta_2-2\pi i,\theta_3,\theta_4)&=&\left(\frac{\theta_{13}+\pi i}{\theta_{13}-\pi i}\right)\left(\frac{\theta_{24}+\pi i}{\theta_{24}-\pi i}\right)^2F_1^0(\theta_1,\theta_2,\theta_3,\theta_4),\nonumber\\
F_2^0(\theta_1-2\pi i,\theta_2-2\pi i,\theta_3,\theta_4)&=&\left(\frac{\theta_{14}+\pi i}{\theta_{14}-\pi i}\right)\left(\frac{\theta_{23}+\pi i}{\theta_{23}-\pi i}\right)\nonumber\\
&&\times\left(\frac{\theta_{24}+\pi i}{\theta_{24}-\pi i}\right)F_2^0(\theta_1,\theta_2,\theta_3,\theta_4),\nonumber\\
F_3^0(\theta_1-2\pi i,\theta_2-2\pi i,\theta_3,\theta_4)&=&\left(\frac{\theta_{13}+\pi i}{\theta_{13}-\pi i}\right)\left(\frac{\theta_{23}+\pi i}{\theta_{23}-\pi i}\right)\nonumber\\
&&\times\left(\frac{\theta_{24}+\pi i}{\theta_{24}-\pi i}\right)F_3^0(\theta_1,\theta_2,\theta_3,\theta_4),\nonumber\\
F_4^0(\theta_1-2\pi i,\theta_2-2\pi i,\theta_3,\theta_4)&=&\left(\frac{\theta_{14}+\pi i}{\theta_{14}-\pi i}\right)\left(\frac{\theta_{23}+\pi i}{\theta_{23}-\pi i}\right)^2F_4^0(\theta_1,\theta_2,\theta_3,\theta_4),\nonumber
\eeq
\beq
F_5^0(\theta_1-2\pi i,\theta_2-2\pi i,\theta_3,\theta_4)&=&\left(\frac{\theta_{14}+\pi i}{\theta_{14}-\pi i}\right)^2\left(\frac{\theta_{23}+\pi i}{\theta_{23}-\pi i}\right)F_5^0(\theta_1,\theta_2,\theta_3,\theta_4),\nonumber\\
F_6^0(\theta_1-2\pi i,\theta_2-2\pi i,\theta_3,\theta_4)&=&\left(\frac{\theta_{14}+\pi i}{\theta_{14}-\pi i}\right)\left(\frac{\theta_{13}+\pi i}{\theta_{13}-\pi i}\right)\nonumber\\
&&\times\left(\frac{\theta_{24}+\pi i}{\theta_{24}-\pi i}\right)F_6^0(\theta_1,\theta_2,\theta_3,\theta_4),\nonumber\\
F_7^0(\theta_1-2\pi i,\theta_2-2\pi i,\theta_3,\theta_4)&=&\left(\frac{\theta_{13}+\pi i}{\theta_{13}-\pi i}\right)^2\left(\frac{\theta_{24}+\pi i}{\theta_{24}-\pi i}\right)F_7^0(\theta_1,\theta_2,\theta_3,\theta_4),\nonumber\\
F_8^0(\theta_1-2\pi i,\theta_2-2\pi i,\theta_3,\theta_4)&=&\left(\frac{\theta_{14}+\pi i}{\theta_{14}-\pi i}\right)\left(\frac{\theta_{13}+\pi i}{\theta_{13}-\pi i}\right)\nonumber\\
&&\times\left(\frac{\theta_{23}+\pi i}{\theta_{23}-\pi i}\right)F_8^0(\theta_1,\theta_2,\theta_3,\theta_4).\nonumber
\eeq
The solution that satisfies (\ref{watsonandsmirnovoneandtwo}), (\ref{watsonsix}) and (\ref{watsonseven}) is
\beq
F_1^0(\theta_1,\theta_2,\theta_3,\theta_4)&=&\frac{g_1(\theta_1,\theta_2,\theta_3,\theta_4)}{(\theta_{13}+\pi i)(\theta_{24}+\pi i)^2},\nonumber\\
F_2^0(\theta_1,\theta_2,\theta_3,\theta_4)&=&\frac{g_2(\theta_1,\theta_2,\theta_3,\theta_4)}{(\theta_{14}+\pi i)(\theta_{23}+\pi i)(\theta_{24}+\pi i)},\nonumber\\
F_3^0(\theta_1,\theta_2,\theta_3,\theta_4)&=&\frac{g_2(\theta_1,\theta_2,\theta_4,\theta_3)}{(\theta_{13}+\pi i)(\theta_{23}+\pi i)(\theta_{24}+\pi i)},\nonumber\\
F_4^0(\theta_1,\theta_2,\theta_3,\theta_4)&=&\frac{g_1(\theta_1,\theta_2,\theta_4,\theta_3)}{(\theta_{14}+\pi i)(\theta_{23}+\pi i)^2},\nonumber\\
F_5^0(\theta_1,\theta_2,\theta_3,\theta_4)&=&\frac{g_1(\theta_2,\theta_1,\theta_3,\theta_4)}{(\theta_{14}+\pi i)^2(\theta_{23}+\pi i)},\nonumber\\
F_6^0(\theta_1,\theta_2,\theta_3,\theta_4)&=&\frac{g_2(\theta_2,\theta_1,\theta_3,\theta_4)}{(\theta_{14}+\pi i)(\theta_{13}+\pi i)(\theta_{24}+\pi i)},\nonumber\\
F_7^0(\theta_1,\theta_2,\theta_3,\theta_4)&=&\frac{g_1(\theta_2,\theta_1,\theta_4,\theta_3)}{(\theta_{13}+\pi i)^2(\theta_{24}+\pi i)},\nonumber\\
F_8^0(\theta_1,\theta_2,\theta_3,\theta_4)&=&\frac{g_2(\theta_2,\theta_1,\theta_4,\theta_3)}{(\theta_{14}+\pi i)(\theta_{13}+\pi i)(\theta_{23}+\pi i)},\label{solutionwatsonandsmirnovoneandtwo}
\eeq
where the functions $g_{1,\dots,8}(\theta_1,\theta_2,\theta_3,\theta_4)$ are periodic under $\theta_{1,2}\to\theta_{1,2}-2\pi i$.

Instead of the analyis of the previous paragraph, we could have combined (\ref{watsonandsmirnovthree}) and (\ref{watsonandsmirnovfour}) to obtain
\beq
&&\overleftrightarrow{M}_4(\theta_1,\theta_4-2\pi i)\overleftrightarrow{M}_3(\theta_1,\theta_3-2\pi i)\overleftrightarrow{M}_2(\theta_2,\theta_4-2\pi i)\nonumber\\
&&\times\overleftrightarrow{M}_1(\theta_2,\theta_3-2\pi i)\vec{F}^0(\theta_1,\theta_2,\theta_3-2\pi i,\theta_4-2\pi i)\nonumber\\
&&=\vec{F}^0(\theta_1,\theta_2,\theta_3,\theta_4).\label{watsonandsmirnovthreeandfour}
\eeq
The condition (\ref{watsonandsmirnovthreeandfour}) is equivalent to (\ref{watsonandsmirnovoneandtwo}). The solution of (\ref{watsonandsmirnovthreeandfour}) is 
(\ref{solutionwatsonandsmirnovoneandtwo})

The functions $g_{1,\dots 8}(\theta_1,\theta_2,\theta_3,\theta_4)$ are fixed by the annihilation pole axiom. Any of the two particles of the four-excitation form factor can annihilate with any of the two antiparticles. This means there must be annihilation poles at $\theta_{24},\theta_{14},\theta_{23},\theta_{13}=-\pi i$. For the $\theta_{24}=-\pi i$ pole, the annihilation-pole axiom implies
\beq
{\rm Res}|_{\theta_{24}=-\pi i}\langle0\!\!\!\!\!&|&\!\!\!\!\! j^L_\mu(0)_{a_0c_0}\mathfrak{A}_A^\dag(\theta_1)_{b_1a_1}\mathfrak{A}_P^\dag(\theta_3)_{a_3b_3}\mathfrak{A}_A^\dag(\theta_2)_{b_2a_2}\mathfrak{A}_P^\dag(\theta_4)_{a_4b_4}|0\rangle\nonumber\\
&=&2i\left\{\langle0| j^L_\mu(0)_{a_0c_0}\mathfrak{A}_A^\dag(\theta_1)_{b_1a_1}\mathfrak{A}_P^\dag(\theta_3)_{a_3b_3}|0\rangle\delta_{a_2a_4}\delta_{b_2b_4}\right.\nonumber\\
&&-\langle0|j^L_\mu(0)_{a_0c_0}\mathfrak{A}_A^\dag(\theta_1)_{b_1'a_1'}\mathfrak{A}_P(\theta_3)_{a_3'b_3'}|0\rangle\nonumber\\
&&\times\left.\delta_{a_2'a_4}\delta_{b_2'b_4}S_{AA}(\theta_{12})^{b_2'a_2';b_1'a_1'}_{d_1c_1;b_1a_1}S_{AP}(\theta_{23})^{d_1c_1;a_3'b_3'}_{a_3b_3;b_2a_2}\right\}.\label{annihilationfourparticles}
\eeq
We substitute the two-particle form factor into the right-hand side of 
(\ref{annihilationfourparticles}) to find
\beq
\langle0\!\!\!\!\!&|&\!\!\!\!\!j^L_\mu(0)_{a_0c_0}\mathfrak{A}_A^\dag(\theta_1)_{b_1a_1}\mathfrak{A}_P^\dag(\theta_3)_{a_3b_3}|0\rangle\delta_{a_2a_4}\delta_{b_2b_4}\nonumber\\
-\langle0\!\!\!\!\!&|&\!\!\!\!\!\mathcal{O}_{a_0c_0}\mathfrak{A}_A^\dag(\theta_1)_{b_1'a_1'}\mathfrak{A}_P(\theta_3)_{a_3'b_3'}|0\rangle\delta_{a_2'a_4}\delta_{b_2'b_4}S_{AA}(\theta_{12})^{b_2'a_2';b_1'a_1'}_{d_1c_1;b_1a_1}S_{AP}(\theta_{23})^{d_1c_1;a_3'b_3'}_{a_3b_3;b_2a_2}\nonumber\\
&&=-\epsilon_{\mu\nu}(p_1+p_3)^\nu\tanh\left(\frac{\theta_{13}}{2}\right)\nonumber\\
&&\times\frac{2\pi}{(\theta_{13}+\pi i)}\left\{\frac{2\pi i}{N\hat{\theta}_{23}}\left(\delta_{a_0a_4}\delta_{a_2a_3}\delta_{c_0a_1}\delta_{b_1b_3}\delta_{b_2b_4}-\frac{1}{N}\delta_{a_0c_0}\delta_{a_1a_4}\delta_{a_2a_3}\delta_{b_1b_3}\delta_{b_2b_4}\right)\right.\nonumber\\
&&+\frac{1}{N}\left(\frac{-2\pi i}{\hat{\theta}_{23}}+\frac{2\pi i}{\theta_{12}}-\frac{4\pi^2}{\theta_{12}\hat{\theta}_{23}}\right)\nonumber\\
&&\times\left(\delta_{a_0a_3}\delta_{a_2a_4}\delta_{a_1c_0}\delta_{b_2b_3}\delta_{b_1b_4}-\frac{1}{N}\delta_{a_0c_0}\delta_{a_1a_3}\delta_{a_2a_4}\delta_{b_2b_3}\delta_{b_1b_4}\right)\nonumber\\
&&\left.\frac{-2\pi i}{N\theta_{12}}\left(\delta_{a_0a_3}\delta_{a_1a_4}\delta_{a_2c_0}\delta_{b_1b_3}\delta_{b_2b_4}-\frac{1}{N}\delta_{a_0c_0}\delta_{a_2a_3}\delta_{a_1a_4}\delta_{b_1b_3}\delta_{b_2b_4}\right)\right\}.\nonumber
\eeq
We can repeat this for every other annihilation pole, which fixes the functions:
\beq
g_1(\theta_1,\theta_2,\theta_3,\theta_4)&=&0\nonumber\\
g_2(\theta_1,\theta_2,\theta_3,\theta_4)&=&8\pi^2 i\tanh\left(\frac{\theta_{13}}{2}\right)\nonumber\\
g_3(\theta_1,\theta_2,\theta_3,\theta_4)&=&8\pi^2 i\tanh\left(\frac{\theta_{14}}{2}\right)\nonumber\\
g_4(\theta_1,\theta_2,\theta_3,\theta_4)&=&0\nonumber\\
g_5(\theta_1,\theta_2,\theta_3,\theta_4)&=&0\nonumber\\
g_6(\theta_1,\theta_2,\theta_3,\theta_4)&=&8\pi^2 i\tanh\left(\frac{\theta_{23}}{2}\right)\nonumber\\
g_7(\theta_1,\theta_2,\theta_3,\theta_4)&=&0\nonumber\\
g_8(\theta_1,\theta_2,\theta_3,\theta_4)&=&8\pi^2 i\tanh\left(\frac{\theta_{24}}{2}\right).\nonumber
\eeq

The minimal four-particle form factor satisfying all of Smirnov's axioms for large $N$ is 
\beq
\langle 0\!\!\!\!\!&&\!\!\!\!\!|j_\mu^L(0)_{a_0c_0}|A,\theta_1,b_1,a_1;A,\theta_2,b_2,a_2;P,\theta_3,a_3,b_3;P,\theta_4,a_4,b_4\rangle\nonumber\\
&=&-\epsilon_{\mu\nu}(p_1+p_2+p_3+p_4)^\nu\frac{8\pi^2 i}{N}\nonumber\\
&&\times\left\{\frac{\tanh\left(\frac{\theta_{13}}{2}\right)}{(\theta_{14}+\pi i)(\theta_{23}+\pi i)(\theta_{24}+\pi i)}\right.\nonumber\\
&&\times\left(\delta_{a_0a_3}\delta_{a_1c_0}\delta_{a_2a_4}\delta_{b_1b_4}\delta_{b_2b_3}-\frac{1}{N}\delta_{a_0c_0}\delta_{a_1a_3}\delta_{a_2a_4}\delta_{b_1b_4}\delta_{b_2b_3}\right)\nonumber\\
&&+\frac{\tanh\left(\frac{\theta_{14}}{2}\right)}{(\theta_{13}+\pi i)(\theta_{23}+\pi i)(\theta_{24}+\pi i)}\nonumber\\
&&\times\left(\delta_{a_0a_4}\delta_{a_1c_0}\delta_{a_2a_3}\delta_{b_1b_3}\delta_{b_2b_4}-\frac{1}{N}\delta_{a_0c_0}\delta_{a_1a_4}\delta_{a_2a_3}\delta_{b_1b_3}\delta_{b_2b_4}\right)\nonumber\\
&&+\frac{\tanh\left(\frac{\theta_{23}}{2}\right)}{(\theta_{14}+\pi i)(\theta_{13}+\pi i)(\theta_{24}+\pi i)}\nonumber\\
&&\left(\delta_{a_0a_3}\delta_{a_1a_4}\delta_{a_2c_0}\delta_{b_1b_3}\delta_{b_2b_4}-\frac{1}{N}\delta_{a_0c_0}\delta_{a_2a_3}\delta_{a_1a_4}\delta_{b_1b_3}\delta_{b_2b_4}\right)\nonumber\\
&&+\frac{\tanh\left(\frac{\theta_{24}}{2}\right)}{(\theta_{14}+\pi i)(\theta_{13}+\pi i)(\theta_{23}+\pi i)}\nonumber\\
&&\left.\left(\delta_{a_0a_4}\delta_{a_1a_3}\delta_{a_2c_0}\delta_{b_1b_4}\delta_{b_2b_3}-\frac{1}{N}\delta_{a_0c_0}\delta_{a_2a_4}\delta_{a_1a_3}\delta_{b_1b_4}\delta_{b_2b_3}\right)\right\}.\label{finalanswerfourparticle}
\eeq

\section{Form factors of an arbitrary number of particles}
\setcounter{equation}{0}
\renewcommand{\theequation}{4.4.\arabic{equation}}

In this section we generalize our results to find the form factor with $M$ particles and $M$ antiparticles. We introduce the permutation $\sigma\in S_{M+1}$ which takes the set of numbers $0,1,\dots,M$ to $\sigma(0),\sigma(1),\dots,\sigma(M)$, respectively, and the permutation $\tau\in S_{M}$ which takes the set of numbers $1,2,\dots,M$ to $\tau(1),\tau(2),\dots,\tau(M)$, respectively. 

The form factor of the current operator with $2M$ excitations is
\beq
\langle 0\!\!\!&\!\!\!\! \vert \!\!\! &\!\!\!\! j_\mu^L(x)_{a_{0} \,a_{\!\,_{{2M+1}}}} \;\vert \; A,\theta_{1},b_{1},a_{1};\dots\,\nonumber\\
&& ;A,\theta_M,b_M,a_M;P,\theta_{M+1},a_{M+1},b_{M+1};\dots\, ;P,\theta_{2M},a_{2M},b_{2M}\rangle_{\rm in}\nonumber\\
&=&\langle 0\vert j_\mu^L(x)_{a_{0} a_{\!\,_{\!\,_{2M+1}}}}\mathfrak{A}^\dagger_A(\theta_{1})_{b_{1}a_{1}}\dots\mathfrak{A}^\dagger_A(\theta_M)_{b_{\!\,_{M}} a_{\!\,_{M}}}\nonumber\\
&&\times\mathfrak{A}^\dagger_{P}(\theta_{M+1})_{a_{\!\,_{M+1}} b_{\!\,_{M+1}}}\dots\mathfrak{A}^\dagger_{P}(\theta_{2M})_{a_{\!\,_{2M}} b_{\!\,_{2M}}}\vert 0\rangle\nonumber\\
&=&\frac{-\epsilon_{\mu\nu}}{N^{M-1}}
\left(p_{1}+\cdots +p_{2M}\right)^{\nu}\;
\sum_{\sigma,\tau} F_{\sigma \tau}(\theta_{1},\dots,\theta_{2M})\;  e^{-ix\cdot \sum_{j=1}^{2M}p_{j}}   \nonumber\\
&\times\!\!\!&\!\!\!\left[\prod_{j=0}^M\delta_{a_{j} a_{\sigma(j)+M}}\prod_{k=1}^{M}\delta_{b_{k} b_{\tau(k)+M}}\right.\nonumber\\
&&\left.-\frac{1}{N}\delta_{a_{\!\,_{0}} a_{\!\,_{{2M+1}}}}
\delta_{a_{\!\,_{l_{\sigma}}} 
a_{\sigma(0)+M}}\prod_{j=1,\,j\neq l_{\sigma}}^M \delta_{a_{j} a_{\sigma(j)+M}}\prod_{k=1}^M\delta_{b_{k} b_{\tau(k)+M}}\right]
,\label{generalformfactor}
\eeq
where $l_{\sigma}$ is defined by $\sigma(l_{\sigma})+M=2M+1$. This is the most general expression consistent with Lorentz invariance, a conserved traceless current (guaranteed by the second term in square brackets) and crossing. 

To simplify our terminology, we say that excitation $h$ is the particle or antiparticle with rapidity $\theta_{h}$ and left and right indices $a_{h}, b_{h}$, respectively. 

We expand the functions $F_{\sigma \tau}(\theta_{1},\dots,\theta_{2M})$ in powers of $1/N$:
\beq
F_{\sigma \tau}(\theta_{1},\dots,\theta_{2M})&=&
F^0_{\sigma \tau}(\theta_{1},\dots,\theta_{2M})\nonumber\\
&&+\frac{1}{N}F^1_{\sigma \tau}(\theta_{1},\dots,\theta_{2M})+
\frac{1}{N^{2}}F^{2}_{\sigma\tau}(\theta_{1},\dots,\theta_{2M})+\cdots\nonumber
\eeq
keeping only the first term.

We now apply the scattering axiom on Eq.(\ref{generalformfactor}):
\beq
\langle 0\!\!\!\!\!&\vert&\!\!\! \!\! j_\mu^L(x)_{a_{0} a_{\!\,_{2M+1}}} \mathfrak{A}^\dagger_{I_{1}}(\theta_{1})_{C_{1}}\dots\mathfrak{A}^\dagger_{I_{i}}(\theta_{i})_{C_{i}}\mathfrak{A}^\dagger_{I_{i+1}}(\theta_{i+1})_{C_{i+1}}\dots\mathfrak{A}^\dagger_{I_{2M}}(\theta_{2M})_{C_{2M}}\vert 0\rangle\nonumber\\
&=&S_{I_{i+1} I_{i}}(\theta_{i}-\theta_{i+1})^{C^{\prime}_{i+1};C^{\prime}_{i}}_{C_{i};C_{i+1}}\nonumber\\
&&\times\langle 0\vert j_\mu^L(x)_{a_{0} a_{\!\,_{2M+1}}} \mathfrak{A}^\dagger_{I_{1}}(\theta_{1})_{C_{1}}\dots\nonumber\\
&&\times\mathfrak{A}^\dagger_{I_{i+1}}(\theta_{i+1})_{C^{\prime}_{i+1}}
\mathfrak{A}^\dagger_{I_{i}}(\theta_{i})_{C^{\prime}_{i}}\dots\mathfrak{A}^\dagger_{I_{2M}}(\theta_{2M})_{C_{2M}}\vert 0\rangle,\label{watsonstheorem}
\eeq
where, for each $k$, $I_{k}=P$ for a particle or $I_{k}=A$ for an antiparticle, and $C_{k}$ is the ordered set of indices $C_{k}=(a_{k} ,b_{k})$ for $I_{k}=P$, or $C_{k}=(b_{k}, a_{k})$ for $I_{k}=A$. 
We use (\ref{watsonstheorem}) to interchange the creation operator of the excitation $h$ with the creation operator of the excitation $i$ in (\ref{generalformfactor}). 
There are four different ways the function $F^0_{\sigma \tau}(\theta_{1},\dots,\theta_{2M})$ can be affected by interchanging the excitations  $h$ and $i$, for a given $\sigma$ and $\tau$. If excitation $h$ and excitation $i$ are both particles or both antiparticles, then the rapidities $\theta_{h}$ and $\theta_{i}$ are interchanged in the function $F^0_{\sigma \tau}(\theta_{1},\dots,\theta_{2M})$. If excitation $h$ is a particle, excitation $i$ is an antiparticle, and $\sigma(i)+M\neq h, \tau(i)+M\neq h$, then the function $F^0_{\sigma \tau}(\theta_{1},\dots,\theta_{2M})$ is unchanged. If 
excitation $h$ is a particle, excitation $i$ an antiparticle, and either 
$\sigma(i)+M=h$, $\tau(i)+M\neq h$, or $\sigma(i)+M\neq h$, $\tau(i)+M=h$, then we multiply 
$F^0_{\sigma \tau}(\theta_{1},\dots,\theta_{2M})$ by the pure phase
$\frac{\theta_{ih}+\pi {\rm i}}{\theta_{ih}-\pi {\rm i}}$. If excitation $h$ is a particle, excitation $i$ is an antiparticle and $\sigma(i)+M=h$, $\tau(i)+M=h$, then we multiply the function 
$F^0_{\sigma \tau}(\theta_{1},\dots,\theta_{2M})$ by the pure phase $\left(\frac{\theta_{ih}+\pi {\rm i}}{\theta_{ih}-\pi {\rm i}}\right)^{2}$. 
The rules for interchanging creation operators described in the previous paragraph suggest an underlying Abelian structure for the large-$N$ limit. The pure phase we use in the scattering axiom, namely 
$1$, $\frac{\theta_{ih}+\pi {\rm i}}{\theta_{ih}-\pi {\rm i}}$ or $\left(\frac{\theta_{ih}+\pi {\rm i}}{\theta_{ih}-\pi {\rm i}}\right)^{2}$ is similar to the S-matrix element of 
a theory of colorless particles.

Smirnov's periodicity axiom  states
\beq
\langle 0\!\!\!\!\!&\vert&\!\!\!\!\! j_\mu^L(x)_{a_{\!\,_{0}} a_{{\!\,_{2M+1}}}}\mathfrak{A}^\dagger_{I_{1}}(\theta_{1})_{C_{1}}\mathfrak{A}^\dagger_{I_{1}}(\theta_{2})_{C_{2}}\dots 
\mathfrak{A}^\dagger_{I_M}(\theta_M)_{C_M}\vert 0\rangle\nonumber\\
&=&\langle 0\vert j_\mu^L(x)_{a_{\!\,_{0}} a_{\!\,_{{2M+1}}}}\mathfrak{A}^\dagger_{I_M}(\theta_M-2\pi {\rm i})_{C_M}\nonumber\\
&&\times\mathfrak{A}^\dagger_{I_{1}}(\theta_{1})_{C_{1}}\dots 
\mathfrak{A}^\dagger_{I_{M-1}}(\theta_{M-1})_{C_{M-1}}\vert 0\rangle. \label{perio}
\eeq
In terms of the function $F^0_{\sigma \tau}(\theta_{1},\dots,\theta_{2M})$, (\ref{perio}) is
\beq
F^0_{\sigma \tau}\!\!\!\!\!&(&\!\!\!\!\!\theta_{1},\dots,\theta_{2M}) \nonumber\\
&=& F^0_{\sigma \tau}(\theta_{2M}-2\pi {\rm i},\theta_{1},\dots,\theta_{2M-1})\nonumber\\
&=&F^0_{\sigma\tau}(\theta_{2M-1}-2\pi {\rm i},\theta_{2M}-2\pi {\rm i},\theta_{1},\dots,\theta_{2M-2})\nonumber\\
&=&\cdots .\label{periodicityaxiom}
\eeq
The general solution of (\ref{watsonstheorem}) and (\ref{periodicityaxiom}) is
\beq
F^0_{\sigma\tau}\!\!\!\!\!&(&\!\!\!\!\!\theta_{1},\dots,\theta_{2M})\nonumber\\
&=&\!\!\!\!\!\frac{H_{\sigma \tau}(\theta_{1},\dots,\theta_{2M})}{\prod_{j=1; j\neq l_{\sigma}}^M\left(\theta_{j}-\theta_{\sigma(j)+M}+\pi {\rm i}\right)\prod_{k=1}^M\left(\theta_{k}-\theta_{\tau(k)+M}+\pi {\rm i}\right)}\;,\label{solutionform}
\eeq
where $\sigma(l_{\sigma})+M=2M+1$, and the functions $H_{\sigma \tau}(\theta_{1},\dots,\theta_{2M})$ are holomorphic and periodic in $\theta_{j}$, with
period $2\pi {\rm i}$, for each $j=1,\dots,2M$.

The annihilation-pole axiom states that the $2M+2$-excitation form factor:
\beq
\langle 0\!\!\!&\vert&\!\!\!j_\mu^L(0)_{a_{0} a_{\!\,_{2M+3}}}\left[\prod_{j=1}^M\mathfrak{A}^\dagger_A(\theta_{j})_{b_{j}a_{j}}\right]\left[\prod_{k=M+1}^{2M}\mathfrak{A}^\dagger_P(\theta_{k})_{a_{k}b_{k}}\right]\nonumber\\
&&\times\mathfrak{A}^\dagger_A(\theta_{{2M+1}})_{b_{\!\,_{2M+1}}a_{\!\,_{2M+1}}}\mathfrak{A}^\dagger_{P}(\theta_{\!\,_{2M+2}})_{a_{\!\,_{2M+2}}b_{\!\,_{2M+2}}}\vert 0\rangle\nonumber\\
&=&-\epsilon_{\mu\nu}(p_{1}+\cdots+p_{{2M+2}})^\nu\;\nonumber\\
&&\times\mathcal{F}(\theta_{1},\dots,\theta_{\!\,_{2M+2}})_{a_{0}a_{2}\dots a_{\!\,_{2M+3}};b_{1}\dots b_{\!\,_{2M+2}}},\nonumber
\eeq
has a pole at $\theta_{2M+1}-\theta_{2M+2}=-\pi i$, with a residue proportional to the form factor of $2M$ excitations, such that
\beq
 {\rm Res}\vert_{\theta_{2M+1}-\theta_{2M+2}=-\pi i}\!\!\!\!\!\!&&\!\!\! 
{\mathcal F}(\theta_{1},\dots,\theta_{\!\,_{2M+2}})_{a_{0}\dots a_{\!\,_{2M+3}};\; b_{1}\dots b_{\!\,_{2M+2}}}\nonumber\\
&=&2i \mathcal{F}(\theta_{1},\dots,\theta_{2M})_{a_{0} a^{\prime}_{1} \dots a^{\prime}_{2M}a_{\!\,_{2M+3}};\;  b^{\prime}_{1}\;\cdots \;b^{\prime}_{2M}}
\delta_{a^{\prime}_{\!\,_{2M+1}}a_{\!\,_{2M+2}}}
\delta_{b^{\prime}_{\!\,_{2M+1}}b_{\!\,_{2M+2}}}\nonumber\\
&& \times \left[\delta_{a^{\prime}_{1} a_{1}}\delta_{b^{\prime}_{1} b_{1}}
\;\cdots\;
\delta_{a^{\prime}_{\!\,_{2M+1}}a_{\!\,_{2M+1}}}\delta_{b^{\prime}_{\!\,_{2M+1}}b_{\!\,_{2M+1}}}\right.\nonumber\\
&&\left.-S_{AA}(\theta_{1\,2M+1})^{b^{\prime}_{\!\,_{2M+1}}
a^{\prime}_{\!\,_{2M+1}};b^{\prime}_{1} a^{\prime}_{1}}_{d_{1}c_{1};b_{1}a_{1}} \;\cdots \;S_{AA}(\theta_{M\,2M+1})^{d_{M-1}c_{M-1};b^{\prime}_{M}   
a^{\prime}_{M}}_{d_Mc_M;b_Ma_M}\right.\nonumber\\
&& \left.\times S_{AP}(\theta_{2M+1\,M+1})^{d_Mc_M;a^{\prime}_{M+1}b^{\prime}_{M+1}}_{c_{M+1}d_{M+1};a_{M+1}b_{M+1}}\;\cdots \;\right.\nonumber\\
&&\left.\times S_{AP}(\theta_{2M+1\,2M})^{c_{2M-1}d_{2M-1};a^{\prime}_{2M}b^{\prime}_{2M}}_{c_{2M}d_{2M};a_{2M}b_{2M}}\right]\,.     \label{annihilationpole}
\eeq
Equation (\ref{annihilationpole}) fixes the functions:
\beq
H_{\sigma \tau}=\left\{ \begin{array}{ccc}2\pi {\rm i}(4\pi)^{M-1}\tanh\left(\frac{\theta_{l_{\sigma}}-\theta_{\sigma(0)+M}}{2}\right) &,&\,\,{\rm if\;}\sigma(j)\neq\tau(j),\,{\rm for\,all}\, j   \\
0&,&{\rm otherwise}\end{array}\right.  \;\;. \label{Hconstant}
\eeq
This concludes our derivation of 
all the form factors of the current operator. They are completely specified in (\ref{generalformfactor}), (\ref{solutionform}) and (\ref{Hconstant}).

\section{Vacuum expectation values of products of current operators }
\setcounter{equation}{0}
\renewcommand{\theequation}{4.5.\arabic{equation}}

Once we have calculated the general form factor for the Noether current operator, we can calculate its two-point correlation function. The current-current correlation function is
\beq
W^{j}_{\mu\nu}(x)_{a_{0} c_{0} ;e_{0}  f_{0}}=\langle 0 \vert j^{L}_{\mu}(x)_{a_{0}c_{0}} \; j^{L}_{\nu}(0)_{e_{0}f_{0}} \vert 0 \rangle
=\sum_{M=1}^{\infty} W_{\mu\nu}^{2M}(x)_{a_{0}c_{0}e_{0}f_{0}}, \label{currcorr}
\eeq
where the contribution from the $2M$-excitation form factor is given by
\beq
W_{\mu\nu}^{2M}\!\!\!\!\!&(&\!\!\!\!\!x)_{a_{0}c_{0}e_{0}f_{0}}=\frac{1}{N(M!)^{2}}\int\frac{d\theta_{1}\dots d\theta_{2M}}{(2\pi)^{2M}}
\;e^{-ix\cdot\sum_{j=1}^{2M} p_{j} }\nonumber\\
&&\times \langle 0 \vert j_\mu^L(0)_{a_{0}c_{0}}\vert A,\theta_{1},b_{1},a_{1};\dots;\nonumber\\
&&\,\,\,\,\,\,\,\,\,\,A,\theta_M,b_Ma_M;P,\theta_{M+1},a_{M+1},b_{M+1};\dots;P,\theta_{2M},a_{2M},b_{2M}\rangle_{\rm in}\nonumber\\
&&\times\langle0\vert j_\nu^L(0)_{e_{0}f_{0}}\vert  A,\theta_{1},b_{1},a_{1};\dots;\nonumber\\
&&\,\,\,\,\,\,\,\,\,\,A,\theta_M,b_Ma_M;P,\theta_{M+1},a_{M+1},b_{M+1};\dots;P,\theta_{2M},a_{2M},b_{2M}\rangle_{\rm in}^*.\nonumber
\eeq
Substituting the form factors (\ref{generalformfactor}), (\ref{solutionform}), (\ref{Hconstant}), we find
\beq
W_{\mu\nu}^{2M}\!\!\!\!\!&(&\!\!\!\!\!x)_{a_{0}c_{0}e_{0}f_{0}}=\frac{1}{(M!)^{2}}\int \prod_{j=1}^{2M}\frac{d\theta_{j}}{4\pi} \;\;
e^{-ix\cdot \sum_{j=1}^{2M}p_{j}}    \nonumber\\
&&\times
\epsilon_{\mu\alpha}\epsilon_{\nu\beta}(p_{1}+\cdots+p_{2M})^\alpha
(p_{1}+\cdots+p_{2M})^\beta      \nonumber\\
&&\times\left[\sum_{\sigma,\tau\in S_M}\frac{\vert H_{\sigma \tau}(\theta_1,\dots,\theta_{2M})\vert^{2}    (\delta_{a_{0}e_{0}}\delta_{c_{0}f_{0}}-\frac{1}{N}\delta_{a_{0}c_{0}}\delta_{e_{0}f_{0}})}
{\prod_{j=1;\,j\neq l_{\sigma}}^{M}\vert\theta_{j}-\theta_{\sigma(j)+M}+\pi {\rm i}\vert^{2}\prod_{k=1}^{M}
\vert \theta_{k}-\theta_{\tau(k)+M}+\pi {\rm i} \vert^{2}}\right.
\nonumber\\
&&\left.+\mathcal{O}\left(\frac{1}{N^{2}}\right)\right],\label{wightmanone}
\eeq
where we have used
\beq
\sum_{a_{1},\dots,a_{2M},b_{1},\dots,b_{2M}}\!\!\!\!\!&&\!\!\!\!\!\left[\prod_{j=0}^M\delta_{a_{j}a_{\sigma(j)+M}}\prod_{k=1}^M\delta_{b_{k}b_{\tau(k)+M}}\right.\nonumber\\
&&\,\,\,\,\,\,\,\,\,\,-\left.\frac{1}{N}\delta_{a_{0}a_{\!\,_{2M+1}}}\delta_{a_{l_{\sigma}}a_{\sigma(0)+M}}\prod_{j=1;\,j\neq l_{\sigma}}^M\delta_{a_{j}a_{\sigma(j)+M}}\prod_{k=1}^M\delta_{b_{k}b_{\tau(k)+M}}\right]\nonumber\\
\times&&\!\!\!\!\!\left[\prod_{j=0}^M\delta_{a^{\prime}_{j} a^{\prime}_{\omega(j)+M}}\prod_{k=1}^M\delta_{b_{k}b_{\varphi(k)+M}}\right.\nonumber\\
&&\,\,\,\,\,\,\,\,\,\,-\left.\frac{1}{N}\delta_{a^{\prime}_{0} 
a^{\prime}_{\!\,_{2M+1}}}\delta_{a^{\prime}_{l_\omega}a^{\prime}_{\omega(0)}}\prod_{j=1\,j\neq l_{\omega}}^M\delta_{a^{\prime}_{j} a^{\prime}_{\omega(j)+M}}\prod_{k=1}^M\delta_{b_{k}b_{\varphi(k)+M}}\right]\nonumber\\ \nonumber \\
&=&N^{2M-1}
(\,
 \delta_{a_{0} e_{0}}\delta_{c_{0} f_{0}} - 
\delta_{a_{0}c_{0}}\delta_{e_{0} f_{0}}/N \,)
\left[ \delta_{\sigma\omega}\delta_{\tau\varphi}+\mathcal{O}\left(\frac{1}{N^{2}}\right) \right],\nonumber
\eeq
where $\{a_{j} \}=a_{0},a_{1},a_{2},\dots,a_{2M},c_{0}$ and $\{a^{\prime}_{j}\}=e_{0},a_{1},a_{2},
\dots,a_{2M},f_{0}$. 

The contribution to (\ref{wightmanone}) from each pair $\sigma,\tau$ is the same (because there is no change if the integration variables are interchanged). There are $(M!)^{2}$ pairs $\sigma,\tau$ that satisfy $H_{\sigma \tau}\neq0$, by (\ref{Hconstant}). We choose the contribution from one pair $\sigma,\tau$ in 
(\ref{wightmanone}) and multiply it by $(M!)^{2}$. We choose $\tau(j)=j,$ for $j=1,\dots,M$, and 
$\sigma(1)=2M+1, \sigma(j)=j-1,$ for $j=2,\dots,M,$ such that
\beq
W_{\mu\nu}^{2M}(x)_{a_{0}c_{0}e_{0}f_{0}}&=&\int
\prod_{j=1}^{2M}\frac{d\theta_{j}}{4\pi}
\;\;e^{-ix\cdot\sum_{j=1}^{2M}p_{j}} 
\;4\pi^{2}(4\pi)^{2M-2}\;
\;\nonumber\\
&&\times\left(\delta_{a_{0}e_{0}}\delta_{c_{0}f_{0}}-\frac{1}{N}\delta_{a_{0}c_{0}}\delta_{e_{0}f_{0}}\right)
 \nonumber\\
&\times&
\epsilon_{\mu\alpha}\,\epsilon_{\nu\beta}(p_{1}+\cdots+p_{2M})^\alpha
(p_{1}+\cdots+p_{2M})^\beta
\nonumber\\
&&\times  \frac{1}{(\theta_{1}-\theta_{M+1})^{2}+\pi^{2}} \frac{1}{(\theta_{2}-\theta_{M+2})^{2}+\pi^{2}} \dots\nonumber\\
&&\times \frac{1}{(\theta_M-\theta_{2M})^{2}+\pi^{2}}  
\nonumber\\
&&\times    \frac{1}{(\theta_{2}-\theta_{M+1})^{2}+\pi^{2}} \frac{1}{(\theta_{3}-\theta_{M+2})^{2}+\pi^{2}} \dots\nonumber\\
&&\times \frac{1}{(\theta_M-\theta_{2M-1})^{2}+\pi^{2}}\tanh^2\left(\frac{\theta_1-\theta_{2M}}{2}\right)
+O\left(\frac{1}{N^{2}}\right)\nonumber
\eeq
We further relabel the integration variables as $\theta_{1}\to\theta_{1}, \theta_{2}\to\theta_{3}, \theta_{3}\to\theta_5,\dots,\theta_M\to\theta_{2M-1};\theta_{M+1}\to\theta_{2}, \theta_{M+2}\to\theta_{4},\dots,\theta_{2M}\to\theta_{2M}$. This yields the expression for the non-time-ordered correlation function of two current operators:
\beq
W_{\mu\nu}^{j}(x)_{a_{0}c_{0}e_{0}f_{0}}
&=&\left(\delta_{a_{0}e_{0}}\delta_{c_{0}f_{0}}-
\frac{1}{N}\delta_{a_{0}c_{0}}\delta_{e_{0}f_{0}}\right)
\nonumber\\
&\times &\sum_{M=1}^{\infty}\frac{1}{4}\int  \prod_{j=1}^{2M}d\theta_{j}\;\;
\;e^{-ix\cdot \sum_{j=1}^{2M}p_{j} }\;\nonumber\\
&&\times\epsilon_{\mu\alpha}\,\epsilon_{\nu\beta}(p_1+\cdots+p_{2M})^{\alpha}(p_1+\cdots+p_{2M})^\beta\nonumber\\
&&\times\prod_{j=1}^{2M-1}
\frac{\tanh^2\left(\frac{\theta_1-\theta_{2M}}{2}\right)}{(\theta_{j}-\theta_{j+1})^{2}+\pi^{2}}\;+\;O\left(\frac{1}{N^{2}}\right).   \label{finalMcurr}
\eeq

\section{Form factors of the stress-energy-momentum tensor}
\setcounter{equation}{0}
\renewcommand{\theequation}{4.6.\arabic{equation}}

The stess- energy momentum tensor of the principal chiral sigma model is given by
\beq
T_{\mu\nu}(x)=\frac{1}{2\pi}(\delta_\mu^\alpha\delta_\nu^\beta+\delta_\mu^\beta\delta_\nu^\alpha-\eta_{\mu\nu}\eta^{\alpha\beta})\,{\rm Tr}\,\partial_\alpha U(x)^\dag\partial_\beta U(x)+\lambda\eta_{\mu\nu},\nonumber
\eeq
where $\lambda$ is chosen to normal order this operator, so that the vacuum energy is zero. This operator is explicitly conserved $(\partial^\mu T_{\mu\nu}=0)$, symmetric $(T_{\mu\nu}=T_{\nu\mu})$
and it is an $SU(N)\times SU(N)$-color singlet. The fact that the bare field is unitary means that there is less freedom when defining  an stress-energy-momentum tensor, quadratic in derivatives, than in ordinary scalar field theories. There is no color-singlet total divergence term of dimension two we can add to the $T_{\mu\nu}$.

This stress-energy-momentum tensor operator is invariant under ${\rm SU}(N)\!\times \!{\rm SU}(N)$ transformations. Thus the only non-vanishing form factors have equal number of particles and antiparticles in the in-state ket. The general form factor with $M$ particles and $M$ antiparticles, which respects energy-momentum conservation, is
\beq
\langle 0\vert\;T_{\mu\nu}(0)\!\!\!&\!\!\!\vert\!\!\!&\!\!\!A,\theta_{1},b_{1},a_{1};\dots;A,\theta_M,b_M,a_M;P,\theta_{M+1},a_{M+1},b_{M+1};\nonumber\\
&&\dots;P,\theta_{2M},a_{2M},b_{2M}\rangle_{\rm in}\nonumber\\
&=&\left[(p_{1}+\cdots+p_{2M})_\mu (p_{1}+\cdots+p_{2M})_\nu-\eta_{\mu\nu}(p_{1}+\cdots+p_{2M})^2\right]\nonumber\\
&&\times\frac{1}{N^{M-1}}\sum_{\sigma,\tau\in S_M}F_{\sigma\tau}(\theta_{1},\dots,\theta_{2M})\nonumber\\
&&\times\prod_{j=1}^M\delta_{a_{j}a_{\sigma(j)+M}}\prod_{k=1}^M\delta_{b_{k}b_{\tau(k)+M}},\label{stressformfactor}
\eeq
where $\sigma, \tau\in S_M$ are permutations of the integers $1,2,\dots,M$ (this is different from the convention in Section III. Recall that there the permutation $\sigma$ was defined as an element of $S_{M+1}$). 

We expand the function 
$F_{\sigma,\tau}(\theta_{1},\dots,\theta_{2M})$ in powers of $1/N$, {\em i.e.}, as $F^0_{\sigma,\tau}(\theta_{1},\dots,\theta_{2M})+\frac{1}{N}F^1_{\sigma,\tau}(\theta_{1},\dots,\theta_{2M})+\cdots$, keeping only the first term.

The form factors in (\ref{stressformfactor}) behave the same way as the current-operator form factors under Watson's theorem and the periodicity axiom. These two axioms give us the solution
\beq
F^0_{\sigma\tau}\!\!\!\!\!&(&\!\!\!\!\!\theta_{1},\dots,\theta_{2M})\nonumber\\
&=&\frac{H_{\sigma \tau}(\theta_{1},\dots,\theta_{2M})}{\prod_{j=1}^M(\theta_{j}-\theta_{\sigma(j)+M}+\pi {\rm i})\prod_{k=1}^M(\theta_{k}-\theta_{\tau(k)+M}+\pi {\rm i})}.\label{SEMFF}
\eeq
The minimal choice is to make $H_{\sigma,\tau}(\theta_{1},\dots,\theta_{2M})=H_{\sigma \tau}$ constants. These constants can be fixed by the annihilation pole axiom, once we fix the constant for the two-particle form factor.

For $M=1$, Equation (\ref{stressformfactor}) becomes
\beq
\langle 0\!\!\!\!\!&\vert&\!\!\!\!\! T_{\mu\nu}(x)\vert  A,\theta_{1},b_{1},a_{1};P,\theta_{2},a_{2},b_{2}\rangle_{\rm in}\nonumber\\
&=&[(p_{1}+p_{2})_\mu(p_{1}+p_{2})_\nu-\eta_{\mu\nu}(p_1+p_2)^2] \frac{g}{(\theta_{12}+\pi {\rm i})^{2}}
e^{-ix(p_{1}+p_{2})}\delta_{a_{1}a_{2}}\delta_{b_{1}b_{2}}\nonumber\\
&&+O\left(\frac{1}{N}\right).\label{twoparticlestress}
\eeq
We fix the constant $g$ by requiring that 
\beq
\int\,dx^1 \,T_{00}(x)\vert  A,\theta_{1},b_{1},a_{1}\rangle_{\rm in}=m\cosh\theta_{1}\vert  A,\theta_{1},b_{1},a_{1}\rangle_{\rm in}.\label{stressconstantfixing}
\eeq
Notice that the pole in (\ref{twoparticlestress}) has vanishing residue. Therefore, by the annihilation-pole axiom, the vacuum energy
is zero. 

We next apply crossing, changing one of the incoming particles in (\ref{twoparticlestress}) to an outgoing antiparticle:
\beq
\langle A,\theta_2,b_2,a_2\!\!\!\!\!&\vert&\!\!\!\!\! T_{\mu\nu}(x)\vert A,\theta_1,b_1,a_1\rangle\nonumber\\
&=&[(p_1-p_2)_\mu(p_1-p_2)_\nu-\eta_{\mu\nu}(p_1-p_2)^2]\frac{g}{(\theta_{12}+2\pi i)^2}\nonumber\\
&&\times e^{-ix\cdot(p_1-p_2)}\delta_{a_1a_2}\delta_{b_1b_2}.\nonumber
\eeq
The matrix element of the energy density is then
\beq
\langle A,\theta_2,b_2,a_2\!\!\!\!\!&\vert&\!\!\!\!\! T_{00}(x)\vert A,\theta_1,b_1,a_1\rangle\nonumber\\
&=&[(p_{1\,0}p_{1\,0}+p_{2\,0}p_{2\,0}-2p_{1\,0}p_{2\,0}-2m^2+2p_1\cdot p_2]\nonumber\\
&&\times\frac{g}{(\theta_{12}+2\pi i)^2}e^{-ix\cdot(p_1-p_2)}\delta_{a_1a_2}\delta_{b_1b_2}.\nonumber\\
&=&[(p_{1\,0}+p_{2\,0})^2-2m^2+2p_1\cdot p_2]\nonumber\\
&&\times\frac{g}{(\theta_{12}+2\pi i)^2}e^{-ix\cdot(p_1-p_2)}\delta_{a_1a_2}\delta_{b_1b_2}.\nonumber
\eeq
We integrate over the spatial coordinate $x^1$, yielding
\beq
\int dx^1\,\langle A,\theta_2,b_2,a_2\!\!\!\!\!&\vert&\!\!\!\!\! T_{00}(x)\vert A,\theta_1,b_1,a_1\rangle\nonumber\\
&=&[(m\cosh \theta_1+m \cosh\theta_2)^2-2m^2+2p_1\cdot p_2]\nonumber\\
&&\times\frac{g}{(\theta_{12}+2\pi i)^2}\frac{2\pi}{m\cosh\theta_2}\nonumber\\
&&\times\delta(\theta_1-\theta_2)\delta_{a_1a_2}\delta_{b_1b_2}.\nonumber\\
&=&m\cosh\theta_1\frac{g}{2\pi^2}4\pi\delta(\theta_1-\theta_2)\delta_{a_1a_2}\delta_{b_1b_2}.\nonumber
\eeq
The condition (\ref{stressconstantfixing}) imples $g=-2\pi^2$.

The constants $H_{\sigma \tau}$ for the $2M$-particle form factor are fixed by the annihilation pole axiom (Equation (\ref{annihilationpole})), which gives the values
\beq
H_{\sigma \tau}=\left\{\begin{array}{ccc}(-2\pi^{2})(4\pi)^{M-1}\!\!\!&,&{\rm for}\,\sigma(j)\neq\tau(j),\,{\rm for\,all}\,j\\
0\;&,&\!\!\!\!\!\!\!\!\!\!{\rm otherwise}\end{array}\right.   \;. \label{SEMH}
\eeq
The $2M$-particle form factor has a total of $(M!)^{2}/2$ non-vanishing terms.

To summarize the results of this section, (\ref{stressformfactor}), (\ref{SEMFF}), (\ref{SEMH}) determine all the form factors of the stress-energy-momentum tensor.

\section{Correlation function of the stress-energy-momentum tensor}
\setcounter{equation}{0}
\renewcommand{\theequation}{4.7.\arabic{equation}}

In this section we obtain the vacuum expectation value of the product of two stress-energy-momentum-tensor 
operators. In other words, we find
\beq
W^{T}_{\mu\nu\alpha\beta}(x)=\frac{1}{N^{2}}\langle 0\vert T_{\mu\nu}(x)T_{\alpha\beta}(0)\vert 0\rangle
=\sum_{M=1}^\infty W_{\mu\nu\alpha\beta}^{2M}(x)\,,\label{SEMTsum}
\eeq
where the terms in the sum over $M$ are defined as
\beq
W_{\mu\nu\alpha\beta}^{2M}(x)&=&\frac{1}{N^{2}}\frac{1}{(M!)^{2}}\int\prod_{j=1}^{2M}\frac{d\theta_{j}}{4\pi}
\;e^{-ix\cdot\sum_{j=1}^{2M}p_{j}}\nonumber\\
&&\times\langle 0\vert T_{\mu\nu}(0)
\vert A,\theta_{1},b_{1},a_{1};\dots;A,\theta_M,b_M,a_M;\nonumber\\
&&\,\,\,\,\,\,\,\,\,\,\,\,\,\,\,\,\,P,\theta_{M+1},a_{M+1},b_{M+1};\dots;P,\theta_{2M},a_{2M},b_{2M}\rangle_{\rm in}.\nonumber\\
&&\times\langle 0\vert T_{\alpha\beta}(0)\vert  A,\theta_{1},b_{1},a_{1};\dots;A,\theta_M,b_M,a_M;\nonumber\\
&&\,\,\,\,\,\,\,\,\,\,\,\,\,\,\,\,\,P,\theta_{M+1},a_{M+1},b_{M+1};\dots;P,\theta_{2M},a_{2M},b_{2M}\rangle_{\rm in}^*. \nonumber
\eeq

Substituting our form factors (\ref{stressformfactor}), (\ref{SEMFF}), (\ref{SEMH}) gives
\beq
W_{\mu\nu\alpha\beta}^{2M}(x)&=&\frac{1}{(M!)^{2}}\int
\prod_{j=1}^{2M}\frac{d\theta_{j}}{4\pi}
\;e^{-ix\cdot\sum_{j=1}^{2M}p_{j}}\nonumber\\
&&\times \left[(p_{1}+\cdots+p_{2M})_\mu(p_{1}+\cdots+p_{2M})_\nu-\eta_{\mu\nu}(p_1+\cdots+p_{2M})^2\right]\nonumber\\
&&\times \left[(p_{1}+\cdots+p_{2M})_\alpha(p_{1}+\cdots+p_{2M})_\beta-\eta_{\alpha\beta}(p_1+\cdots+p_{2M})^2\right]\nonumber\\
&&\times\sum_{\sigma,\tau\in S_M}\frac{\vert H_{\sigma \tau}\vert^{2}}{\prod_{j=1}^M\left(\theta_{j}-\theta_{\sigma(j)+M}+\pi {\rm i}\right)^{2}\prod_{k=1}^{M}\left(\theta_{k}-\theta_{\tau(k)+M}+\pi {\rm i}\right)^{2}}\nonumber\\
&&+\mathcal{O}\left(\frac{1}{N}\right),\nonumber
\eeq
where we have used
\beq
\sum_{a_{1},\dots,a_{2M},b_{1},\dots,b_{2M}}\!\!\!\!\!&&\!\!\!\!\!\left[\prod_{j=1}^M\prod_{k=1}^M\delta_{a_{j}a_{\sigma(j)+M}}\delta_{b_{k}b_{\tau(k)+M}}\right]\left[\prod_{j=1}^M\prod_{k=1}^M\delta_{a_{j}a_{\omega(j)+M}}\delta_{b_{k}b_{\varphi(k)+M}}\right]\nonumber\\
&=&N^{2M}\left[\delta_{\sigma\omega}\delta_{\tau\varphi}+\mathcal{O}\left(\frac{1}{N}\right)\right].\nonumber
\eeq

The contribution to $W_{\mu\nu\alpha\beta}^{2M}(x)$ from each pair $\sigma,\tau$ is the same. There are $\frac{(M!)^{2}}{2}$ possible  pairs $\sigma,\tau$. We write the contribution from just one of these pairs, and multiply by the factor $\frac{(M!)^{2}}{2}$. We choose $\sigma(j)=j$ for $j=1,\dots,M$, and $\tau(1)=M,\,\tau(j)=j-1$ for $j=2,\dots,M$. Then we have
\beq
W_{\mu\nu\alpha\beta}^{2M}(x)&=&\frac{1}{2}\int
\prod_{j=1}^{2M}\frac{d\theta_{j}}{4\pi}\;
e^{-ix\cdot\sum_{j=1}^{2M}p_{j}}4\pi^4(4\pi)^{2M-2}\nonumber\\
&&\times \left[(p_{1}+\cdots+p_{2M})_\mu(p_{1}+\cdots+p_{2M})_\nu-\eta_{\mu\nu}(p_1+\cdots+p_{2M})^2\right]\nonumber\\
&&\times \left[(p_{1}+\cdots+p_{2M})_\alpha(p_{1}+\cdots+p_{2M})_\beta-\eta_{\alpha\beta}(p_1+\cdots+p_{2M})^2\right]\nonumber\\&&\times
\frac{1}{\left(\theta_{1}-\theta_{M+1}\right)^{2}+\pi^{2}} \frac{1}{\left(\theta_{2}-\theta_{M+2}\right)^{2}+\pi^{2}} \cdots 
\frac{1}{\left(\theta_M-\theta_{2M}\right)^{2}+\pi^{2}}\nonumber\\
&&\times\frac{1}{\left(\theta_{1}-\theta_{2M}\right)^{2}+\pi^{2}} 
\frac{1}{\left(\theta_{2}-\theta_{M+1}\right)^{2}+\pi^{2}}
\cdots  \frac{1}{\left(\theta_M-\theta_{2M-1}\right)^{2}+\pi^{2}}\nonumber\\
&&+O\left(\frac{1}{N}\right).
\label{SEMCF}
\eeq

Finally, we relabel the integration variables by $\theta_{1}\to\theta_{1},\;\theta_{2}\to\theta_{3},\;\theta_{3}\to\theta_5,\;\dots,\;\theta_M\to\theta_{2M-1},\;\theta_{M+1}\to\theta_{2},\;
\theta_{M+2}\to\theta_{4},\;\dots,\;\theta_{2M}\to\theta_{2M}$. This gives the expression for the non-time-ordered correlation function
\beq
W^{T}_{\mu\nu\alpha\beta}(x)&=&\frac{\pi^{2}}{8}\sum_{M=1}^{\infty}\int\ \prod_{j=1}^{2M}d\theta_{j}\;
e^{-ix\cdot\sum_{j=1}^{2M}p_{j}}
\nonumber\\
&&\times \left[(p_{1}+\cdots+p_{2M})_\mu(p_{1}+\cdots+p_{2M})_\nu-\eta_{\mu\nu}(p_1+\cdots+p_{2M})^2\right]\nonumber\\
&&\times \left[(p_{1}+\cdots+p_{2M})_\alpha(p_{1}+\cdots+p_{2M})_\beta-\eta_{\alpha\beta}(p_1+\cdots+p_{2M})^2\right]\nonumber\\
&&\times
\frac{1}{(\theta_{1}-\theta_{2M})^{2}+\pi^{2}}\prod_{j=1}^{2M-1}\frac{1}{(\theta_{j}-\theta_{j+1})^{2}+\pi^{2}}
+O\left(\frac{1}{N}\right)
\;.
\label{SEMCF2}
\eeq

\chapter{The spectrum of (1+1)-dimensional massive Yang-Mills theory}
\setcounter{equation}{0}
\renewcommand{\theequation}{5.\arabic{equation}}

\section{Massive Yang-Mills in 1+1 dimensions}
\setcounter{equation}{0}
\renewcommand{\theequation}{5.1.\arabic{equation}}

This chapter contains material previously published in \cite{massiveYM}.

Yang-Mills theory in $1+1$ dimensions has no local degrees of freedom. Introducing
an explicit mass $\mathcal M$ gives a theory of longitudinally-polarized gluons at tree level. It may
seem intuitively obvious, for small gauge coupling, that a particle is either a vector 
Boson, with a mass roughly equal to $\mathcal M$,
or a bound state of such
vector Bosons. This intuition, however, is wrong.We show in this chapter that the massive
Yang-Mills theory describes an infinite number of particles, with masses that 
are much less than $\mathcal M$. We call this, dynamical mass reduction.

The massive Yang-Mills model
can be thought of as a gauge field, coupled to a principal chiral nonlinear sigma model. The equivalence is seen
by choosing the unitary gauge condition. The sigma model behaves like a Higgs field, which gives mass to the vector gluons. It was shown by Bardeen and Shizuya that this model is renormalizable \cite{bardeen}


The tree-level description fails because the excitations of the sigma model
(without the gauge field)
are not Goldstone Bosons. These excitations are massive. The gauge field produces a confining force between these excitations. There is no Higgs or Coulomb phase. There is 
only a confined phase. In this respect, the model is similar to a non-Abelian gauge theory with massive adjoint fermions \cite{KMTX}.

A quantum field theory of an $SU$($N$) gauge field, coupled minimally to an adjoint matter field, can have distinct
Higgs and confinement phases \cite{FradShenk}, separated by a phase boundary, for space-time dimension greater than two. If this
dimension is two, however, there 
is only the confined phase. In the confined phase, the excitations are bound states of the 
massive particles of the 
sigma model. These massive particles are color multiplets of degeneracy $N^{2}$ \cite{wiegmann}.

The action of the massive $SU$($N$) Yang-Mills field in $1+1$ dimensions is
\beq
S=\int d^2x \left(-\frac{1}{4}{\rm Tr}\,F_{\mu\nu}F^{\mu\nu} +\frac{e^2}{2g_0^2}{\rm Tr} \,A_\mu A^\mu
\right),\label{unitarygauge}
\eeq
where $A_{\mu}$ is Hermitian and 
$F_{\mu\nu}=\partial_\mu A_\nu-\partial_\nu A_\mu -{\rm i}e[A_\mu,A_\nu]$ with $\mu,\nu=0,1$ and
indices are raised by $\eta^{\mu\nu}$, where $\eta^{00}=-\eta^{11}=1,\,\eta^{01}=\eta^{10}=0$. If we drop the cubic and
quartic terms from (\ref{unitarygauge}), the particles are gluons with mass ${\mathcal M}=e/g_{0}$.

The action (\ref{unitarygauge}) is equivalent to that of a principal chiral sigma model with one of its $SU(N)$ symmetries gauged by a Yang-Mills field.
We promote the left-handed ${ SU}(N)$ global symmetry of the sigma model to a local symmetry, by 
introducing the covariant derivative $D_\mu=\partial_\mu-{\rm i} e A_{\mu}$,
where $A_{\mu}$ is a new Hermitian vector field that 
transforms as $A_{\mu}\to V_{L}^{\dag}(x) A_{\mu} V_{L}(x)-\frac{i}{e}V_L^{\dag}(x) \partial_{\mu} V_{L}(x)$. We do not 
gauge the right-handed symmetry. The action is now 
\beq
S=\int d^2x \left[-\frac{1}{4}{\rm Tr} \, F_{\mu\nu} F^{\mu\nu} +\frac{1}{2g_0^2} {\rm Tr}\,(D_\mu U)^{\dag} D^\mu U\right].
\label{gaugedaction}
\eeq
In the unitary gauge, with $U(x)=1$ everywhere, this action (\ref{gaugedaction}) reduces to (\ref{unitarygauge}). In the remainder 
of this chapter, however, we will study (\ref{gaugedaction}) 
in the axial gauge.  

It is best to think of the left-handed symmetry as (confined) color-SU($N$) and the
right-handed symmetry as flavor-SU($N$). Confinement 
of left-handed color means that only singlets of the left-handed color group exist in the spectrum. There
are ``mesonic" bound states, as well as ``baryonic" bound states. The mesonic bound states have one elementary
particle of the sigma model and one elementary antiparticle. The simplest baryonic bound states consist
of $N$ of these elementary particles, with no antiparticles. There are also more complicated bound states, which
exist because there are excitations in the sigma model (with no gauge
field) 
transforming as higher representations of the color group \cite{wiegmann2}, \cite{integrablePCSM}. In this thesis, we
only discuss the mesonic states in detail.

A mesonic bound state, in the
axial gauge, is a sigma-model particle-antiparticle pair, confined by a linear potential. The string tension is
\beq
\sigma={e^2}C_N,\label{stringtension}
\eeq
where $C_N$ is the smallest eigenvalue of the Casimir operator of ${\rm SU}(N)$. The mass gap is 
\beq
M=2m+E_{0}\ll{\mathcal M},\nonumber
\eeq
where $E_{0}$ is the smallest (positive) binding energy, and $m$ is the mass of a sigma-model elementary excitation. This mass $M$ is finite, for
fixed $m$, as the ultraviolet cutoff is removed. In contrast, the bare Yang-Mills mass
$\mathcal M$, which is proportional to $1/g_{0}$, diverges.

Our approach is similar to that of Ref. \cite{glueball}. We find the wave function of an unbound particle-antiparticle pair, taking into account scattering
at the origin. Next, we generalize this to the wave function of the pair, confined by a 
linear potential. The method is 
inspired by the determination of the spectrum of the two-dimensional Ising model in 
an external magnetic field \cite{mccoy}. More sophisticated approaches to this and other two-dimensional models of confinement
\cite{DMS}, \cite{DM}, \cite{fonseca}, including fine structure (form factors) of the fundamental excitations, have been developed. We do not take into account decays or corrections to
the spectrum from matrix elements with more fundamental excitations \cite{DGM} in this chapter. 
For a general review, see Ref. \cite{tsvelik}.

\section{The axial gauge Hamiltonian and confinement}
\setcounter{equation}{0}
\renewcommand{\theequation}{5.2.\arabic{equation}}
In this section we derive the massive Yang-Mills Hamiltonian in the axial gauge. We show that this action describes sigma-model particles and antiparticles confined by a linear potential.

 If the axial
gauge $A_{1}=0$, is chosen, the action (\ref{gaugedaction}) is
\beq
S=\int d^{2}x \!\!\!\!\!\!\!\!\!\!&&\left[\frac{1}{2}{\rm Tr}\,(\partial_{1}A_{0})^{2}+
\frac{1}{2g_{0}^{2}}{\rm Tr}\,(\partial_{0}U^{\dagger}+{\rm i}eU^{\dagger}A_{0})(\partial_{0}U-{\rm i}eA_{0}U)
\right.\nonumber\\
&&\,\,\,\,\,\,\,\,\,\left.-\frac{1}{2g_{0}^{2}}{\rm Tr}\,\partial_{1}U^{\dagger}\partial_{1}U
 \right]\,. \nonumber
\eeq
 If we 
naively eliminate $A_{0}$, by its equation of motion (or integrate $A_{0}$ from the functional integral), we obtain the effective
action
\beq
S=\int d^{2}x \left(\frac{1}{2g_0^2} {\rm Tr}\,\partial_\mu U^{\dag} \partial^\mu U +\frac{1}{2} \,{j_{0}^{L}}_{a}\,\frac{1}{-\partial_{1}^{2}+e^{2}/g_{0}^{2}}
\, {j_{0}^{L}}_{a}\right), \label{screened}
\eeq
where $j_\mu^{L}(x)_{b}=-{\rm i} {\rm Tr}\,t_{b} \partial_\mu U(x) U^\dag(x)$
is the Noether current of the left-handed ${\rm SU}(N)$ symmetry. The potential induced on the color-charge density,
in the second term of (\ref{screened}), indicates that charges are screened, instead of confined. This conclusion, however, 
is based on the fact that $U^{\dagger}U=1$. In
the renormalized theory, $U$ is not a physical field. The physical scaling field of the principal chiral nonlinear sigma model is not
a unitary matrix. This fact is discussed more explicitly in Refs. \cite{renormalizedfield}, in the limit $N\rightarrow\infty$, with $g_{0}^{2}N$ 
fixed. The actual excitations of the principal
chiral model are massive, with a left and right color charge \cite{wiegmann}, so that no screening takes place.

A more careful approach is to first find the Hamiltonian in the temporal gauge $A_{0}=0$. Gauge invariance, or Gauss' law, must be imposed on
physical states. The Hamiltonian is
\beq
H&=&\int dx^{1} \,\left\{\frac{g_{0}^{2}}{2} [j^{L}_{0}(x^{1})_{b}]^{2}+ \frac{1}{2g_{0}^{2}}[j^{L}_{1}(x^{1})_{b}]^{2}+ \frac{1}{2}[E(x^{1})_{b}]^{2}\right.
\nonumber\\
&&\,\,\,\,\,\,\,\,\,\,\,\,\,\,\,\,\,\left.+\frac{e}{g_{0}^{2}} j^{L}_{1}(x^{1})_{b} A_{1}(x^{1})_{b}\right\},    \label{axialgauge}
\eeq
where $A_{1}(x^{1})_{b}={\rm Tr}\,t_{b}A$ and $E_{a}$ is the electric field, obeying $[E(x^{1})_{a}, A_{1}(y^{1})_{b}]=-{\rm i}\delta_{ab}\delta(x^{1}-y^{1})$. The Hamiltonian
(\ref{axialgauge}) must be 
supplemented by Gauss' law $G(x^{1})_{a}\Psi=0$, for any physical state $\Psi$, where $G(x^{1})_{a}$ is the generator of spatial gauge transformations:
\beq
G(x^{1})_{a}=\partial_{1}E(x^{1})_{a}+e f_{abc}A_{1}(x^{1})^{b}E(x^{1})_{c}-\frac{e}{g_{0}^{2}} j^{L}_{0}(x^{1})_{a}\,.
\eeq
If we require that the electric field vanishes at the boundaries $x^{1}=\pm l/2$, Gauss' law may be explicitly solved \cite{2+1}, to yield the expression for the electric field:
\beq
E(x^{1})_{a}=\int_{-l/2}^{x^{1}}dy^{1} \,\left\{ {\mathcal P}\exp\left[ ie\int_{-l/2}^{y^{1}} dz^{1}{\mathcal A}_{1}(z^{1})\right]\right\}_{a}^{\;\;\;\;\;b}\;\;
\frac{e}{g_{0}^{2}} j^{L}_{0}(y^{1})_{b}, \label{electricequiv}
\eeq
where ${\mathcal A}_{1}(x^{1})_{a}^{\;\;\;b}={\rm i}f_{abc}A_{1}(x^{1})_{c}$ is the gauge field in the adjoint representation. There remains a global gauge invariance, which must be satisfied by physical states, {\em i.e.}, $\Gamma_{a}\Psi=0$, where
\beq
\Gamma_{a}= \int_{-l/2}^{l/2}dy^{1} \,\left\{ {\mathcal P}\exp\left[ ie\int_{-l/2}^{y^{1}} dz^{1}{\mathcal A}_{1}(z^{1})\right]\right\}_{a}^{\;\;\;\;\;b}\;\;
\frac{e}{g_{0}^{2}} j^{L}_{0}(y^{1})_{b} .
\label{residual}
\eeq
Now we are free to chose $A_{1}(x^{1})_{b}=0$, which simplifies (\ref{electricequiv}) and (\ref{residual}). The solution for the
electric field yields the Hamiltonian
\beq
H&=&\int dx^{1} \,\left\{\frac{g_{0}^{2}}{2} [j^{L}_{0}(x^{1})_{b}]^{2}+ \frac{1}{2g_{0}^{2}}[j^{L}_{1}(x^{1})_{b}]^{2}\right\}\nonumber
\\
&&\,\,\,\,\,\,\,\,\,\,\,\,\,\,\,\,\,\,-\frac{e^{2}}{2g_{0}^{4}} \int dx^{1} \!\!\int dy^{1}\; \vert x^{1}-y^{1} \vert \; j^{L}_{0}(x^{1})_{b} \; j^{L}_{0}(y^{1})_{b}
,    \label{finalaxialgauge}
\eeq
where in the last step, we have taken the size $l$ of the system to infinity. The last term is a linear potential which confines left-handed color. The Hamiltonian (\ref{finalaxialgauge}) is not bounded from below on the full Hilbert space. This is because of the last, nonlocal term; the energy can be lowered by
adding pairs of colored particles (or antiparticles) and by separating them. The residual Gauss-law
condition $\Gamma_{a}\Psi=0$, forces the global left-handed color to be a singlet, thereby removing the instability,

\section{The free particle-antiparticle wave function: $N>2$}
\setcounter{equation}{0}
\renewcommand{\theequation}{5.3.\arabic{equation}}
The wave function of a free antiparticle at $x^1$ and a free particle at $x^{2}$, with momenta $p_1$ and $p_2$, respectively, is 
\beq
\Psi_{p_1,\,p_2}(x^1,y^1)_{a_1a_2;b_1b_2}=\left\{\begin{array}{cc}
e^{{\rm i}p_1x^1+{\rm i}p_2y^1}A_{a_1a_2;b_1b_2},\;&{\rm for} \;x^1<y^1,\\ \\
e^{{\rm i}p_2x^1+{\rm i}p_1y^1}S_{AP}(\theta)_{a_1b_1;b_2a_2}^{d_2c_2;c_1d_1}A_{c_1c_2;d_1d_2},\; &{\rm for}\; x^1>y^1.\end{array}
\right. \nonumber\\
\label{freewavefunction}
\eeq
where $A_{a_1a_2;b_1b_2}$ is set of arbitrary complex numbers, and  $S_{AP}(\theta)_{a_1b_1;b_2a_2}^{d_2c_2;c_1d_1}$ is the principal chiral sigma model S-matrix.

The residual Gauss' law in the axial gauge, $\Gamma_{a}\Psi=0$, restricts physical states to those which are invariant under global
left-handed ${\rm SU}(N)$ color transformations.  This means that the 
particle-antiparticle state of the form (\ref{freewavefunction}) must be projected to a global left-color singlet. A left-color-singlet 
wave function is
\beq
\!\!\!\!\!\!\!\!\!\!\!\!
\Psi_{p_1p_2} (x^1,y^1)_{b_1b_2}=\delta^{a_1a_2}\Psi_{p_1,\,p_2}(x^1,y^1)_{a_1a_2b_1b_2}.
\label{leftsinglet}
\eeq

There are states of degeneracy $N^{2}-1$, which resemble massive gluons. These transform as the adjoint representation
of the right-handed color symmetry. The wave function of such a state 
is traceless in the right-handed color indices:
\beq
\delta^{b_1b_2}\Psi_{p_1p_2}\!\!\!&(&\!\!\!x^1,y^1)_{b_1b_2}=0.\label{tracelesscondition}
\eeq
We use a non-relativistic approximation $p_{1,2}\ll m$. The wave function in this limit becomes
\beq
\Psi_{p_1p_2}(x^1,y^1)_{b_1b_2}=\left\{\begin{array}{c}
e^{{\rm i}p_1x^1+{\rm i}p_2y^1}A_{b_1b_2},\,\,\,\,\,\,\,\,\,\,\,\,\,{\rm for}\,\,x^1<y^1,\\
\,\\
e^{{\rm i}p_2x^1+{\rm i}p_1y^1}\exp ({\rm i}\pi -\frac{i h_{N}}{\pi m}\vert p_1-p_2\vert )A_{b_1b_2},\,\,\,\,\,\,{\rm for}\,\,x^1>y^1.\end{array}\right.
\label{nonrelativisticwave}
\eeq
where ${\rm Tr}\, A=0$, and
\beq
h_{N}&=&2\int_0^\infty \frac{d\xi}{\sinh \xi}\left[2(e^{2\xi/N}-1)-\sinh(2\xi/N)\right]\nonumber\\
&=&-4\gamma-\psi\left(\frac{1}{2}+\frac{1}{N}\right)-3\psi\left(\frac{1}{2}-\frac{1}{N}\right)-4\ln 4,
\label{N>2}
\eeq
where $\gamma$ is the Euler-Mascheroni constant, and 
$\psi(x)={d}\ln \Gamma(x)/dx$ is the digamma function. The 
expression in (\ref{nonrelativisticwave}) must be equal to the wave function of two confined particles for
sufficiently small $\vert x^1-y^1\vert$. To compare the two expressions, it is convenient to use center-of-mass coordinates, $X,\,x$, and their respective momenta $P,\,p$. Explicitly, 
$X=x^{1}+y^{1}$, $x=y^{1}-x^{1}$, $P=p_{1}+p_{2}$ and $p=p_{2}-p_{1}$. In these coordinates, the wave function is
\beq
\Psi_p(x)_{b_1b_2}=\left\{\begin{array}{c}
\cos(px+\omega)A_{b_1b_2},\,\,\,\,\,\,\,\,\,\,\,{\rm for}\,\,x>0,\\
\,\\
\cos[-px+\omega-\phi(p)]A_{b_1b_2},\,\,\,\,\,\,\,{\rm for}\,\,x<0,\end{array}\right.\label{comcoordinates}
\eeq
for some constant $\omega$, with the phase shift $\phi(p)=\pi-\frac{h_N}{\pi m}\vert p\vert$.

Another type of mesonic state is the right-handed color singlet, with $A_{b_1b_2}=\delta_{b_1b_2}$. The non-relativistic limit 
of the wave function in this case is
\beq
\Psi_p(x)_{\rm singlet}=\left\{\begin{array}{c}
\cos(px+\omega),\,\,\,\,\,\,\,\,\,\,\,{\rm for}\,\,x>0,\\
\,\\
\cos[-px+\omega-\chi(p)],\,\,\,\,\,\,\,{\rm for}\,\,x<0,\end{array}\right.\label{singletstate}
\eeq
where $\chi(p)=-\frac{h_N}{\pi m}\vert p\vert.$

\section{Mesonic states of massive Yang-Mills theory: $N>2$}
\setcounter{equation}{0}
\renewcommand{\theequation}{5.4.\arabic{equation}}

The wave function of a particle-antiparticle pair, confined by string tension $\sigma$, satisfies the Schroedinger equation
\beq
-\frac{1}{m}\frac{d^2}{dx^2}\Psi(x)_{b_1b_2}+\sigma \left\vert x \right\vert \,\Psi(x)_{b_1b_2}=E\Psi(x)_{b_1b_2},\label{schroedinger}
\eeq
where $E$ is the binding energy \cite{mccoy}. The solution to Equation (\ref{schroedinger}) is
\beq
\Psi(x)_{b_1b_2}=\left\{\begin{array}{c}
C {\rm Ai}\left[(m\sigma)^{\frac{1}{3}}\left(x-\frac{E}{\sigma}\right)\right]A_{b_1b_2},\,\,\,\,\,\,\,\,\,{\rm for}\,\,x>0\\
\,\\
C^\prime {\rm Ai}\left[(m\sigma)^{\frac{1}{3}}\left(-x-\frac{E}{\sigma}\right)\right]A_{b_1b_2},\,\,\,\,\,\,\,\,{\rm for}\,\,x<0,\end{array}\right.\label{airy}
\eeq
where ${\rm Ai}(x)$ is the Airy function of the first kind, and $C,\,C^\prime$ are constants.

For $\vert x\vert\ll (m\sigma)^{-1/3}$, the potential energy in (\ref{schroedinger}) is sufficiently small that the wave function is 
(\ref{comcoordinates}), with $\vert p\vert=(mE)^{\frac{1}{2}}$. The wave function (\ref{airy}) is approximated in this region by
\beq
\Psi(x)_{b_1b_2}=\left\{\begin{array}{c}
C\frac{1}{\left(x+\frac{E}{\sigma}\right)^{\frac{1}{4}}}\cos \left[\frac{2}{3}(m\sigma)^{\frac{1}{2}}\left(-x+\frac{E}{\sigma}\right)^{\frac{3}{2}}-\frac{\pi}{4}\right] A_{b_1b_2},\,\,\,\,\,\,\,{\rm for}\,\,x>0,\\
\,\\
C^\prime\frac{1}{\left(x+\frac{E}{\sigma}\right)^{\frac{1}{4}}}\cos \left[-\frac{2}{3}(m\sigma)^{\frac{1}{2}}
\left(x+\frac{E}{\sigma}\right)^{\frac{3}{2}}+\frac{\pi}{4}\right]A_{b_1b_2},\,\,\,\,\,\,\,{\rm for}\,\,x<0.\end{array}\right.\nonumber
\eeq

Let us now consider the $(N^{2}-1)$-plet of mesonic states.  The wave functions (\ref{comcoordinates}) and (\ref{airy}) should be the same for $x\downarrow0$, yielding
\beq
\frac{C}{(\frac{E}{\sigma})^{\frac{1}{4}}}\cos\left[\frac{2}{3}(m\sigma)^{\frac{1}{2}}\left(\frac{E}{\sigma}\right)^{\frac{3}{2}}-\frac{\pi}{4}\right]=\cos(\omega).\label{conditionabove}
\eeq
Equation (\ref{conditionabove}) implies
\beq
C=\left(\frac{E}{\sigma}\right)^{\frac{1}{4}},\,\,\,\,\,\,\,\omega=\frac{2}{3}(m\sigma)^{\frac{1}{2}}\left(\frac{E}{\sigma}\right)^{\frac{3}{2}}-\frac{\pi}{4}.\nonumber
\eeq
The wave functions (\ref{comcoordinates}) and (\ref{airy}) should also be the same for $x\uparrow0$, yielding
\beq
\frac{C^\prime}{\left(\frac{E}{\sigma}\right)^{\frac{1}{4}}}\cos\left[-\frac{2}{3}(m\sigma)^{\frac{1}{2}}\left(\frac{E}{\sigma}\right)^{\frac{3}{2}}+\frac{\pi}{4}\right]=\cos \left[\omega-\pi+\frac{h_N}{\pi m}(mE)^{\frac{1}{2}}\right],\label{fromabove}
\eeq
hence $C^\prime=C=\left(\frac{E}{\sigma}\right)^{\frac{1}{4}}$. The arguments of the cosine on each side of (\ref{fromabove}) must be the same, modulo $2\pi$:
\beq
-\frac{2}{3}(m\sigma)^{\frac{1}{2}}\left(\frac{E}{\sigma}\right)^{\frac{3}{2}}+\frac{\pi}{4}+2\pi n=\frac{2}{3}(m\sigma)^{\frac{1}{2}}\left(\frac{E}{\sigma}\right)^{\frac{3}{2}}-\frac{5\pi}{4}+\frac{h_N}{\pi m}(m E)^{\frac{1}{2}},\nonumber
\eeq
for $n=0,1,2,\dots$. We simplify this to
\beq
\frac{4}{3}(m\sigma)^{\frac{1}{2}}\left(\frac{E}{\sigma}\right)^{\frac{3}{2}}+\frac{h_N}{\pi m}(m E)^{\frac{1}{2}}-\left(n+\frac{3}{4}\right)2\pi=0.\label{quantization}
\eeq

An analysis which is similar to that of the previous paragraph yields the quantization condition for the right-handed singlet state (\ref{singletstate}). This is 
\beq
\frac{4}{3}(m\sigma)^{\frac{1}{2}}\left(\frac{E}{\sigma}\right)^{\frac{3}{2}}+\frac{h_N}{\pi m}(m E)^{\frac{1}{2}}-\left(n+\frac{1}{4}\right)2\pi=0.\label{quantizationsinglet}
\eeq

Equations (\ref{quantization}) and (\ref{quantizationsinglet}) are depressed cubic equations of the variable $Z_n=E_n^{\frac{1}{2}}$. These cubic equations have only one real solution for each value of $n$, because ${h_{N}}/({\pi m^{\frac{1}{2}}})>0$. The solution of Equations (\ref{quantization}) and (\ref{quantizationsinglet}) is
\beq
E_n=\left\{\left[\epsilon_n+\left(\epsilon_n^2+\beta_N^3\right)^{\frac{1}{2}}\right]^{\frac{1}{3}}+\left[\epsilon_n-\left(\epsilon_n^2+\beta_N^3\right)^{\frac{1}{2}}\right]^{\frac{1}{3}}\right\}^{\frac{1}{2}},\label{spectrum1} \eeq
where
\beq
\epsilon_n=\frac{3\pi}{4}\left(\frac{\sigma}{m}\right)^{\frac{1}{2}}\left(n+\frac{1}{2}\pm \frac{1}{4}\right),\,\,\,\,\,\,\,\,\,\,\,\,\,
\beta_N=\frac{h_N\sigma^{\frac{1}{2}}}{4\pi m}, \label{spectrum2}
\eeq
where $\pm=+$ for the $(N^2-1)$-plet, and $\pm=-$ for the singlet.

We show in the next section that the expressions (\ref{spectrum1}) and (\ref{spectrum2}) remain valid for the SU($2$) case, with 
$h_{2}=-4\ln 2+2$ and, significantly, with a reversal of the sign in (\ref{spectrum2}). For $N=2$ {\em only} we must take
$\pm=-$ for the $(N^{2}-1)$-plet (the triplet) and $\pm=+$ for the singlet. 

Another interesting special case is the 't~Hooft limit, $N\rightarrow \infty$. The mass gap of the sigma model should be fixed in this
limit. The string tension $\sigma$ will be fixed as well \cite{'t1+1}, provided $e^{2}N$ is fixed.  In this limit $h_{N}\rightarrow 0$, and we find
\beq
E_n=\left[\frac{3\pi}{2}\left(\frac{\sigma}{m}\right)^{\frac{1}{2}}\left(n+\frac{1}{2}\pm \frac{1}{4}\right) \right]^{1/3}.
\label{spectrumlargeN}
\eeq

\section{The $N=2$ case}
\setcounter{equation}{0}
\renewcommand{\theequation}{5.5.\arabic{equation}}

The exponential expression for the S-matrix (\ref{afunction}) is only correct for $N>2$. The principal chiral model with $SU(2)\times SU(2)$ symmetry is equivalent to the $O(4)$-symmetric nonlinear sigma model. We will express the S-matrix, first found in Ref. \cite{zamolodchikov}, by an exponential expression \cite{karowski}.

A state with one excitation has a left-handed color index $a=1,2$ and a right-handed color index $b=1,2$. In the $O(4)$ formulation, excitations have a single species index $j=1,2,3,4$. The $SU(2)\times SU(2)$-symmetric states are related to the $O(4)$-symmetric states by
\beq
\vert P, \theta, a, b\rangle_{\rm in}&=&\sum_j \frac{1}{\sqrt{2}} \left(\delta^{j 4}\delta_{ab}-i\sigma_{ab}^j\right) \vert \theta, j\rangle_{\rm in},\nonumber\\
\vert A, \theta, a, b\rangle_{\rm in}&=&\sum_j \frac{1}{\sqrt{2}} \left(\delta^{j 4}\delta_{ab}-i\sigma_{ab}^j\right)^* \vert \theta, j\rangle_{\rm in},\nonumber
\eeq
where $\sigma^j$ with $j=1,2,3$ are the Pauli matrices. The $O(4)$ two-excitation S-matrix,  $S(\theta)^{j_1 j_2}_{j^\prime_1 j^\prime_2} $ is given by
\beq
\,_{\rm out}\langle \theta^\prime_1, j^\prime_1;\theta^\prime_2, j^\prime_2\vert \theta_1, j_1;\theta_2,j_2\rangle_{\rm in}\nonumber=S(\theta)^{j_1 j_2}_{j^\prime_1 j^\prime_2}\, 4\pi \delta(\theta_1-\theta^\prime_1)\,4\pi \delta(\theta_2-\theta^\prime_2),\nonumber
\eeq
where \cite{karowski}
\beq
S(\theta)^{j_1 j_2}_{j^\prime_1 j^\prime_2} = \left[\frac{\theta+\pi i}{\theta-\pi i}(P^0)^{j_1 j_2}_{j_1^\prime j_2^\prime}+\frac{\theta-\pi i}{\theta+\pi i} (P^{+})^{j_1 j_2}_{j_1^\prime j_2^\prime}+(P^{-})^{j_1 j_2}_{j_1^\prime j_2^\prime}\right] Q(\theta),\nonumber
\eeq
\beq
Q(\theta)=\exp 2\int_0^\infty \frac{d\xi}{\xi}\frac{e^{-\xi}-1}{e^\xi+1}\sinh\left(\frac{\xi\theta}{\pi {\rm i}}\right),\nonumber
\eeq
and $P^0,\,P^{+},$ and $P^{-}$ are the singlet, symmetric-traceless, and antisymmetric projectors, which are
\beq
(P^0)^{j_1 j_2}_{j_1^\prime j_2^\prime}=\frac{1}{4}\delta^{j_1j_2}\delta_{j_1^\prime j_2^\prime}&,&\,\,\,
(P^+)^{j_1 j_2}_{j_1^\prime j_2^\prime}=\frac{1}{2}(\delta^{j_1}_{j_1^\prime}\delta^{j_2}_{j_2^\prime}+\delta^{j_1}_{j_2^\prime}\delta^{j_2}_{j_1^\prime})
-\frac{1}{4}\delta^{j_1j_2}\delta_{j_1^\prime j_2^\prime},\nonumber\\
(P^-)^{j_1 j_2}_{j_1^\prime j_2^\prime}&=&\frac{1}{2}(\delta^{j_1}_{j_1^\prime}\delta^{j_2}_{j_2^\prime}-\delta^{j_1}_{j_2^\prime}\delta^{j_2}_{j_1^\prime}),\nonumber
\eeq
respectively.

We write the left-color-singlet wave function for a free particle and antiparticle:
\beq
\Psi_{p_1,p_2}(x^1,y^1)_{b_1b_2}&=&D_{b_1b_2}^{j_1j_2}\left\{\begin{array}{c}
e^{ip_1 x^1+ip_2y^1}A_{j_1 j_2},\,\,\,\,{\rm for}\,\,x^1>y^1\\
\,\\
e^{ip_2x^1+ip_1y^1}S(\theta)^{j^\prime_1 j^\prime_2}_{j_1 j_2} A_{j^\prime_1 j^\prime_2},\,\,\,\,{\rm for}\,\,x^1<y^1,
\end{array}\right.\label{ofourwave}
\eeq
where
\beq
D_{b_1b_2}^{j_1j_2}=\frac{1}{2}\delta^{a_1a_2}\left(\delta^{j_1 4}\delta_{a_1b_1}-i\sigma_{a_1b_1}^{j_1}\right)^*\left(\delta^{j_2 4}\delta_{a_2b_2}-i\sigma_{a_2b_2}^{j_2}\right)\,.\nonumber
\eeq
There is a triplet of degenerate states and one singlet state. The triplet satisfies
\beq
\delta^{b_1b_2}\Psi_{p_1,p_2}(x^1,y^1)_{b_1b_2}=0.\label{tracelessofour}
\eeq
Substituting (\ref{ofourwave}) into (\ref{tracelessofour}) gives the condition
\beq
\delta^{b_1b_2}\,D_{b_1b_2}^{j_1j_2}\,A_{j_1j_2}=\delta^{j_1 j_2}A_{j_1j_2}=0\,.\nonumber
\eeq
The traceless matrix $A_{j_1j_2}$ can be split into a symmetric and an antisymmetric part, $A^{+}_{j_1j_2}=(A_{j_1j_2}+A_{j_2j_1})/2$ and $A^{-}_{j_1j_2}=(A_{j_1j_2}-A_{j_2j_1})/2$, respectively. The matrix $A^{+}_{j_1j_2}$, however, does not contribute to the wave function (\ref{ofourwave}), because
\beq
D^{j_1j_2}_{b_1b_2}A^{+}_{j_1j_2}=\frac{1}{2}\delta_{b_1b_2}{\rm Tr}\,A^{+}=0.\nonumber
\eeq
The matrix $A^{-}_{j_1 j_2}$ satisfies \cite{zamolodchikov}, \cite{karowski}:
\beq
S(\theta)^{j_{1} j_{2}}_{j_{1}^{\prime} j_{2}^{\prime}}A^{-}_{j_{1}j_{2}}=Q(\theta)A^{-}_{j_{1}^{\prime}j_{2}^{\prime}}.\label{antisymmetric}
\eeq
Substituting (\ref{antisymmetric}) into (\ref{ofourwave}), in center-of-mass coordinates and the non-relativistic limit, we find
\beq
\Psi_{p}(x)_{b_1b_2}=D_{b_1b_2}^{j_1j_2}\left\{\begin{array}{c}
\cos(px+\omega)A_{j_1j_2},\,\,\,\,\,\,\,\,\,\,\,{\rm for}\,\,x>0,\\
\,\\
\cos[-px+\omega-\phi(p)]A_{j_1j_2},\,\,\,\,\,\,\,{\rm for}\,\,x<0,\end{array}\right.\label{nonrelativisticofour}
\eeq
where 
$\phi(p)=-\frac{i h_2}{\pi m}\vert p\vert$, where
\beq
h_2=2\int_0^\infty d\xi\,\frac{e^{-\xi}-1}{e^\xi+1}=-4\ln 2+2.\label{N=2}
\eeq

The wave function of the right-color-singlet bound state is
\beq
\Psi_{p_1,p_2}^{\rm singlet}(x^1,y^1)=
\left\{\begin{array}{c}
e^{ip_1 x^1+ip_2y^1}
,\,\,\,\,{\rm for}\,\,x^1>y^1,\\
\,\\
e^{ip_2x^1+ip_1y^1}\frac{\theta+\pi {\rm i}}{\theta-\pi {\rm i}}Q(\theta)
,\,\,\,\,{\rm for}\,\,x^1<y^1.
\end{array}\right.\label{singletofourwave}
\eeq
In center-of-mass coordinates, in the non-relativistic approximation, this becomes
\beq
\Psi_{p}^{\rm singlet}(x)=
\left\{\begin{array}{c}
\cos(px+\omega),\,\,\,\,\,\,\,\,\,\,\,{\rm for}\,\,x>0,\\
\,\\
\cos[-px+\omega-\chi(p)],\,\,\,\,\,\,\,{\rm for}\,\,x<0,\end{array}\right.\label{nonrelativisticofoursinglet}
\eeq
where $\chi(p)=\pi -\frac{i h_2}{\pi m}\vert p\vert$.

From this point onward, the analysis is similar to what we've presented in the last two sections. We obtain (\ref{spectrum1}), (\ref{spectrum2}),
except that $h_{N}$ (defined in (\ref{N>2})) is replaced with $h_{2}$ (defined in (\ref{N=2})), with one important difference; we
have $\pm=+$ for the singlet and $\pm=-$ for the triplet in Eq. (\ref{spectrum2}).

\section{Relativistic corrections}
\setcounter{equation}{0}
\renewcommand{\theequation}{5.6.\arabic{equation}}

In the future, we would like to find relativistic corrections to the mass spectrum. This was done in Ref. \cite{fonseca} for the Ising model in 
an external magnetic field. The goal would be to find mesonic eigenstates of the Hamiltonian (\ref{finalaxialgauge}) of the form:
\beq
\vert \Psi_B\rangle_{b_1b_2}=\vert \Psi_B^{(2)}\rangle_{b_1b_2}+\vert \Psi_B^{(4)}\rangle_{b_1b_2}+\vert \Psi_B^{(6)}\rangle_{b_1b_2}+\dots,\nonumber
\eeq
where the state $\vert \Psi_B^{(2M)}\rangle_{b_1b_2}$ contains $M$ particles and $M$ antiparticles. The multi-particle contributions are included
because an electric string may break \cite{DGM}, producing pairs of sigma-model excitations. Nonetheless, for
small gauge coupling, the ``two-quark" approximation is valid. In the this approximation, the bound state is treated as
\beq
\vert \Psi_B\rangle_{b_1b_2}\approx\vert \Psi_B^{(2)}\rangle_{b_1b_2}&=&\frac{1}{2}\int
\frac{d\theta_{1}}{4\pi}
\frac{d\theta_{2}}{4\pi}
\Psi(p_1,p_2)_{a_2a_2}\vert A,\theta_1,b_1,a_1;P,\theta_2,a_2,b_2\rangle,\;\nonumber\\
&&
{\rm where},\nonumber \\ 
\Psi(p_1,p_2)_{a_1a_2}&=&S(\theta)\left[\delta_{a_1}^{c_1}\delta_{a_2}^{c_2}-\frac{2\pi {\rm i}}{N(\pi {\rm i}-\theta)}\delta_{a_1a_2}\delta^{c_1c_2}\right]\Psi(p_2,p_1)_{c_1c_2}.\nonumber\\
\label{twoquark}\
\eeq
The spectrum of masses $\Delta$, of the states (\ref{twoquark}) is found from 
the Bethe-Salpeter equation $(H-\Delta)\vert \Psi^{(2)}_B\rangle_{b_1b_2}=0$. Acting with the Hamiltonian (\ref{finalaxialgauge}) on this state, yields
\beq
&&\left(m\cosh\theta_1+m\cosh\theta_2-\Delta\right)\Psi(p_1^\prime,p_2^\prime)_{c_1c_2}\delta_{b_1d_1}\delta_{b_2d_2}\nonumber\\
&&\,\,\,\,\,\,\,\,\,\,=\frac{e^2}{4g_0^4}\int
\frac{d\theta_{1}}{4\pi}
\frac{d\theta_{2}}{4\pi}
\Psi(p_1,p_2)_{a_1a_2}\int dx^1 dy^1\vert x^1-y^1\vert\nonumber\\
&&\,\,\,\,\,\,\,\,\,\,\,\,\,\,\times\langle A, \theta_1^\prime, d_1,c_1;P,\theta_2^\prime,c_2,d_2\vert {\rm Tr}\,\left[j_0^L(x^1)j_0^L(y^1)\right]\vert A,\theta_1, b_1, a_1;P, \theta_2,a_2,b_2\rangle,\nonumber\\
\label{bethesalpeter}
\eeq
where the operator ${\rm Tr}\,\left[j_0^L(x^1)j_0^L(y^1)\right]$ is not time-ordered. The matrix element
\beq
\langle A, \theta_1^\prime, d_1,c_1;P,\theta_2^\prime,c_2,d_2\vert {\rm Tr}\,\left[j_0^L(x^1)j_0^L(y^1)\right] \vert A,\theta_1, b_1, a_1;P, \theta_2,a_2,b_2\rangle\nonumber
\eeq
is obtained by inserting a complete set of states between the current operators and using the exact 
form factors of the currents of the principal chiral sigma model. For finite $N$, only the leading two-particle form factors of currents are known from the previous chapter,
and only a vacuum insertion can be made. The complete matrix element is known at large $N$, which should
help in finding the relativistic corrections to the eigenvalues of Eq. (\ref{bethesalpeter}).

\chapter{(2+1)-dimensional Yang-Mills theory and form factor perturbation theory}
\setcounter{equation}{0}
\renewcommand{\theequation}{6.\arabic{equation}}

\section{The low lying glueball mass spectrum}
\setcounter{equation}{0}
\renewcommand{\theequation}{6.1.\arabic{equation}}

This chapter contains material previously published in \cite{twoplusoneform}

In this chapter we turn our attention to the anisotropic (2+1)-dimensional Yang-Mills theory described in Chapter 2. In the highly anisotropic limit, this theory becomes an array principal chiral sigma models, weakly coupled in the $x^2$ direction. We calculate physical quantities of the anisotropic theory using exact, non-perturbative information from the sigma model. The physical quantities we calculate are the low lying glueball masses, and the string tensions for quarks separated either in the $x^1$, or $x^2$ direction.

I should clarify that these calculations have been done before by P. Orland, in References \cite{glueball},\cite{horizontal},\cite{vertical}, for the $SU(2)$ gauge group. The glueball masses and string tensions in these references were calculated using the $O(4)$-nonlinear sigma model S-matrix and form factors (similar to our calculation from Section 5.5). In this chapter we merely generalize these calculations for the $SU(N)$ group, using our new results from Chapter 4. The reason this has not been done before is that the form factors of the principal chiral sigma model were not known until now. 

In this section we calculate the masses of the lightest particles in the anisotropic Yang-Mills theory (which we call glueballs). We recall the Hamiltonian of the anisotropic theory (in axial gauge):
\beq
H=H_0+\lambda^2 H_1,\nonumber
\eeq
with 
\beq
H_0=\sum_{x^2}H_{\rm PCSM}(x),\nonumber
\eeq
and
\beq
H_1&=&\sum_{x^2}\int dx^1\int dy^1\frac{1}{4g_0^2 a}\vert x^1-y^1\vert\nonumber\\
&&\times\left[j_0^L(x^1,x^2)-j_0^R(x^1,x^2-a)\right]\nonumber\\
&&\times\left[j_0^L(y^1,x^2)-j_0^R(y^1,x^2-a)\right].\label{perturbationhamiltonianagain}
\eeq
After axial gauge fixing, there is the residual (global) Gauss's law constraint on physical states:
\beq
\int dx^1\left[ j_0^L(x^1,x^2)_b-j_0^R(x^1,x^2-a)_b\right]\Psi=0,\label{globalgausslaw}
\eeq

The constraint (\ref{globalgausslaw}) requires that there be an equal number of particles and antiparticles in each $x^2$ layer. Furthermore, it requires that the excitations form left- and right-color singlets. If the sigma model at $x^2$ has a particle with a left color index, $a_1$, this index has to be  contracted with either the left-color index of an antiparticle in the $x^2$ layer, or the right color index of a particle in the $(x^2+a)$ layer. A glueball in this theory consists of several sigma-model excitations, forming a color-singlet bound state.

The simplest and lightest glueball is one composed of only one particle and one antiparticle, at the same value of $x^2$. The gauss law constraint requires that their left and right handed color indices be contracted. The interaction Hamiltonian (\ref{perturbationhamiltonianagain}) provides a confining linear potential, with string tension
\beq
\sigma=2\sigma^H=2\lambda^2\, \frac{g_0^2}{a^2}C_N.\label{stringhorizontal}
\eeq
The factor of $2$ comes the fact that both the left and right color charges are confined.

The problem is now essentially (1+1)-dimensional. We can calculate the massive gluon spectrum doing exactly the same calculation as in Chapter 5, but using the new string tension, $\sigma$, from Eq. (\ref{stringhorizontal}).

The glueball masses are given by $M_n=2m+E_n$, where $m$ is the mass of the sigma model particles, and $E_n$ is the binding energy. Because both the left and right handed colors are confined, The glueball masses correspond to the singlet mesons from last chapter. For $N>2$, the binding energy is then given by
\beq
E_n=\left\{\left[\epsilon_n+\left(\epsilon_n^2+\beta_N^3\right)^{\frac{1}{2}}\right]^{\frac{1}{3}}+\left[\epsilon_n-\left(\epsilon_n^2+\beta_N^2\right)^{\frac{1}{2}}\right]^{\frac{1}{3}}\right\}^{\frac{1}{2}},\label{glueballmass}
\eeq
where 
\beq
\epsilon_n=\frac{3\pi}{4}\left(\frac{\sigma}{m}\right)^{\frac{1}{2}}\left(n+\frac{1}{4}\right),\,\,\,\,\,\,\,\beta_N=\frac{h_N\sigma^{\frac{1}{2}}}{4\pi m},\nonumber
\eeq
for $n=0,1,2,\dots$, where $\sigma$ is defined in Eq. (\ref{stringhorizontal}) and $h_N$ is defined in Eq. (\ref{N>2}). 

The spectrum (\ref{glueballmass}) generalizes the result from Ref. \cite{glueball} for general $N>2$. The result found in Ref. \cite{glueball} corresponds to our $N=2$ spectrum from Section 5.5. The $N=2$ low lying glueball spectrum is given by Eq. (\ref{glueballmass}) but substituting 
\beq
\epsilon_n=\frac{3\pi}{4}\left(\frac{\sigma}{m}\right)^{\frac{1}{2}}\left(n+\frac{5}{4}\right),\nonumber
\eeq
and $h_2=-4\ln 2+2$.

\section{The horizontal string tension}
\setcounter{equation}{0}
\renewcommand{\theequation}{6.2.\arabic{equation}}

In this section we compute quantum corrections to the string tension $\sigma^H$. This calculation has been done before, in Reference \cite{horizontal}, for $N=2$ using the form factors of the $O(4)$ sigma model. In this section we generalize these results for $N>2$, using the form factors from Chapter 4.

It is convenient to rewrite the Hamiltonian (\ref{perturbationhamiltonianagain}) by reintroducing the auxiliary field $\Phi=-A_0$, such that
\beq
H_1&=&\sum_{x^2}\int dx^1\left\{\frac{g_0^2\, a^2}{4}\partial_1 \Phi(x^1,x^2)\partial_1 \Phi(x^1,x^2)\right.\nonumber\\
&&\left.-j_0^L(x^1,x^2)\Phi(x^1,x^2)-j_0^{R}(x^1,x^2)\Phi(x^1,x^2+a)\right\}.\label{withauxiliary}
\eeq
By integrating out the auxiliary field, $\Phi$, we see the Hamiltonians, (\ref{withauxiliary}) and (\ref{perturbationhamiltonianagain}) are equivalent.

We can easily introduce static quarks into the Hamiltonian (\ref{withauxiliary}) by coupling them to the auxiliary field, $\Phi$. Our goal is to find the potential energy of a quark-antiquark pair separated only in the $x^1$ direction. By integrating out the sigma model degrees of freedom, we can find the quantum corrections to the string tension $\sigma^H$. The Hamiltonian with a quark of charge $q$ at the space point $(u^1,u^2)$, and an antiquark of charge $q^\prime$ at the space-time point $(v^1,v^2)$, is 
\beq
H_1&=&\sum_{x^2}\int dx^1\left\{\frac{g_0^2\, a^2}{4}\partial_1 \Phi(x^1,x^2)\partial_1 \Phi(x^1,x^2)\right.\nonumber\\
&&-j_0^L(x^1,x^2)\Phi(x^1,x^2)-j_0^{R}(x^1,x^2)\Phi(x^1,x^2+a)\nonumber\\
&&+\left.g_0^2\,q\,\Phi (u^1,u^2)-g_0^2\,q^\prime\Phi(v^1,v^2)\right\}.\label{withquarks}
\eeq
With these static quarks, the residual gauss law on physical states is modified to:
\beq
&&\int dx^1\left[j_0^L(x^1,x^2)_b-j_0^R(x^1,x^2-a)_b\right.\nonumber\\
&&\,\,\,\,\,\,\,\,\,\,+\left.q_b\delta(x^1-u^1)\delta_{x^2 u^2}-q^\prime_b\delta(x^1-v^1)\delta_{x^2 v^2}\right]\Psi=0.\label{modifiedresidualgauss}
\eeq

To find the string tension, $\sigma^H$, we set $u^2=v^2$, and integrate out the sigma model field, $U$. We obtain an effective action, $S_{\rm eff}(\Phi)$, by
\beq
e^{iS_{\rm eff}(\Phi)}=\langle 0\vert \mathcal{T}\, e^{i\int dx^0 \lambda^2 H_1}\vert 0\rangle,\label{integrateout}
\eeq
where $\mathcal{T}$ stands for time ordering. The field $\Phi$ in (\ref{integrateout}) is treated as a background classical field. Expanding (\ref{integrateout}) in powers of $\lambda$, up to quartic order, we find
\beq
i\!\!\!\!\!\!&&\!\!\!\!\!\!\!\!S_{\rm eff}(A_0)\approx -i\lambda^2\sum_{x^2}\int d^2x\,\frac{g_0^2a^2}{4}\Phi\partial_1^2\Phi+i\lambda^4S^{(2)}(\Phi)+\mathcal{O}(\lambda^6)\nonumber\\
&-&\lambda^2\sum_{x^2}\int d^2x\left[ g_0^2\,q(x^0)\,\Phi (x^0,u^1,u^2)-g_0^2\,q^\prime(x^0)\Phi(x^0,v^1,v^2)\right]\!\!\!,\label{connectedgraph}
\eeq
where
\beq
iS^{(2)}\equiv-\frac{1}{2}\sum_{x^2}\int d^2x d^2y \,D(x^0, x^1,y^0,y^1,x^2)_{acef} \Phi(x^0,x^1,x^2)_{ac}\Phi(y^0,y^1,x^2)_{ef},\nonumber
\eeq
where
\beq 
D(x^0, x^1,y^0,y^1,x^2)_{acef}\equiv\langle 0\vert \mathcal{T} j_0^L(x^0,x^1,x^2)_{ac}\,j_0^L(y^0,y^1,x^2)_{ef}\vert 0\rangle.\label{twopointtimeordered}
\eeq

We compute the correlation function (\ref{twopointtimeordered}) by introducing a complete set of intermediate states between the two operators. The non-time-ordered correlation function is given by 
\beq
&&\langle 0\vert  j_0^L(x)_{ac}\,j_0^L(y)_{ef}\vert 0\rangle=\sum_{M=1}^{\infty}\frac{1}{N(M!)^2}\int\frac{d\theta_1\dots d\theta_{2M}}{(2\pi)^{2M}}e^{-i(x-y)\cdot\left[\sum_{j=1}^{2M} p_j\right]}\nonumber\\
&&\,\,\,\,\,\times\langle0|j_\mu^L(0)_{a_0c_0}|A,\theta_1,b_1,a_1;\dots;\nonumber\\
&&\,\,\,\,\,\,\,\,\,\,\,A,\theta_M,b_Ma_M;P,\theta_{M+1},a_{M+1},b_{M+1};\dots;P,\theta_{2M},a_{2M},b_{2M}\rangle\nonumber\\
&&\,\,\,\,\,\times\left[\langle0|j_\nu^L(0)_{e_0f_0}|A,\theta_1,b_1,a_1;\dots;\right.\nonumber\\
&& \,\,\,\,\,\,\,\,\,\,\,\left.A,\theta_M,b_Ma_M;P,\theta_{M+1},a_{M+1},b_{M+1};\dots;P,\theta_{2M},a_{2M},b_{2M}\rangle\right]^*.\nonumber
\eeq
The correlation function (\ref{twopointtimeordered}) can be found exactly at large $N$ using the form factors from Chapter 4. For general $N<\infty$, we can only calculate a large-distance approximation, using the two-particle form factor (also found in Chapter 4). At large distances, it is sufficient to compute only the first intermediate state, with one particle and one antiparticle. 

We recall, from Section 4.2, the form factor with one particle and one antiparticle is
\beq
\langle 0\vert j_\mu^L(x)_{ac}\!\!\!&\vert&\!\!\! A,\theta_1,b_1,a_1;P,\theta_2,a_2,b_2\rangle\nonumber\\
&=&(p_1-p_2)_\mu\left( \delta_{a_0a_2}\delta_{c_0a_1}\delta_{b_1b_2}-\frac{1}{N}\delta_{a_0c_0}\delta_{a_1a_2}\delta_{b_1b_2}\right)e^{-ix\cdot (p_1+p_2)}\nonumber\\
&&\times\frac{2\pi i}{(\theta+\pi i)}\exp \int_0^\infty \frac{d\xi}{\xi}\left[\frac{-2\sinh\left(\frac{2\xi}{N}\right)}{\sinh \xi}\right.\nonumber\\
&&\left.+\frac{4e^{-\xi}\left(e^{2\xi/N}-1\right)}{1-e^{-2\xi}}\right]\frac{\sin^2[\xi(\pi i-\theta)/2\pi]}{\sinh \xi}.\label{twoparticlerecall}
\eeq
Inserting (\ref{twoparticlerecall}) into (\ref{twopointtimeordered}) and time ordering, we find
\beq
D(x,y)_{acef}&=&\int\frac{d\theta_1 \, d\theta_2}{(2\pi)^2}m^2(\cosh\theta_1-\cosh\theta_2)^2 \nonumber\\
&&\times\left(\delta_{a a_2}\delta_{c a_1}-\frac{1}{N}\delta_{a c}\delta_{a_1 a_2}\right)\left(\delta_{e a_2}\delta_{f a_1}-\frac{1}{N}\delta_{e f}\delta_{a_1 a_2}\right)\nonumber\\
&&\times\exp\left\{-im\,{\rm sgn}(x^0-y^0)\left[(x^0-y^0)(\cosh\theta_1+\cosh\theta_2)\right.\right.\nonumber\\
&&\,\,\,\,\,\,\,\,\,\,\,\,\,\,\,\left.\left.-(x^1-y^1)(\sinh\theta_1+\sinh\theta_2)\right]\right\}\nonumber\\
&&\times\left\{\frac{2\pi i}{(\theta+\pi i)}\exp \int_0^\infty \frac{d\xi}{\xi}\left[\frac{-2\sinh\left(\frac{2\xi}{N}\right)}{\sinh \xi}\right.\right.\nonumber\\
&&\left.\left.\,\,\,\,\,\,\,+\frac{4e^{-\xi}\left(e^{2\xi/N}-1\right)}{1-e^{-2\xi}}\right]\frac{\sin^2[\xi(\pi i-\theta)/2\pi]}{\sinh \xi}\right\}^2.\label{timeorderedtwopoint}
\eeq
The color factor in (\ref{timeorderedtwopoint}) is
\beq
\left(\delta_{a a_2}\delta_{c a_1}-\frac{1}{N}\delta_{a c}\delta_{a_1 a_2}\right)\left(\delta_{e a_2}\delta_{f a_1}-\frac{1}{N}\delta_{e f}\delta_{a_1 a_2}\right)=\delta_{ae}\delta_{ef}-\frac{1}{N}\delta_{ac}\delta_{ef}.\label{deltas}
\eeq
The term in the right-hand side of (\ref{deltas}) proportional to $\frac{1}{N}$ does not contribute when we plug (\ref{timeorderedtwopoint}) back into (\ref{connectedgraph}), because the field $\Phi$ is traceless, so we will ignore this term from now on.

We evaluate $iS^{(2)}(\Phi)$ using coordinates $X^\mu,\,r^\mu$, defined by $x^\mu=X^\mu+\frac{1}{2}r^\mu,$ and $y^\mu=X^\mu-\frac{1}{2}r^\mu$. We then use the derivative expansion for $X\gg r$:
\beq
\Phi(x)&=&\Phi(X)+\frac{r^\mu}{2}\partial_\mu \Phi(X)+\frac{r^\mu r^\nu}{8}\partial_\mu\partial_\nu \Phi(X)+\dots,\nonumber\\
\Phi(y)&=&\Phi(X)-\frac{r^\mu}{2}\partial_\mu \Phi(X)+\frac{r^\mu r^\nu}{8}\partial_\mu \partial_\nu \Phi(X)\pm\dots,\label{derivativeexpansion}
\eeq
where $\partial_\mu$ denotes $\partial/\partial X^\mu$. This derivative expansion is valid at large distances. The quadratic contribution to the effective action is 
\beq
iS^{(2)}&=&-\frac{i}{2}\int d^2Xd^2r\,D\left(X+\frac{r}{2},X-\frac{r}{2}\right)_{acef}\nonumber\\
&&\,\,\,\,\,\,\,\,\,\,\,\,\,\times\Phi\left(X+\frac{r}{2}\right)_{ac}\,\Phi\left(X-\frac{r}{2}\right)_{ef}.\label{integrater}
\eeq

We substitute (\ref{derivativeexpansion}) into (\ref{integrater}) and find
\beq
iS^{(2)}&=&-\frac{i}{2}\int d^2Xd^2r\,\int\frac{d\theta_1 \, d\theta_2}{(2\pi)^2}m^2(\cosh\theta_1-\cosh\theta_2)^2 \delta_{a e}\delta_{c f}\nonumber\\
&&\times\exp\left\{-im\,{\rm sgn}(r^0)\left[(r^0)(\cosh\theta_1+\cosh\theta_2)\right.\right.\nonumber\\
&&\left.\left.\,\,\,\,\,\,\,\,\,\,\,\,\,\,\,\,\,\,\,\,-(r^1)(\sinh\theta_1+\sinh\theta_2)\right]\right\}\nonumber\\
&&\times\left\{\frac{2\pi i}{(\theta+\pi i)}\exp \int_0^\infty \frac{d\xi}{\xi}\left[\frac{-2\sinh\left(\frac{2\xi}{N}\right)}{\sinh \xi}\right.\right.\nonumber\\
&&\left.\left.\,\,\,\,\,\,\,\,\,+\frac{4e^{-\xi}\left(e^{2\xi/N}-1\right)}{1-e^{-2\xi}}\right]\frac{\sin^2[\xi(\pi i-\theta)/2\pi]}{\sinh \xi}\right\}^2\nonumber\\
&&\times \left(\Phi(X)_{ac}+\frac{r^\mu}{2}\partial_\mu \Phi(X)_{ac}+\frac{r^\mu r^\nu}{8}\partial_\mu\partial_\nu \Phi(X)_{ac}\right)\nonumber\\
&&\times\left(\Phi(X)_{ef}-\frac{r^\mu}{2}\partial_\mu \Phi(X)_{ef}+\frac{r^\mu r^\nu}{8}\partial_\mu \partial_\nu \Phi(X)_{ef}\right).\label{explicitintegrate}
\eeq
We keep only terms quadratic in $r$ in (\ref{explicitintegrate}) and then integrate out the $r$ variable. Only the terms proportional to $(r^1)^2$ give a non-vanishing contribution in (\ref{explicitintegrate}). 
 After this integration, we find the effective action:
\beq
S_{\rm eff}(\Phi)&=&\int d^2x\frac{1}{2}\Phi\partial_1^2\Phi\nonumber\\
&&\times\left\{1-\lambda^2\frac{Nm}{2(2\pi)^2}\int d\theta_1 d\theta_2 \frac{\sinh^2\left(\frac{\theta_1+\theta_2}{2}\right)\sinh^2\left(\frac{\theta_1-\theta_2}{2}\right)}{\cosh \left(\frac{\theta_1+\theta_2}{2}\right)\cosh\left(\frac{\theta_1-\theta_2}{2}\right)}\right.\nonumber\\
&&\times\left.\delta^{\prime\prime}\left(2m\cosh\left(\frac{\theta_1+\theta_2}{2}\right)\sinh\left(\frac{\theta_1-\theta_2}{2}\right)\right)\right.\nonumber\\
&&\left.\times\frac{4\pi^2}{\left(\theta_1-\theta_2\right)^2+\pi^2}\exp 2\int_0^\infty \frac{d\xi}{\xi}\left[\frac{-2\sinh\left(\frac{2\xi}{N}\right)}{\sinh \xi}\right.\right.\nonumber\\
&&\,\,\,\,\,\,\,\left.\left.+\frac{4e^{-\xi}\left(e^{2\xi/N}-1\right)}{1-e^{-2\xi}}\right]\frac{\sin^2[\xi(\pi i-(\theta_1-\theta_2))/2\pi]}{\sinh \xi}\right\}\nonumber\\
&-&\lambda^2\sum_{x^2}\int d^2x\left[ g_0^2\,q(x^0)\,\Phi (x^0,u^1,u^2)-g_0^2\,q^\prime(x^0)\Phi(x^0,v^1,v^2)\right].\nonumber
\eeq

We can now read off the renormalized string tension $\sigma^H$, by integrating out the auxiliary field $\Phi$:
\beq
\sigma^H&=&\lambda^2\, \frac{g_0^2}{a^2}C_N\left\{1-\left[\lambda^2\frac{Nm}{2(2\pi)^2}\int d\theta_1 d\theta_2 \frac{\sinh^2\left(\frac{\theta_1+\theta_2}{2}\right)\sinh^2\left(\frac{\theta_1-\theta_2}{2}\right)}{\cosh \left(\frac{\theta_1+\theta_2}{2}\right)\cosh\left(\frac{\theta_1-\theta_2}{2}\right)}\right.\right.\nonumber\\
&&\,\,\,\,\,\,\,\,\,\,\,\,\,\,\,\,\,\,\,\,\left.\times\delta^{\prime\prime}\left(2m\cosh\left(\frac{\theta_1+\theta_2}{2}\right)\sinh\left(\frac{\theta_1-\theta_2}{2}\right)\right)\right.\nonumber\\
&&\,\,\,\,\,\,\,\,\,\,\,\,\,\,\,\,\,\,\,\,\,\left.\times\frac{4\pi^2}{\left(\theta_1-\theta_2\right)^2+\pi^2}\exp 2\int_0^\infty \frac{d\xi}{\xi}\left[\frac{-2\sinh\left(\frac{2\xi}{N}\right)}{\sinh \xi}\right.\right.\nonumber\\
&&\,\,\,\,\,\,\,\,\,\,\,\,\,\,\,\,\,\,\,\,\left.\left.\left.+\frac{4e^{-\xi}\left(e^{2\xi/N}-1\right)}{1-e^{-2\xi}}\right]\frac{\sin^2[\xi(\pi i-(\theta_1-\theta_2))/2\pi]}{\sinh \xi}\right]\right\}^{-1}\nonumber.
\eeq
After the integration over $\theta_1$ and $\theta_2$, the string tension is
\beq
\sigma^H&=&\lambda^2\, \frac{g_0^2}{a^2}C_N\left\{1-\lambda^2\frac{N}{3m^3 (2\pi)^2}\exp2\int_0^\infty\frac{d\xi}{\xi}\left[\frac{-2\sinh\left(\frac{2\xi}{N}\right)}{\sinh \xi}\right.\right.\nonumber\\
&&\left.\left.\,\,\,\,\,\,\,\,\,\,\,\,\,\,\,\,\,\,\,+\frac{4e^{-\xi}\left(e^{2\xi/N}-1\right)}{1-e^{-2\xi}}\right]\frac{\sin^2\left(\frac{i\xi}{2}\right)}{\sinh\xi}\right\}^{-1}.\label{horizontaltension}
\eeq
The string tension (\ref{horizontaltension}) generalizes the result from Refence \cite{horizontal} from $N=2$, to general $N>2$. 

In the next section we compute the string tension for a quark-antiquark pair separated in the $x^2$ direction, rather than $x^1$. We call this the vertical string tension $\sigma^V$.

\section{The vertical string tension}
\setcounter{equation}{0}
\renewcommand{\theequation}{6.3.\arabic{equation}}

In this section we calculate the string tension, $\sigma^V$, for a quark-antiquark pair separated only in the $x^2$ direction. This calculation has been done before in Reference \cite{vertical} for the $SU(2)$ gauge group. We show here how to generalize this result for $N>2$ using the form factors from Chapter 4.

If we place a static quark at the space point $u^1,u^2$, and an antiquark at $u^1,v^2$, with $u^2>v^2$, The residual Gauss's Law (\ref{modifiedresidualgauss}) requires that there be at least one sigma model particle in each $x^2$ layer, for $u^2>x^2>v^2$. The left-handed color index of a particle at $x^2$ is contracted with the right-handed color of the particle at $x^2+a$. The left-handed color index of the particle at $u^2-a$ and the right-handed color of the particle at $v^2+a$ are contracted with the color indices of the quark at $u^2$, and the antiquark at $v^2$, respectively. The physical state then resembles a color-singlet string of sigma model particles, whose endpoints are the quarks.
The vertical string tension is obtain by calculating the energy of this string,
\beq
\sigma^V=\lim_{\vert u^2-v^2\vert\to\infty} \frac{E_{\rm string}}{\vert u^2-v^2\vert}.\nonumber
\eeq
The first approximation is obtained by assuming the energy of the string is only given by the mass of the sigma model particles, such that $E_{\rm string}=\frac{m}{a}\vert u^2-v^2\vert$, so $\sigma^V=m/a$.

Corrections to the vertical string tension are found by calculating the contributions to the energy of the string from the Hamiltonian $\lambda^2H_1$. As in Reference \cite{vertical}, we will use a non relativistic approximation, where the sigma model particles have momentum much smaller than their mass, so we will ignore any creation or annihilation of particles.

The projection of the Hamiltonian onto the non relativistic string state is
\beq
H=\sum_{x^2=v^2}^{u^2}\left\{m+\int\frac{dp}{2\pi}\frac{p^2}{2m}\mathfrak{A}_P^\dag(p)_{a b}\mathfrak{A}_P(p)_{a b}\right\}+\lambda^2H_1,\nonumber
\eeq
where  $\mathfrak{A}_P^\dag(p)_{a b}$, and $\mathfrak{A}_P(p)_{a b}$ are the sigma model particle creation and annihilation operators, respectively, and 
\beq
&&H_1=\sum_{x^2}\int dx^1\int dy^1\frac{1}{4g_0^2 a}\vert x^1-y^1\vert\nonumber\\
&&\,\,\,\,\times\left[j_0^L(x^1,x^2)-j_0^R(x^1,x^2-a)+q_b\delta(x^1-u^1)\delta_{x^2 u^2}-q^\prime_b\delta(x^1-u^1)\delta_{x^2 v^2}\right]\nonumber\\
&&\,\,\,\,\times\left[j_0^L(y^1,x^2)-j_0^R(y^1,x^2-a)+q_b\delta(y^1-u^1)\delta_{x^2 u^2}-q^\prime_b\delta(y^1-u^1)\delta_{x^2 v^2}\right],\nonumber\\
\label{stringhamiltonian}
\eeq
where we have again eliminated the auxiliary field, $\Phi$.

We now now need to find the expectation value 
\beq
\langle {\rm string}\vert H_1\vert {\rm string}\rangle,\label{expectationstring}
\eeq
 where the state $\vert {\rm string}\rangle$ has a sigma-model particle for every $x^2$, whose center of mass is located at $x^1=z(x^2)$. To evaluate (\ref{expectationstring}), we need matrix elements of the form
 \beq
 &&\langle P, z_1,a_1,b_1\vert j_0^{C}(x)_{ac}\vert P,z_2,a_2,b_2\rangle=\int \frac{dp_1}{2\pi}\frac{1}{\sqrt{2E_1}}\int\frac{dp_2}{2\pi}\frac{1}{\sqrt{2E_2}}\nonumber\\
 &&\,\,\,\,\,\,\,\,\times e^{-ip_1\cdot(z_1-x)+ip_2\cdot(z_2-x)}\langle P, \theta_1,a_1,b_1\vert j_0^C(x)_{ac}\vert P,\theta_2,a_2,b_2\rangle,\label{fourierform}
 \eeq
where the matrix element on the right hand side of (\ref{fourierform}) is the two particle form factor from Section 4.2 (with the incoming antiparticle crossed to an outgoing particle), and $C=L,R$. By applying crossing symmetry on the form factor (\ref{finitenform}), we find
\beq
&&\langle P, \theta_1,a_1,b_1\vert j_0^C(x)_{ac}\vert P,\theta_2,a_2,b_2\rangle\nonumber\\
&&\,\,\,\,=(p_1+p_2)_0\mathfrak{D}^C_{a\,c\,a_1a_2b_1b_2}\nonumber\\
&&\,\,\,\,\times\frac{2\pi i}{\theta+2\pi i}\exp\int_0^\infty\frac{d\xi}{\xi}\left[\frac{-2\sinh\left(\frac{2\xi}{N}\right)}{\sinh\xi}+\frac{4 e^{-\xi}\left(e^{\frac{2\xi}{N}}-1\right)}{1-e^{-2\xi}}\right]\frac{\sin^2[\xi\theta/2\pi]}{\sinh\xi},\nonumber
\eeq
where
\beq
\mathfrak{D}^L_{a\,c\,a_1a_2b_1b_2}&=&\delta_{a\,a_2}\delta_{c\,a_1}\delta_{b_1b_2}-\frac{1}{N}\delta_{a\,c\,}\delta_{a_1a2}\delta_{b_1b_2},\nonumber\\
\mathfrak{D}^R_{a\,c\,a_1a_2b_1b_2}&=&\delta_{a\,b_2}\delta_{c\,b_1}\delta_{a_1a_2}-\frac{1}{N}\delta_{a\,c}\delta_{a_1a_2}\delta_{b_1b_2}.\nonumber
\eeq
Taking the non-relativistic limit, we find
\beq
&&\frac{1}{\sqrt{2E_1}}\frac{1}{\sqrt{2E_2}}\langle P, \theta_1,a_1,b_1\vert j_0^C(x)_{a\,c}\vert P,\theta_2,a_2,b_2\rangle\nonumber\\
&&\,\,\,\,\approx\mathfrak{D}^C_{a\,c\,a_1a_2b_1b_2}\exp-\frac{A_N}{m^2}(p_1-p_2)^2.\nonumber
\eeq
where
\beq
A_N&=&\int_0^\infty\frac{d\xi}{4\pi^2}\frac{\xi}{\sinh\xi}\left[\sinh\left(\frac{2\xi}{N}\right)-2\left(e^{2\xi/N}-1\right)\right]\nonumber\\
\nonumber\\
&=&\frac{1}{16}\pi^2\left[2\pi^2-3\psi^{(1)}\left(\frac{1}{2}-\frac{1}{N}\right)-\psi^{(1)}\left(\frac{1}{2}+\frac{1}{N}\right)\right],\nonumber
\eeq
for $N>2$, where $\psi^{(n)}(x)=d^{n+1}\ln \Gamma(x)/dx^{n+1}$ is the $n$-th polygamma function.

The matrix element $(\ref{fourierform})$ is then
\beq
&&\langle P, z_1,a_1,b_1\vert j_0^{C}(x)_{ac}\vert P,z_2,a_2,b_2\rangle\nonumber\\
&&\,\,\,\,\,\,\,\,\,\,=\sqrt{\frac{m^2}{2\pi A_N}}\mathfrak{D}^C_{a\,c\,a_1a_2b_1b_2}\exp\left[-\frac{m^2}{4A_N}\left(\frac{z_1+z_2}{2}-y\right)^2\right]\delta(z_1-z_2).\nonumber\\\label{smearedcolor}
\eeq
This means that the color of a particle is a Gaussian distribution in the non relativistic limit. In this sense, they are not point-like particles, but the color is smeared over space.

We now use (\ref{smearedcolor}) to write the effective Hamiltonian of the non-relativistic string. This is given by the projection of the Hamiltonian (\ref{stringhamiltonian}) onto the state $\vert {\rm string}\rangle$, which has a sigma-model particle at each $x^2$ layer, located at the point $z^{1}(x^2)$, for  $u^2>x^2>v^2$, a static quark at $u^1,u^2$, and an antiquark at $u^1,v^2$. The string Hamiltonian is
\beq
H_{\rm string}=\frac{m}{a}(v^2-u^2)-\frac{1}{2m}\sum_{x^2=v^2}^{u^2-a}\frac{\partial^2}{\partial z^1(x^2)^2}+\lambda^2V_{\rm bulk}+\lambda^2V_{\rm ends},\nonumber
\eeq
where
\beq
V_{\rm bulk}&=&-\frac{m^2}{8\pi A_N}\frac{1}{g_0^2 a^2}\sum_{x^2=v^2+a}^{u^2-a}\int dx^1\,dy^1\vert x^1-y^1\vert\nonumber\\
&&\times\left\{e^{-\frac{m^2}{4A_N}\left[z^1(x^2)-x^1\right]^2}\mathfrak{D}^L(x^2)_{a\,c\,a_1a_2b_1b_2}\right.\nonumber\\
&&\left.-e^{-\frac{m^2}{4A_N}\left[z^1(x^2-a)-x^1\right]^2}\mathfrak{D}^R(x^2-a)_{a\,c\,a_1a_2b_1b_2}\right\}\nonumber\\
&&\times\left\{e^{-\frac{m^2}{4A_N}\left[z^1(x^2)-y^1\right]^2}\mathfrak{D}^L(x^2)_{a\,c\,a_2a_1b_2b_1}\right.\nonumber\\
&&\left.-e^{-\frac{m^2}{4A_N}\left[z^1(x^2-a)-y^1\right]^2}\mathfrak{D}^R(x^2-a)_{a\,c\,a_2a_1b_2b_1}\right\}\label{bulk},
\eeq
and
\beq
V_{\rm ends}&=&-\frac{1}{4g_0^2 a^2}\int dx^1 dy^1\vert x^1-y^1\vert\left\{\sqrt{\frac{m^2}{2\pi A_N}}e^{-\frac{m^2}{4A_N}\left[z^1(v^2)-x^1\right]^2}\mathfrak{D}^R(v^2)_{a\,c\,a_1a_2b_1b_2}\right.\nonumber\\
&&\,\,\,\,\,\,\,\,\,\left.+\delta(x^2-v^1)q^\prime_{a\,c}4\pi\delta_{a_1a_2}\delta_{b_1b_2}\right\}\nonumber\\
&&\times\left\{\sqrt{\frac{m^2}{2\pi A_N}}e^{-\frac{m^2}{4A_N}\left[z^1(v^2)-y^1\right]^2}\mathfrak{D}^R(v^2)_{a\,c\,a_1a_2b_1b_2}\right.\nonumber\\
&&\left.\,\,\,\,\,\,\,\,\,+\delta(y^1-u^1)q^\prime_{a\,c}4\pi\delta_{a_1a_2}\delta_{b_1b_2}\right\}\nonumber\\
&&-\frac{1}{4g_0^2 a^2}\int dx^1 dy^1\vert x^1-y^1\vert\nonumber\\
&&\times\left\{\sqrt{\frac{m^2}{2\pi A_N}}e^{-\frac{m^2}{4A_N}\left[z^1(u^2-a)-x^1\right]^2}\mathfrak{D}^L(u^2-a)_{a\,c\,a_1a_2b_1b_2}\right.\nonumber\\
&&\,\,\,\,\,\,\,\,\,\left.+\delta(x^2-u^1)q^\prime_{a\,c}4\pi\delta_{a_1a_2}\delta_{b_1b_2}\right\}\nonumber\\
&&\times\left\{\sqrt{\frac{m^2}{2\pi A_N}}e^{-\frac{m^2}{4A_N}\left[z^1(u^2-a)-y^1\right]^2}\mathfrak{D}^L(u^2-a)_{a\,c\,a_1a_2b_1b_2}\right.\nonumber\\
&&\left.\,\,\,\,\,\,\,\,\,+\delta(y^2-u^1)q^\prime_{a\,c}4\pi\delta_{a_1a_2}\delta_{b_1b_2}\right\}\label{ends}.
\eeq

Enforcing the residual Gauss's law (\ref{modifiedresidualgauss}) on (\ref{bulk}) and (\ref{ends}), implies
\beq
&&\int dx^1\left\{-\sqrt{\frac{m^2}{2\pi A_N}}e^{-\frac{m^2}{4A_N}\left[z^1(x^2)-x^1\right]^2}\mathfrak{D}^L(x^2)_{a\,c\,a_1a_2b_1b_2}\right.\nonumber\\
&&\left.+\sqrt{\frac{m^2}{2\pi A_N}}e^{-\frac{m^2}{4A_N}\left[z^1(x^2-a)-x^1\right]^2}\mathfrak{D}^R(x^2-a)_{a\,c\,a_1a_2b_1b_2}\right\}\Psi=0,\label{gaussconstraintone}
\eeq
for $u^2>x^2>v^2$, and
\beq
&&\int dx^1\sqrt{\frac{m^2}{2\pi A_N}}\left\{e^{-\frac{m^2}{4A_N}\left[z^1(v^2)-x^1\right]^2}\mathfrak{D}^R(v^2)_{a\,c\,a_1a_2b_1b_2}\right.\nonumber\\
&&\,\,\,\,\,\,\,\,\,\,\,\,\,\,\,\left.-q^\prime_{a\,c}\delta(x^1-u^1)4\pi 
\delta_{a_1a_2}\delta_{b_1b_2}\right\}\Psi=0,\nonumber\\
&&\int dx^1\sqrt{\frac{m^2}{2\pi A_N}}\left\{e^{-\frac{m^2}{4A_N}\left[z^1(u^2-a)-x^1\right]^2}\mathfrak{D}^L(u^2-a)_{a\,c\,a_1a_2b_1b_2}\right.\nonumber\\
&&\,\,\,\,\,\,\,\,\,\,\,\,\,\,\,\left.-q_{a\,c}\delta(x^1-u^1)4\pi 
\delta_{a_1a_2}\delta_{b_1b_2}\right\}\Psi=0,\label{gaussconstrainttwo}
\eeq
respectively. The constraint (\ref{gaussconstraintone}) is satisfied by identifying $\mathfrak{D}^L(x^2)_{a\,c\,a_1a_2b_1b_2}=\mathfrak{D}^R(x^2-a)_{a\,c\,a_1a_2b_1b_2}$. The constraint (\ref{gaussconstrainttwo}) is satisfied by identifying $\mathfrak{D}^R(v^2)_{a\,c\,a_1a_2b_1b_2}=q^\prime_{a\,c}4\pi\delta_{a_1a_2}\delta_{b_1b_2}$, and $\mathfrak{D}^L(u^2-a)_{a\,c\,a_1a_2b_1b_2}=q_{a\,c}4\pi\delta_{a_1a_2}\delta_{b_1b_2}$. Using this, we can eliminate the color degrees of freedom from (\ref{bulk}) and (\ref{ends}).

Now we want to integrate out the variables $x^1$ and $y^1$ from equations (\ref{bulk}) and (\ref{ends}). The integrals involved are:
\beq
\int dx^1 dy^1\vert x^1-y^1\vert e^{-\frac{m^2}{4A_N}\left[(x^1)^2+(y^1)^2\right]}&=&\frac{4\sqrt{2\pi}A_N^{3/2}}{m^3},\nonumber\\
\int dx^1 dy^1\vert x^1-y^1\vert e^{-\frac{m^2}{4A_N}\left[(x^1+r)^2+(y^1)^2\right]}&=&\frac{4\sqrt{2\pi}A_N^{3/2}}{m^3}P(r),\nonumber\\
\int dx^1 \vert x^1-u^1\vert e^{-\frac{m^2}{4A_N}\left[x^1-z^1(U^2)\right]^2}&=&\frac{2A_N}{m^2}P\left[\sqrt{2}z^1(u^2)-\sqrt{2}u^1\right],\nonumber
\eeq
Where $P(r)$ is a function for which we do not have an exact analytic expression, but its behavior for small and large $r$ is
\beq
P(r)=\left\{\begin{array}{cc}
1+\frac{m^2r^2}{4A_N},&r<<\frac{1}{m},\\
\sqrt{\frac{\pi}{2A_N}}m\vert r\vert,& r>>\frac{1}{m}.\end{array}\right.\label{limitspfunction}
\eeq

After integrating out $x^1$, and $y^1$, the string Hamiltonian is
\beq
H_{\rm string}&=&\frac{m}{a}(u^2-v^2)-\frac{1}{2m}\sum_{x^2=v^2}^{u^2-a}\frac{\partial^2}{\partial z^1(x^2)^2}\nonumber\\
&&-\frac{\lambda^2N(N^2-1)}{m\,g_0^2 a^2}\sqrt{\frac{A_N}{2\pi}}\sum_{x^2=v^2+a}^{u^2}\left\{1-P\left[z^1(x^2)-z^1(x^2-a)\right]\right\}\nonumber\\
&&-\frac{\lambda^2N(N^2-1)}{m\,g_0^2 a^2}\sqrt{\frac{A_N}{2\pi}}\left(1+P\left\{\sqrt{2}\left[z^1(v^2)-u^1\right]\right\}\right.\nonumber\\
&&\,\,\,\,\,\,\,\,\,\,\,\,\,\,\,\,\,\,\,\,\,\,\,\,\,\,\,\,\,\,\,\,\,\,\,\,\,\,\,\,\,\left.+P\left\{\sqrt{2}\left[z^1(u^2-a)-u^1\right]\right\}\right),\label{afterintegratex}
\eeq
where we have used
\beq
\left(\mathfrak{D}^{C}\right)^2=N\left(N^2-1\right).\nonumber
\eeq
The potential energy between a static quark-antiquark pair is then determined by finding the ground state of the Hamiltonian (\ref{afterintegratex}).

We further simplify the Hamiltonian (\ref{afterintegratex}) using the small-gradient approximation. That is, in the non-relativistic limit (when the sigma model mass gap is taken to be very large), we expect that the sigma-model particles in two adjacent $x^2$ layers are close to each other in the $x^1$ direction. Specifically, we assume $\vert z^1(x^2)-z^1(x^2-a)\vert<< m^{-1}$. At the endpoints of the string, we also assume $\vert z^1(v^2)-u^1\vert<< m^{-1}$, and $\vert z^1(u^2-a)-u^1\vert<< m^{-1}$.
Using Eq. (\ref{limitspfunction}), the small-gradient approximation gives the Hamiltonian
\beq
H_{\rm string}&=&\frac{\lambda^2N(N^2-1)}{m\,g_0^2 a^2}\sqrt{\frac{A_N}{2\pi}}+\frac{m}{a}(u^2-v^2)-\frac{1}{2m}\sum_{x^2=v^2}^{v^2-a}\frac{\partial^2}{\partial z^1(x^2)^2}\nonumber\\
&&+\frac{\lambda^2N(N^2-1)}{4 m g_0^2 a^2}\sqrt{\frac{1}{2\pi A_N}}\sum_{x^2=v^2+a}^{u^2-a}\left[z^1(x^2)-z^1(x^2-a)\right]^2\nonumber\\
&&+\frac{\lambda^2N(N^2-1)}{2 m g_0^2 a^2}\sqrt{\frac{1}{2\pi A_N}}\left\{\left[z^1(v^2)-u^1\right]^2+\left[z^1(u^2-a)-u^1\right]^2\right\}.\nonumber\\\label{smallgradientapprox}
\eeq

The first term in the Hamiltonian (\ref{smallgradientapprox}) is just a constant with no physical significance, so we will ignore it from now on. The Hamiltonian (\ref{smallgradientapprox}) is equivalent $Q=\left(u^2-v^2\right)/a$ coupled harmonic oscillators. The ground-state energy is then given by
\beq
E_0=mQ-\frac{\lambda\sqrt{N(N^2-1)}}{g_0a}\left(\frac{1}{2\pi A_N}\right)^{\frac{1}{4}}\sum_{q=0}^Q\sin\frac{\pi q}{2Q}.\label{groundstatestring}
\eeq
Using the Euler summation formula, for large $Q$:
\beq
\sum_{q=0}^QF\left(\frac{q}{Q}\right)&=&Q\int_0^1 dxF(x)-\frac{1}{2}\left[F(1)-F(0)\right]\nonumber
\\
&&+\frac{1}{12Q}\left[F^\prime(1)-F^\prime(0)\right]+\mathcal{O}\left(\frac{1}{Q^2}\right), \nonumber
\eeq
the ground-state energy (\ref{groundstatestring}) becomes (dropping any constants that do not depend on $Q$)
\beq
E_0&=&\left[\frac{m}{a}-\frac{2\lambda\sqrt{N(N^2-1)}}{\pi g_0a^2}\left(\frac{1}{2\pi A_N}\right)^{\frac{1}{4}}\right]L\nonumber\\
&&+\frac{\pi}{24}\frac{\lambda\sqrt{N(N^2-1)}}{ g_0}\left(\frac{1}{2\pi A_N}\right)^{\frac{1}{4}}\frac{1}{L}+\mathcal{O}\left(\frac{1}{L^2}\right).\label{transversegroundstate}
\eeq
where the distance between the quark and antiquark is $L=Qa$. 

We can easily read the vertical string tension off (\ref{transversegroundstate}):
\beq
\sigma^V=\frac{m}{a}-\frac{2\lambda\sqrt{N(N^2-1)}}{\pi g_0a^2}\left(\frac{1}{2\pi A_N}\right)^{\frac{1}{4}}.\label{verticalstringtension}\eeq
There is also a Coulomb-like term in the quark-antiquark potential, which is proportional to $1/L$.

\chapter{Anisotropic renormalization group for gauge theories}

\setcounter{equation}{0}
\renewcommand{\theequation}{7.\arabic{equation}}

\section{Classical longitudinal rescaling }
\setcounter{equation}{0}
\renewcommand{\theequation}{7.1.\arabic{equation}}

This chapter contains material previously published in \cite{longQED} and \cite{longQCD}.

The longitudinal rescaling of coordinates described in Chapter 2 is completely classical. In a quantum field theory, a rescaling of coordinates changes the energy scales of the theory, and a renormalization group procedure is needed. We consider the effect of longitudinal rescaling of (3+1)-dimensional quantum electrodynamics (QED) in Sections (7.1-7.5). We study the longitudinal rescaling of (3+1)-dimensional  QCD in Sections (7.6-7.9).

 In QED, the Abelian gauge field with Lorentz components $A_{\mu}$, $\mu=0,1,2,3$,
transforms as $A^{0,3}\to\lambda^{-1}A^{0,3}$ and $A^{1,2}\to A^{1,2}$. The action of the Maxwell gauge field is $S_{G}=-\frac{1}{4g^{2}}\int d^{4}xF_{\mu\nu}F^{\mu\nu}$, where $F_{\mu\nu}=\partial_{\mu}A_{\nu}-\partial_{\nu}A_{\mu}$, indices are raised with the usual Minkowski metric and $g$ is the bare electric charge.  Under longitudinal
rescaling, the gauge action becomes 
\beq
S_G\longrightarrow \frac{1}{4g^{2}}\int d^{4}x \left(F_{01}^{2}+F_{02}^{2}-F_{13}^{2}-F_{23}^{2}+\lambda^{-2}F_{03}^{2}-\lambda^{2}F_{12}^{2}\right).
\label{rescaledGauge}
\eeq
The massless Dirac action transforms as
 \beq
 S_{\rm Dirac}\longrightarrow i\int d^{4}x\,\bar{\psi}\left[\lambda^{-1}\gamma^0D_{0}+\lambda^{-1}\gamma^{3}D_{3}+\gamma^{1}D_{1}+\gamma^{2}D_{2}\right]\psi, \nonumber
 \eeq
where $D_{\mu} \psi=\partial_{\mu}\psi+iA_{\mu}\psi$ is the covariant derivative of $\psi$.
If we make an additional rescaling of  the spinor field $\psi\to\lambda^{\frac{1}{2}}\psi$ and $\bar{\psi}\to\lambda^{\frac{1}{2}}\bar{\psi}$, we obtain
 \beq
 S_{\rm Dirac}\longrightarrow i\int d^{4}x\,\bar{\psi}\left[\gamma^{0}D_0+\gamma^{3}D_3+\lambda\gamma^{1}D_1+\lambda\gamma^{2}D_2\right]\psi. \label{rescaledDirac}
 \eeq
In the quantum theory, anomalous powers of $\lambda$ will appear in the rescaled action
(\ref{rescaledGauge}) (\ref{rescaledDirac}). 

In the quantum theory, classical rescalings are no longer possible. The best-known is example is
the effect of a dilatation on a classically conformal invariant field theory. Quantum corrections violate this classical symmetry. 

One can imagine cutting off the quantum field theory in the ultraviolet by a cubic lattice, with lattice spacing
$a$. Rescaling changes the lattice spacing of longitudinal coordinates to $\lambda a$, but does not change the lattice spacing of transverse coordinates, making the cutoff anisotropic. We therefore use a 
two-step process, where we first integrate out high-longitudinal momentum degrees of freedom, then restore isotropy with 
longitudinal rescaling. Instead of a lattice, we use a sharp-momentum cutoff and Wilson's renormalization procedure, to integrate out 
high-momentum modes \cite{RG}. This was done in reference \cite{xiao} for pure Yang-Mills theory. 

In the next section we review basic Wilsonian renormalization. Then we examine the renormalization of QED first with a spherical momentum cutoff and then with aspherical cutoffs, which treat longitudinal momenta and transverse momenta differently. We then find the quantum corrections to the QED action. 

\section{Wilsonian renormalization }
\setcounter{equation}{0}
\renewcommand{\theequation}{7.2.\arabic{equation}}

We Wick rotate to obtain the standard Euclidean
metric, so that the action is
\beq
S=\int d^{4}x (\frac{1}{4g^{2}}F_{\mu\nu}F^{\mu\nu}+i {\bar \psi}\,/\!\!\!\!D \,\psi). \nonumber
\eeq
where raising and lowering of indices is done with the Euclidean metric and where the slash on a vector quantity $J_{\mu}$ is $/\!\!\! J=\gamma^{\mu}J_{\mu}$, where 
$\gamma^{\mu}$ are the Euclidean Dirac matrices.

 We choose cutoffs $\Lambda$ and $\tilde \Lambda$ to be real 
positive numbers with units of
$cm^{-1}$ and $b$ and $\tilde b$ to be two dimensionless real numbers, such that
$b\ge 1$ and ${\tilde b}\ge 1$. These quantities satisfy $\Lambda>{\tilde \Lambda}$
and that $\Lambda^{2}/b \ge {\tilde \Lambda}^{2}/{\tilde b}$.  We  
introduce the ellipsoid in momentum space $\mathbb P$, which is the set of points $p$, such that
$bp_{L}^{2}+p_{\perp}^{2}<\Lambda^{2}$. We define the smaller ellipsoid ${\tilde {\mathbb P}}$ 
to
be the
set of points $p$, such that ${\tilde b}p_{L}^{2}+p_{\perp}^{2}<{\tilde \Lambda}^{2}$. Finally,
we define $\mathbb S$ to be the shell between the two ellipsoidal surfaces
 ${\mathbb S}={\mathbb P}-{\tilde{\mathbb P}}$. 

We split our fields into ``slow" and ``fast" pieces:
\beq
\psi(x)=\tilde{\psi}(x)+\varphi (x), \;
\bar{\psi}(x)= \tilde{\bar{\psi}}(x)+\bar{\varphi}(x),\;
A_{\mu}(x) =\tilde{A_{\mu}} (x)+a_{\mu} (x), \label{slow-fast}
\eeq
where the Fourier components of $\psi(x)$, ${\bar \psi}(x)$ and $A_{\mu}(x)$ vanish outside the
ellipsoid $\mathbb P$, the Fourier components of the slow fields  ${\tilde \psi}(x)$, ${\tilde{ \bar{ \psi}}}(x)$ and ${\tilde A}_{\mu}(x)$ vanish outside the
inner ellipsoid $\tilde {\mathbb P}$, and the Fourier components of the fast fields 
$\varphi(x)$, ${\bar \varphi}(x)$ and $a_{\mu}(x)$ vanish outside of the shell $\mathbb S$. Explicitly
\beq
\tilde{\psi}(x)=\int_{\tilde{\mathbb{P}}}\frac{d^{4}p}{(2\pi)^{4}}\psi(p)e^{-ip\cdot x}\!\!\!\!\!&,&\!\!\!\!\! \,\,\varphi(x)
=\int_{\mathbb{S}}\frac{d^{4}p}{(2\pi)^{4}}\psi(p)e^{-ip\cdot x},\;\nonumber\\
\tilde{\bar{\psi}}(x)=\int_{\tilde{\mathbb{P}}}\frac{d^{4}p}{(2\pi)^{4}}\bar{\psi}(p)e^{ip\cdot x}\!\!\!\!\!&,&\!\!\!\!\!\,\,\bar{\varphi}(x)=\int_{\mathbb{S}}\frac{d^{4}p}{(2\pi)^{4}}\bar\psi(p)e^{ip\cdot x}, \nonumber \\
\tilde{A}_{\mu}(x)=
\int_{\tilde{\mathbb{P}}}\frac{d^{4}p}{(2\pi)^{4}}A_{\mu}(p)e^{-ip\cdot x}\!\!\!\!\!&,&\!\!\!\!\!\,\,\,a_{\mu}(x)=\int_{\mathbb{S}}\frac{d^{4}p}{2(\pi)^{4}}A_{\mu}(p)e^{-ip\cdot x}.\nonumber
\eeq
We denote the field strength of the slow fields by ${\tilde F}_{\mu\nu}=
\partial_{\mu}{\tilde A}_{\nu}-\partial_{\nu}{\tilde A}_{\mu}$. The functional integral with the ultraviolet cutoff $\Lambda$ and anisotropy parameter $b$ is
\beq
Z=\int_{\mathbb{P}}\;\mathcal{D}\psi\mathcal{D}\bar{\psi}\mathcal{D}A e^{-S} .\label{originalZ}
\eeq
There is no renormalization of a gauge-fixing parameter, because
we do not impose a gauge condition on the slow gauge field. We do impose a Feynman gauge condition
on the fast gauge field. As we show below, counterterms must be included in the action to maintain gauge invariance. We expect that renormalizability of the the field theory implies that these have the same form at each loop order; we have not proved this, however.

Before integrating over the fast fields, we 
must expand the action to second order in these fields. This expansion is
\beq
S={\tilde S}+S_{0}+S_{1}+S_{2}+S_{3},\nonumber
\eeq
where 
\beq
{\tilde S}\!&\!=\!&\!\int d^{4}x ({\tilde F}_{\mu\nu}{\tilde F}^{\mu\nu}+i {\tilde {\bar \psi}}\,/\!\!\!\!D \,{\tilde \psi}), \nonumber \\
S_{0}\!&\!=\!&\! \int_{\mathbb S} \frac{d^{4}q}{(2\pi)^{4}}\frac{1}{2}q^{2} a_{\mu}(-q) a^{\mu}(q)
+i\int_{\mathbb{S}}\frac{d^{4}q}{(2\pi)^{4}}\bar{\varphi}(-q)\; /\!\!\!q
\;\varphi(q), \nonumber \\
S_{1}\!&\!=\!&\!
-\int_{\mathbb{S}}\frac{d^{4}q}{(2\pi)^{4}}\int_{\tilde{\mathbb{P}}}\frac{d^{4}p}{(2\pi)^{4}}\left[\bar{\varphi}(q)/\!\!\!\!A(p)\varphi(-q-p)\right], \nonumber \\
S_{2} \!&\!=\!&\!  -\int_{\mathbb{S}}\frac{d^{4}q}{(2\pi)^{4}}\int_{\tilde{\mathbb{P}}}\frac{d^{4}p}{(2\pi)^{4}}\left[\bar{\psi}(p)/\!\!\!a(q)\varphi(-q-p)\right], \nonumber \\
S_{3}\!&\!=\!\!&\!\!-\int_{\mathbb{S}}\frac{d^{4}q}{(2\pi)^{4}}\int_{\tilde{\mathbb{P}}}\frac{d^{4}p}{(2\pi)^{4}}\left[\bar{\varphi}(-q-p)/\!\!\!a(q)\psi(p)\right]. \label{expansion}
\eeq
Notice that $S_2^*=S_3$. Henceforth, we drop the tildes on the slow fields, denoting these by 
$\psi,\, \bar{\psi}$ and $A_{\mu}$, but we keep the tilde on the slow action $\tilde S$.

The functional integral (\ref{originalZ}) may be written as
\beq
Z=\int_{\tilde{\mathbb{P}}}{\mathcal D}\;\psi{\mathcal D}\bar{\psi}{\mathcal D}A\;e^{-{\tilde S}}
\;\int_{\mathbb{S}}\;{\mathcal D}\varphi{\mathcal D}\bar{\varphi}{\mathcal D}a\;e^{-S_0}
e^{-S_{I}} \label{functionalintegral}
\eeq
where the integral of the interaction Lagranigian is $S_{I}=S_{1}+S_{2}+S_{3}$. To evaluate \ref{functionalintegral}, 
we use the fast-field propagators
\beq
\langle a_{\mu}(p)a_{\nu}(q)\rangle&=&\frac{g^{2}}{q^{2}}g_{\mu\nu}\delta^{(4)}(p+q)(2\pi)^{4}
\,,\;\;\nonumber\\
\langle\varphi(p)\bar{\varphi}(q)\rangle&=&\frac{-i/\!\!\!q}{q^{2}}\delta^{(4)}(p+q)(2\pi)^{4}\,,
\label{propagators}
\eeq
where the brackets $\langle\,\rangle$ mean
\beq
\langle Q\rangle =\left(
\int_{\mathbb{S}}\;{\mathcal D}\varphi{\mathcal D}\bar{\varphi}{\mathcal D}a\;e^{-S_0}
\right)^{-1}\int_{\mathbb{S}}\;{\mathcal D}\varphi{\mathcal D}\bar{\varphi}{\mathcal D}a\;Q\;e^{-S_0}.
\label{bracket-def}
\eeq
for any quantity $Q$.
We will ignore an overall free-energy renormalization from the first factor in 
(\ref{bracket-def}). We use the connected-graph expansion
\beq
\langle e^{-S_{I}}\rangle&=&\exp [-\langle S_{I}\rangle+\frac{1}{2}(\langle S_{I}^{2}\rangle\nonumber\\
&-&\langle 
S_{I}\rangle^{2})-\frac{1}{3!}(\langle S_{I}^{3}\rangle-3\langle S_{I}^{2}\rangle\langle S_{I}\rangle+\langle S_{I}\rangle ^3)+\cdots], \label{cge}
\eeq
for the interaction $S_{I}$. The terms in the exponent of (\ref{cge}) are straightforward to evaluate using (\ref{propagators}). We find
\beq
\langle S_1\rangle=0\nonumber
\eeq
and
\beq
\frac{1}{2}\langle S_{1}^{2}\rangle=\int_{\tilde{\mathbb{P}}} \frac{d^{4}p}{(2\pi)^{4}}\;\Pi^{\mu\nu}(p)
A_{\mu}(p)A_{\nu}(-p), \nonumber
\eeq
where the polarization tensor $\Pi^{\mu\nu}(p)$ is defined as
\beq
\Pi^{\mu\nu}(p)=\frac{1}{2}\int_{\mathbb{S}}\frac{d^{4}q}{(2\pi)^{4}}{\rm Tr}\,\left[\frac{/\!\!\!q}{q^{2}}\gamma^\mu \frac{/\!\!\!q+/\!\!\!p}{(q+p)^{2}}\gamma^\nu\right].\label{polarizationtensor}
\eeq
Similarly
\beq
\langle S_{2}\rangle=\langle S_{3}\rangle=0\nonumber
\eeq
and
\beq
\langle S_{2}S_{3}\rangle=\langle S_{3}S_{2}\rangle=\int_{\mathbb{S}}\frac{d^{4}q}{(2\pi)^{4}}
\int_{\tilde{\mathbb{P}}}\frac{d^{4}p}{(2\pi)^{4}} \left[\frac{-(/\!\!\!p+/\!\!\!q)}{(p+q)^{2}}\gamma^\mu\frac{g^{2}}{q^{2}}\psi(p)\bar{\psi}(-p)\gamma_{\mu}\right]. \nonumber
\eeq
Thus
\beq
\frac{1}{2}(\langle S_{2}S_{3}\rangle+\langle S_{3}S_{2}\rangle)=\int_{\tilde{\mathbb{P}}}\frac{d^4p}{(2\pi)^4}\Sigma(p)\bar{\psi}(p)
\psi(-p),\nonumber
\eeq
where the self-energy correction $\Sigma(p)$ is
\beq
\Sigma(p)&=&g^{2}\int_{\mathbb{S}}\frac{d^{4}q}{(2\pi)^{4}}\left[\gamma^\mu\frac{i(/\!\!\!p+/\!\!\!q)}{(p+q)^{2}}\gamma_{\mu}\frac{1}{q^{2}}\right]=\nonumber\\
&-&2g^{2}\int_{\mathbb{S}}\frac{d^{4}q}{(2\pi)^{4}}\left[\frac{i(/\!\!\!p+/\!\!\!q)}{q^{2}(q+p)^{2}}\right].\label{selfenergy} 
\eeq

From the cubic term in (\ref{cge}), we find 
\beq
-\frac{1}{3!}(\langle S_{I}^{3}\rangle\!&\!-\!&\! 3\langle S_{I}^{2}\rangle\langle S_{I}\rangle+\langle S_{I}\rangle ^3)
=-\frac{1}{3!}\langle S_{I}^{3} \rangle 
=-\langle S_{1} S_{2}S_{3}\rangle \nonumber \\
&=&\int_{\tilde{\mathbb{P}}}\frac{d^4p}{(2\pi)^4}\int_{\tilde{\mathbb{P}}}\frac{d^4q}{(2\pi)^4}\bar{\psi}(p)\Gamma^{\mu}(p,q)A_{\mu}(q)\psi(-p-q),\nonumber
\eeq
where the vertex correction $\Gamma^\mu(p,q)$ is
\beq
\Gamma^\mu(p,q)=2g^{2}\int_{\mathbb{S}}\frac{d^{4}k}{(2\pi)^{4}}\frac{/\!\!\!k\gamma^\mu(/\!\!\!k+/\!\!\!q)}{(k-p)^{2}(k+q)^{2}k^{2}}. \label{vertex}
\eeq

\section{Spherical momentum cutoffs }
\setcounter{equation}{0}
\renewcommand{\theequation}{7.3.\arabic{equation}}

The cutoffs of our theory become isotropic if $b,\tilde{b}=1$. Then the region $\mathbb{P}$ is a sphere in momentum space, whose elements $p_\mu$ satisfy $p^{2}<\Lambda^2$. The region $\tilde{\mathbb{P}}$ is also a sphere, whose elements $q_\mu$ satisfy $q^2<\tilde{\Lambda}^2$. The region $\mathbb{S}$ is a spherical shell, $\mathbb{S}=\mathbb{P}-\tilde{\mathbb{P}}$.

The polarization tensor (\ref{polarizationtensor}) may be written
\beq
\Pi^{\mu\nu}(p)=tr\left[\frac{1}{2}\gamma^\mu\gamma^\alpha\gamma^\nu\gamma^\beta\int_{\mathbb{S}}\frac{d^{4}q}{(2\pi)^{4}}\frac{q_\alpha(q_\beta+p_\beta)}{q^{2}(q+p)^{2}}\right].\nonumber
\eeq
We expand the integrand in powers of $p$ to second order, to find
\beq
&\Pi^{\mu\nu}(p)=tr\left\{\frac{1}{2}\gamma^\mu\gamma^\alpha\gamma^\nu\gamma^\beta\frac{2\pi^{2}}{(2\pi)^{4}}\left[\int dq\frac{q^3\delta_{\alpha\beta}}{4q^{2}}-\int dq\frac{q^3p^{2}\delta_{\alpha\beta}}{4q^{4}}\right. \right.\nonumber\\
&\left.\left. -\int dq\frac{q^3p_\beta p^\gamma \delta_{\alpha\gamma}}{2q^{4}}+\int dq\frac{q^3}{6q^{4}}p^\gamma p^\delta(\delta_{\alpha\beta}\delta_{\gamma\delta}+\delta_{\alpha\delta}\delta_{\gamma\beta}+\delta_{\alpha\gamma}\delta_{\beta\delta})\right] \right\}.\label{ofour}
\eeq
To obtain (\ref{ofour}), we have used 
\beq
\int_{\mathbb{S}} d^{4}q \,q_\alpha q_\beta=2\pi^{2}\int_{\tilde{\Lambda}}^\Lambda dq\, q^3 \frac{\delta_{\alpha\beta} q^{2}}{4}\nonumber
\eeq
and
\beq
\int_{\mathbb{S}}d^{4}q\, q_\alpha q_\beta q_\gamma q_\delta=\frac{1}{24}\int_{\mathbb{S}} 
d^{4}q \,q^{4}(\delta_{\alpha\beta}\delta_{\gamma\delta}+\delta_{\alpha\delta}\delta_{\gamma\beta}+\delta_{\alpha\gamma}\delta_{\beta\delta}),\nonumber
\eeq
which follow from $\mathcal{O}(4)$ symmetry.
Thus the polarization tensor is
\beq
\Pi^{\mu\nu}(p)&=&tr\left\{\frac{1}{2}\gamma^\mu\gamma^\alpha\gamma^\nu\gamma^\beta\frac{1}{8\pi^{2}}\left[\frac{\delta_{\alpha\beta}}{8}(\Lambda^{2}-\tilde{\Lambda}^{2})\right.\right.\nonumber\\
&-&\left.\left.\frac{1}{12}(\delta_{\alpha\beta}p^{2}+2p_\alpha p_\beta)\ln \left(\frac{\Lambda}{\tilde{\Lambda}}\right)\right]\right\}.\label{logs}
\eeq
We must remove non-gauge-invariant terms, namely those quadratic in the cutoffs, with counterterms. The remaining logarithmically-divergent part of (\ref{logs}) is
\beq
\hat\Pi^{\mu\nu}(p)&=&\Pi_{\mu\nu}(p)-\Pi_{\mu\nu}(0)\nonumber\\
&=& tr\left[\frac{1}{2}\gamma^\mu\gamma^\alpha\gamma^\nu\gamma^\beta \frac{1}{8\pi^{2}}\left(\frac{-1}{12}\right)(\delta_{\alpha\beta}p^{2}+2p_\alpha p_\beta)\ln \left(\frac{\Lambda}{\tilde{\Lambda}}\right)\right]\nonumber\\
&=&\frac{e^{2}}{12\pi^{2}}(g^{\mu\nu}p^{2}-p^\mu p^\nu)\ln\left(\frac{\Lambda}{\tilde{\Lambda}}\right).\nonumber
\eeq
This gauge-invariant contribution satisfies $p^{\mu}\hat\Pi_{\mu\nu}(p)=0$. The contribution to the action associated with this term is
\beq
\frac{1}{2}\langle S_1^2\rangle=\int_{\tilde{\mathbb{P}}} \frac{d^4p}{(2\pi)^4}\,\frac{1}{12\pi^2}\ln\left(\frac{\Lambda}{\tilde{\Lambda}}\right)(g^{\mu\nu}p^2-p^\mu p^\nu)A_{\mu}(-p)A_{\nu}(p).
\label{S1squared}
\eeq
Equation (\ref{S1squared}) gives the effective coupling $\tilde{g}$ for the theory with cutoff 
$\tilde{\Lambda}$:
\beq
\frac{1}{4\tilde{g}^{2}}=\frac{1}{4g^2} +\frac{1}{12\pi^2}\ln\left(\frac{\Lambda}{\tilde{\Lambda}}\right).\nonumber
\eeq

The self-energy correction (\ref{selfenergy}) is 
\beq
\Sigma(p)=-2g^{2}\int_{\mathbb{S}}\frac{d^{4}q}{(2\pi)^{4}}\frac{(/\!\!\!q+/\!\!\!p)}{q^{2}(q+p)^{2}}.\nonumber
\eeq
We expand the integrand of $\Sigma(p)$ in powers of $p$, which gives
\beq
\Sigma(p)=-2g^{2}\frac{\gamma^\alpha}{8\pi^{2}}\int dq\left[\frac{-p_\alpha}{2q}+\frac{p_\alpha}{q}\right]
=-\frac{g^{2}/\!\!\!p}{8\pi^{2}}\ln \left(\frac{\Lambda}{\tilde{\Lambda}}\right). 
\eeq

The vertex correction (\ref{vertex}) is
\beq
\Gamma^\mu(p,q)=2g^{2}\gamma^\alpha\gamma^\mu\gamma^\beta\int_{\mathbb{S}}\frac{d^{4}k}{(2\pi)^{4}}\frac{k_\alpha(k_\beta+q_\beta)}{(k-p)^{2}(k+q)^{2}k^{2}}.\nonumber 
\eeq
Expanding the integrand in powers of $p$,
\beq
\Gamma^\mu(p,q)=2g^{2}\gamma^\alpha\gamma^\mu\gamma^\beta\int_{\mathbb{S}}\frac{d^{4}k}{(2\pi)^{4}}\left[\frac{\delta_{\alpha\beta}}{4k^{4}}-\frac{\delta_{\alpha\beta}p^{2}}{4k^6}-\frac{\delta_{\alpha\beta}q^{2}}{4k^6}+\frac{\delta_{\alpha\gamma}}{2k^6}(q_\beta p^\gamma-q_\beta q^\gamma)\right. \nonumber \\
\left.-\frac{\delta_{\alpha\beta}\delta_{\gamma\delta}+\delta_{\alpha\gamma}\delta_{\beta\delta}+\delta_{\alpha\delta}\delta_{\beta\gamma}}{6k^6}(q^\gamma p^\delta-q^\gamma q^\delta-p^\gamma p^\delta)\right].\nonumber 
\eeq
We retain only the divergent part of $\Gamma^{\mu}(p,q)$, namely the first term:
\beq
\Gamma^\mu(p,q)=2g^{2}\frac{\gamma^\alpha\gamma^\mu\gamma^\beta}{8\pi^{2}}\frac{\delta_{\alpha\beta}}{4}\ln\left(\frac{\Lambda}{\tilde{\Lambda}}\right)=\frac{-g^{2}\gamma^\mu}{8\pi^{2}}\ln\left(\frac{\Lambda}{\tilde{\Lambda}}\right). 
\eeq

\section{Ellipsoidal momentum cutoffs }
\setcounter{equation}{0}
\renewcommand{\theequation}{7.4.\arabic{equation}}

Next we consider the more general ellipsoidal case. The integration over $\mathbb{S}$ is done by changing variables from $q_{\mu}$ to two variables $u$ and $w$, with units of momentum squared, and two angles $\theta$ and $\phi$. These variables are defined by
\beq
q_1=\sqrt{u}\cos\theta,\,\,q_2=\sqrt{u}\sin\theta,\,\,q_3=\sqrt{w-u}\cos\phi,\,\,q_0=\sqrt{w-u}\sin\phi.
\nonumber
\eeq
The integration over these variables is
\beq
\int_{\mathbb{S}}d^{4}q &=&\frac{1}{4}\int_{0}^{2\pi}d\theta\int_{0}^{2\pi}d\phi\left[\int_0^{\tilde{\Lambda}^{2}}du\int_{\tilde{b}^{-1}\tilde{\Lambda}^{2}+(1-\tilde{b}^{-1})u}^{b^{-1}\Lambda^{2}+(1-b^{-1})u}dw\right.\nonumber\\
 &&+\left.\int_{\tilde{\Lambda}^{2}}^{\Lambda^{2}}du\int_{u}^{b^{-1}\Lambda^{2}+(1-b^{-1})u}dw\right].\nonumber 
\eeq
We have a $\mathcal{O}(2)\times\mathcal{O}(2)$ symmetry, generated by the translations $\theta\to \theta+d\theta$ and $\phi\to\phi+d\phi$, rather than $\mathcal{O}(4)$ symmetry.

Our three corrections are expressed in terms of the integrals
\beq
&A_{\alpha\beta}=\int_{\mathbb{S}}\frac{d^{4}q}{(2\pi)^{4}}\frac{q_\alpha q_\beta}{q^{4}}, 
\nonumber \\
&B_{\alpha\beta}=\int_{\mathbb{S}}\frac{d^{4}q}{(2\pi)^{4}}\frac{q_\alpha q_\beta}{q^6},
\nonumber \\
&C_{\alpha\beta\gamma\delta}=\int_{\mathbb{S}}\frac{d^{4}q}{(2\pi)^{4}}\frac{q_\alpha q_\beta q_\gamma q_\delta}{q^8} \nonumber
\eeq
and
\beq
D=\int_{\mathbb{S}}\frac{d^{4}q}{(2\pi)^{4}}\frac{1}{q^{4}}.
\eeq
By inspection we write
\beq
&\Pi^{\mu\nu}(p)=tr\left[\frac{1}{2}\gamma^\mu\gamma^\alpha\gamma^\nu\gamma^\beta\left[A_{\alpha\beta}+4C_{\alpha\beta\gamma\delta}p^\gamma p^\delta-p^{2}B_{\alpha\beta}-2B_{\alpha\gamma}p_\beta p^\gamma\right]\right],
\nonumber \\
&\Sigma(p)=-2g^{2}\gamma^\alpha\left[-2B_{\alpha\gamma}p^\gamma+p_\alpha D\right]
\nonumber
\eeq
and
\beq
\Gamma^\mu=2g^{2}\gamma^\alpha\gamma^\mu\gamma^\beta B_{\alpha\beta}.
\nonumber 
\eeq

We use $C$ and $D$ to denote Lorentz indices taking only the values $1$ and $2$. We use
$\Omega$ and $\Xi$ to denote Lorentz indices taking only the values $3$ and $0$. The integration is straightforward, though tedious. We present only the results: 
\beq
A_{CD}&=&\frac{\delta_{CD}}{32\pi^{2}}\Lambda^{2}\left[1+\frac{b}{(b-1)^{2}}(1-b+\ln b)\right]\nonumber\\
&&-\frac{\delta_{CD}}{32\pi^{2}}\tilde{\Lambda}^{2}\left[1+\frac{\tilde{b}}{(\tilde{b}-1)^{2}}(1-\tilde{b}+\ln \tilde{b})\right],
\nonumber \\
A_{\Omega\Xi}&=&\frac{\delta_{\Omega\Xi}}{32\pi^{2}}\left[\Lambda^{2}\left(\frac{1}{b-1}-\frac{\ln b}{(b-1)^{2}}\right)-\tilde{\Lambda}^{2}\left(\frac{1}{b-1}-\frac{\ln \tilde{b}}{(\tilde{b}-1)^{2}}\right)\right],
\nonumber \\
A_{C\Omega}&=&0,
\nonumber \\
B_{CD}&=&\frac{\delta_{CD}}{32\pi^{2}}\ln\left(\frac{\Lambda}{\tilde{\Lambda}}\right)-\frac{\delta_{CD}}{64\pi^{2}}\left[\frac{b^{2}\ln b}{(b-1)^{2}}-\frac{b}{b-1}\right]\nonumber\\
&&+\frac{\delta_{CD}}{64\pi^{2}}\left[\frac{\tilde{b}^{2}\ln \tilde{b}}{(\tilde{b}-1)^{2}}-\frac{\tilde{b}}{\tilde{b}-1}\right],
\nonumber \\
B_{\Omega\Xi}&=&\frac{\delta_{\Omega\Xi}}{32\pi^{2}}\ln\left(\frac{\Lambda}{\tilde{\Lambda}}\right)-\frac{\delta_{\Omega\Xi}}{64\pi^{2}}\left[\frac{b(b-2)\ln b}{(b-1)^{2}}+\frac{b}{b-1}\right]\nonumber\\
&&+\frac{\delta_{\Omega\Xi}}{64\pi^{2}}\left[\frac{\tilde{b}(\tilde{b}-2)\ln \tilde{b}}{(\tilde{b}-1)^{2}}+\frac{\tilde{b}}{\tilde{b}-1}\right],
\nonumber \\
B_{C\Omega}&=&0,
\nonumber \\
C_{CCCC}&=&\frac{1}{64\pi^{2}}\ln \left(\frac{\Lambda}{\tilde{\Lambda}}\right)-\frac{1}{128\pi^{2}}\frac{b^3}{(b-1)^3}\left(\ln b-\frac{2(b-1)}{b}+\frac{(b-1)(b+1)}{2b^{2}}\right)
\nonumber \\
&&+\frac{1}{128\pi^{2}}\frac{\tilde{b}^3}{(\tilde{b}-1)^3}\left(\ln \tilde{b}-\frac{2(\tilde{b}-1)}{\tilde{b}}+\frac{(\tilde{b}-1)(\tilde{b}+1)}{2\tilde{b}}\right),\nonumber \\
C_{1122}&=&\frac{C_{CCCC}}{3},
\nonumber\\
C_{\Omega\Omega\Omega\Omega}&=&\frac{1}{64\pi^{2}}\ln \left(\frac{\Lambda}{\tilde{\Lambda}}\right)-\frac{1}{128\pi^{2}}\left[\frac{b^3}{(b-1)^3}\left(\ln b-\frac{2(b-1)}{b}+\frac{(b-1)(b+1)}{2b^{2}}\right)\right.\nonumber\\
&&\left.+\frac{3b\ln b}{b-1}-\frac{3b^{2}\ln b}{(b-1)^{2}}+\frac{3b}{b-1}\right]
\nonumber\\
&&+\frac{1}{128\pi^{2}}\left[\frac{\tilde{b}^3}{(\tilde{b}-1)^3}\left(\ln \tilde{b}-\frac{2(\tilde{b}-1)}{\tilde{b}}+\frac{(\tilde{b}-1)(\tilde{b}+1)}{2\tilde{b}^{2}}\right)\right.\nonumber\\
&&\left.+\frac{3\tilde{b}\ln \tilde{b}}{\tilde{b}-1}-\frac{3\tilde{b}^{2}\ln \tilde{b}}{(\tilde{b}-1)^{2}}+\frac{3\tilde{b}}{\tilde{b}-1}\right],
\nonumber
\eeq
\beq
C_{0033}&=&\frac{C_{\Omega\Omega\Omega\Omega}}{3},
\nonumber\\
C_{CC\Omega\Omega}&=&\frac{1}{192\pi^{2}}\ln \left(\frac{\Lambda}{\tilde{\Lambda}}\right)-\frac{1}{384\pi^{2}}\left[\frac{-2b^{2}\ln b}{(b-1)^3}+\frac{b^{2}\ln b+2b}{(b-1)^{2}}\right]\nonumber\\
&&+\frac{1}{384\pi^{2}}\left[\frac{-2\tilde{b}^{2}\ln \tilde{b}}{(\tilde{b}-1)^3}+\frac{\tilde{b}^{2}\ln \tilde{b}+2\tilde{b}}{(\tilde{b}-1)^{2}}\right],
\nonumber\\
D&=&\frac{1}{8}\ln \left(\frac{\Lambda}{\tilde{\Lambda}}\right)-\frac{1}{16\pi^{2}}\left[\frac{b\ln b}{b-1}-\frac{\tilde{b}\ln \tilde{b}}{\tilde{b}-1}\right].\label{results}\eeq

Setting $b=\tilde{b}$ in (\ref{results}), we recover the results from the spherical integration done in Section III.
We simplify by setting $b=1$ and $\tilde{b}\approx 1$, using the expansion $\tilde{b}=1+\ln \tilde{b}+\frac{\ln^{2}\tilde{b}}{2}+\frac{\ln^3\tilde{b}}{3!}+\cdots$ and $\ln \tilde{b}=\ln \tilde{b}-\frac{\ln^{2}\tilde{b}}{2}+\frac{\ln^3 \tilde{b}}{3}-\frac{\ln^{4} \tilde{b}}{4}+\cdots $, dropping terms of second order in $\ln \tilde{b}$.

The vertex correction is
\beq
\Gamma^\mu\!\!\!\!\!\!\!\!\!&&\!\!\!\!\!(p,q)=2g^{2}\gamma^\alpha\gamma^\mu\gamma^\beta B_{\alpha\beta}\nonumber\\
&=&2g^{2}\left(\frac{-\gamma^\mu}{16\pi^{2}}\ln\left(\frac{\Lambda}{\tilde{\Lambda}}\right)-\frac{\gamma^\mu}{16\pi^{2}}\ln\tilde{b}+\frac{g^{C\mu}\gamma_C}{32\pi^{2}}\frac{5}{6}\ln\tilde{b}+\frac{g^{\Omega\mu}\gamma_{\Omega}}{32\pi^{2}}\frac{7}{6}\ln\tilde{b}\right).\,\,\,\,\,\,\,\,\,\,\,
\eeq

The self-energy correction is
\beq
\Sigma(p)&=&-2g^{2}\gamma^\alpha[-2B_{\alpha\beta}p^\beta+p_\alpha D]\nonumber\\
&=&-2g^{2}\left[\frac{\gamma^\mu p_{\mu}}{16\pi^{2}}\ln \left(\frac{\Lambda}{\tilde{\Lambda}}\right)+\frac{1}{32\pi^{2}}\ln \tilde{b}\frac{\gamma^C p_C}{6}-\frac{1}{32\pi^{2}}\ln \tilde{b}\frac{\gamma^\Omega p_\Omega}{6}\right].\,\,\,\,\,\,\,\,\,\,\,
\eeq
In the next section, we show how these affect the effective action.

The most general gauge-field action which is quadratic in $A_{\mu}$, is $\mathcal{O}(2)\times\mathcal{O}(2)$ invariant and gauge invariant, to leading order is
\beq 
S_{\rm quadratic}=\int_{\mathbb{P}}\frac{d^{4}p}{(2\pi)^{4}}A(-p)^T\left[a_{1}M_1(p)+a_{2}M_2(p)+a_{3}M_3(p)\right]A(p),
\nonumber
\eeq
where
\beq
 M_1(p)=\left(\begin{array}{cccc}p_2^{2}&-p_1p_2&0&0\\\noalign{\medskip}-p_1p_2&p_1^{2}&0&0\\\noalign{\medskip}0&0&0&0\\0&0&0&0\end{array}\right),
M_2(p)=\left(\begin{array}{cccc}0&0&0&0\\0&0&0&0\\\noalign{\medskip}0&0&p_0^{2}&-p_3p_0\\\noalign{\medskip}0&0&-p_3p_0&p_3^{2}\end{array}\right),
\nonumber
\eeq
\beq 
M_3(p)=\left(\begin{array}{cccc}p_L^{2}&0&-p_1p_3&-p_1p_0\\\noalign{\medskip}0&p_L^{2}&-p_2p_3&-p_2p_0\\\noalign{\medskip}-p_1p_3&-p_2p_3&p_{\perp}^{2}&0\\\noalign{\medskip}-p_1p_0&-p_2p_0&0&p_{\perp}^{2}\end{array}\right),
\label{mmatrices}
\eeq
and $a_{1},\,a_{2}$ and $a_{3}$ are real numbers.
Any part of the polarization tensor that cannot be expressed in terms of these matrices (i.e. $\int_{\tilde{\mathbb{P}}}\frac{d^4p}{(2\pi)^4}A_\mu(-p)\Pi_{\mu\nu}(p)A_{\nu}(p)-S_{\rm quadratic}$) must be removed with counterterms. After some work we find
\beq
\Pi^{\mu\nu}(p)&=&{\rm tr}\left[\frac{1}{2}\gamma^\mu\gamma^\alpha\gamma^\nu\gamma^\beta\left[A_{\alpha\beta}+4C_{\alpha\beta\gamma\delta}p^\gamma p^\delta-p^{2}B_{\alpha\beta}-2B_{\alpha\gamma}p_\beta p^\gamma\right]\right]
\nonumber\\
&=&\frac{1}{12\pi^{2}}\ln\left(\frac{\Lambda}{\tilde{\Lambda}}\right)(p^{2}\mathbf{1}-pp^T)^{\mu\nu}+\frac{5\ln\tilde{b}}{48\pi^{2}}(p^{2}\mathbf{1}-pp^T)^{\mu\nu}
\nonumber\\
&&+\frac{\ln\tilde{b}}{128\pi^{2}}\left[\frac{8}{9}M_3+\frac{40}{9}M_2-\frac{104}{9}M_1\right.\nonumber\\
&&\left.+\frac{8}{3}\left(\begin{array}{cc}\left(\frac{17}{6}p_\perp^{2}+\frac{4}{3}p_L^{2}\right)\mathbf{1}_{2\times2}&0\\0&-\left(\frac{7}{6}p_L^{2}+\frac{14}{3}p_\perp^{2}\right)\mathbf{1}_{2\times2}\end{array}\right)\right]^{\mu\nu}\!\!\!\!\!.\,\,\,\,\,\,\label{polarization}
\eeq
This determines $a_{1},\,a_{2}$ and $a_{3}$, so that
\beq 
S_{\rm diff}&=&\int_{\tilde{\mathbb{P}}}\frac{d^{4}p}{(2\pi^{2})}A(-p)^TM_{\rm diff}A(p)\nonumber\\
&=&\int_{\tilde{\mathbb{P}}}\frac{d^{4}p}{(2\pi)^{2}}A(-p)^T \Pi A(p)-S_{\rm quadratic}
\eeq
is maximally non-gauge invariant. The matrix $M_{\rm diff}$ is the last diagonal matrix in (\ref{polarization}). The quantity $S_{\rm diff}$ is proportional to the local counterterms to include 
in the action. We find
\beq
a_{1}&=&\frac{1}{12\pi^2}\ln\left(\frac{\Lambda}{\tilde{\Lambda}}\right)+\left(\frac{5}{48\pi^2}-\frac{1}{128\pi^{2}}\frac{104}{9}\right)\ln\tilde{b},\,\,\nonumber\\
a_{2}&=&\frac{1}{12\pi^{2}}\ln\left(\frac{\Lambda}{\tilde{\Lambda}}\right)+\left(\frac{5}{48\pi^{2}}+\frac{1}{128\pi^{2}}\frac{40}{9}\right)\ln\tilde{b},\nonumber
\eeq
and
\beq
a_{3}=\frac{1}{12\pi^{2}}\ln\left(\frac{\Lambda}{\tilde{\Lambda}}\right)+\left(\frac{5}{48\pi^{2}}+\frac{1}{128\pi^{2}}\frac{8}{9}\right)\ln\tilde{b}\nonumber
\eeq
In the next section, we show how the action changes under renormalization. We then rescale to restore the isotropy.

\section{The rescaled effective action}
\setcounter{equation}{0}
\renewcommand{\theequation}{7.5.\arabic{equation}}

We define the effective action $S'$, which contains the effects of integrating out the fast fields, by
\beq
Z=\int_{\tilde{\mathbb{P}}}\mathcal{D}\psi\mathcal{D}\bar{\psi}\mathcal{D}A\,e^{-S'}=\int_{\tilde{\mathbb{P}}}\mathcal{D}\psi\mathcal{D}\bar{\psi}\mathcal{D}A\,e^{-\tilde{S}}\;\int_{\mathbb{S}}\;{\mathcal D}\varphi{\mathcal D}\bar{\varphi}{\mathcal D}a\,e^{-S_0}
e^{-R},\nonumber
\eeq
where $S'=\int d^4x \left[\mathcal{L}_{\rm Fermion}+\mathcal{L}_{\rm vertex}+\mathcal{L}_{\rm gauge}\right]=\int d^4x\left[\mathcal{L}_{\rm Dirac}+\mathcal{L}_{\rm gauge}\right]$. To one loop
\beq
&\mathcal{L}_{\rm Fermion}=\bar{\psi}i (/\!\!\!\partial-\Sigma(\partial))\psi,\nonumber\\
&\mathcal{L}_{\rm vertex}=\bar{\psi}(\gamma^\mu-\Gamma^\mu)A_\mu\psi \nonumber
\eeq
and
\beq
\mathcal{L}_{\rm gauge}=\frac{1}{4g^2}F_{\mu\nu}F^{\mu\nu}+A_{\mu} (\sum_{i=1}^{3}a_{i}M_{i}^{\mu\nu}(\partial))A_\nu.\nonumber
\eeq 
Explicitly, $\mathcal{L}_{\rm vertex}$ is
\beq
&\mathcal{L}_{\rm vertex}=\bar{\psi}\left[\gamma^C\left(1+\frac{g^{2}}{8\pi^{2}}\ln\left(\frac{\Lambda}{\tilde{\Lambda}}\right)+\frac{g^{2}}{8\pi^{2}}\ln\tilde{b}-\frac{5g^{2}}{96\pi^{2}}\ln\tilde{b}\right)A_C\right.\nonumber\\
&\left.+\gamma^\Omega\left(1+\frac{g^{2}}{8\pi^{2}}\ln\left(\frac{\Lambda}{\tilde{\Lambda}}\right)+\frac{g^{2}}{8\pi^{2}}\ln\tilde{b}-\frac{7g^{2}}{96\pi^{2}}\ln\tilde{b}\right)A_\Omega\right]\psi\nonumber\\
&=R\bar{\psi}\left[\gamma^CA_C+\lambda^{\frac{g^{2}}{24\pi^{2}\tilde{R}}}\gamma^\Omega A_\Omega\right]\psi,\nonumber
\eeq
where 
\beq
R=\tilde{R}+\left(\frac{g^{2}}{8\pi^{2}}-\frac{5g^{2}}{96\pi^{2}}\right)\ln\tilde{b}\approx \tilde{R}\tilde{b}^{\frac{7g^{2}}{96\pi^{2}\tilde{R}}}=\tilde{R}\lambda^{-\frac{7g^{2}}{48\pi^{2}\tilde{R}}},\nonumber
\eeq
and
\beq
\tilde{R}=1+\frac{g^{2}}{8\pi^{2}}\ln\left(\frac{\Lambda}{\tilde{\Lambda}}\right),\nonumber
\eeq
for small $\ln\tilde{b}$, where we have identified $\tilde{b}=\lambda^{-2}$.

The term $\mathcal{L}_{\rm Fermion}$, which contains the self-energy correction:
\beq
&\mathcal{L}_{\rm Fermion}=\bar{\psi}i\left[\gamma^C\partial_C\left(1+\frac{g^{2}}{8\pi^{2}}\ln\left(\frac{\Lambda}{\tilde{\Lambda}}\right)+\frac{g^{2}}{8\pi^{2}}\ln\tilde{b}-\frac{5g^{2}}{96\pi^{2}}\ln\tilde{b}-\frac{g^{2}}{16\pi^{2}}\ln\tilde{b}\right)\right.\nonumber\\
&\left.+\gamma^\Omega\partial_\Omega\left(1+\frac{g^{2}}{8\pi^{2}}\ln\left(\frac{\Lambda}{\tilde{\Lambda}}\right)+\frac{g^{2}}{8\pi^{2}}\ln\tilde{b}-\frac{5g^{2}}{96\pi^{2}}\ln\tilde{b}-\frac{g^{2}}{12\pi^{2}}\ln\tilde{b}\right)\right]\psi\nonumber\\
&=R'\bar{\psi}i\left[\gamma^C\partial_C+\lambda^{\frac{g^{2}}{24\pi^{2}\tilde{R}}}\gamma^\Omega\partial_\Omega\right]\psi,\nonumber
\eeq
where 
\beq 
R'=R\tilde{b}^{-\frac{g^{2}}{16\pi^{2}\tilde{R}}}=R\lambda^{\frac{g^{2}}{8\pi^{2}\tilde{R}}}.\nonumber
\eeq
For consistency, we write $\mathcal{L}_{\rm vertex}$ in terms of $R'$,
\beq
\mathcal{L}_{\rm vertex}=R'\lambda^{-\frac{g^{2}}{8\pi^{2}\tilde{R}}}\bar{\psi}\left[\gamma^C A_C+\lambda^{\frac{g^{2}}{24\pi^{2}\tilde{R}}}\gamma^\Omega A_\Omega\right]\psi.\nonumber
\eeq

We must rescale the gauge field by
\beq
\lambda^{\frac{-g^{2}}{8\pi^{2}\tilde{R}}}A_{\mu}\to A_{\mu},\label{Arescaling}
\eeq
to express $\mathcal{L}_{\rm Dirac}=\mathcal{L}_{\rm Fermion}+\mathcal{L}_{\rm vertex}$ in terms of a covariant derivative.
This rescaling also affects $\mathcal{L}_{\rm gauge}$. We now have
\beq
\mathcal{L}_{\rm Dirac}=R'\bar{\psi}i\left[\gamma^C D_C+\lambda^{\frac{g^{2}}{24\pi^{2}\tilde{R}}}\gamma^\Omega D_\Omega\right]\psi.\nonumber
\eeq
Rescaling the spinor field by
\beq 
R'\lambda^{-1+\frac{g^{2}}{24\pi^{2}\tilde{R}}}\bar{\psi}\psi\to\bar{\psi}\psi,\nonumber
\eeq
gives us the form
\beq
\mathcal{L}_{\rm Dirac}=\bar{\psi}i\left[\lambda^{1-\frac{g^{2}}{24\pi^{2}\tilde{R}}}\gamma^CD_C+\gamma^\Omega D_\Omega\right]\psi.
\eeq

Including vacuum-polarization corrections, $\mathcal{L}_{\rm gauge}$ becomes
\beq
&\mathcal{L}_{\rm gauge}=\left(\frac{1}{4g^{2}}+\frac{1}{12\pi^{2}}\ln\left(\frac{\Lambda}{\tilde{\Lambda}}\right)+\frac{1}{9\pi^{2}}\ln\tilde{b}\right)(F_{01}^{2}+F_{02}^{2}+F_{13}^{2}+F_{23}^{2})\nonumber\\
&+\left(\frac{1}{4g^{2}}+\frac{1}{12\pi^{2}}\ln\left(\frac{\Lambda}{\tilde{\Lambda}}\right)+\frac{1}{9\pi^{2}}\ln\tilde{b}+\frac{1}{36\pi^{2}}\ln\tilde{b}\right)F_{03}^{2}\nonumber\\
&+\left(\frac{1}{4g^{2}}+\frac{1}{12\pi^{2}}\ln\left(\frac{\Lambda}{\tilde{\Lambda}}\right)+\frac{1}{9\pi^{2}}\ln\tilde{b}-\frac{7}{72\pi^{2}}\ln\tilde{b}\right)F_{12}^{2}.\nonumber
\eeq
We introduce the effective coupling $g_{\rm eff}$,
\beq
\frac{1}{g_{\rm eff}^{2}}=\frac{1}{\tilde{g}^{2}}+\frac{4}{9\pi^{2}}\ln\tilde{b}\approx \frac{1}{\tilde{g}^{2}}\tilde{b}^{\frac{4\tilde{g}^{2}}{9\pi^{2}}}=\frac{1}{\tilde{g}^{2}}\lambda^{\frac{-8}{9\pi^{2}}\tilde{g}^{2}},\nonumber
\eeq
where 
\beq
\frac{1}{\tilde{g}^{2}}=\frac{1}{g^{2}}+\frac{1}{3\pi^{2}}\ln\left(\frac{\Lambda}{\tilde{\Lambda}}\right).
\nonumber\\
\eeq
Then
\beq
\mathcal{L}_{\rm gauge}=\frac{1}{4g_{\rm eff}^{2}}\left(F_{01}^{2}+F_{02}^{2}+F_{13}^{2}+F_{23}^{2}+\lambda^{-\frac{2}{9\pi^{2}}\tilde{g}^{2}}F_{03}^{2}+\lambda^{\frac{7}{9\pi^{2}}\tilde{g}^{2}}F_{12}^{2}\right).\nonumber
\eeq
We finally rescale the gauge field with the factor from (\ref{Arescaling}), 
\beq
 F_{\mu\nu}^{2}\to\lambda^{\frac{g^{2}}{4\pi^{2}\tilde{R}}}F_{\mu\nu}^{2},\nonumber
\eeq
and define a new effective coupling $g'_{\rm eff}$ that absorbs this factor
\beq
\frac{1}{{g'}_{\rm eff}^{2}}=\frac{1}{\tilde{g}^{2}}\lambda^{-\frac{8}{9\pi^{2}}\tilde{g}^{2}+\frac{g^{2}}{4\pi^{2}\tilde{R}}},\label{effectivecoupling}
\eeq
\beq
\mathcal{L}_{\rm gauge}=\frac{1}{4{g'}_{\rm eff}^{2}}\left(F_{01}^{2}+F_{02}^{2}+F_{13}^{2}+F_{23}^{2}+\lambda^{-\frac{2}{9\pi^{2}}\tilde{g}^{2}}F_{03}^{2}+\lambda^{\frac{7}{9\pi^{2}}\tilde{g}^{2}}F_{12}^{2}\right).\eeq

Our final result, after longitudinal rescaling and Wick-rotating back to Minkowski space-time is
\beq
&\mathcal{L}=\mathcal{L}_{\rm Dirac}+\mathcal{L}_{\rm gauge}=\bar{\psi}i\left[\lambda^{1-\frac{g^{2}}{12\pi^{2}\tilde{R}}}\gamma^CD_C+\gamma^\Omega D_\Omega\right]\psi \nonumber \\
&+\frac{1}{4{g'}_{\rm eff}^{2}}\left(F_{01}^{2}+F_{02}^{2}-F_{13}^{2}-F_{23}^{2}+\lambda^{-2-\frac{2}{9\pi^{2}}\tilde{g}^{2}}F_{03}^{2}-\lambda^{2+\frac{7}{9\pi^{2}}\tilde{g}^{2}}F_{12}^{2}\right).\label{finalaction}
\eeq

\section{Wilsonian renormalization of QCD}
\setcounter{equation}{0}
\renewcommand{\theequation}{7.6.\arabic{equation}}

We now examine the quantum longitudinal rescaling of a non-Abelian gauge theory. We start with the  Euclidean action
\beq
S=S_{\rm gauge}+S_{\rm Dirac},\nonumber
\eeq
where
\beq
S_{\rm gauge}=\frac{1}{4g_{0}^{2}} \int d^{4}x F^{\mu\nu}F_{\mu\nu}\nonumber,
\eeq
\beq
S_{\rm Dirac}=\int d^4x\,\bar{\psi}\gamma^\mu D_\mu \psi,\nonumber
\eeq
where now $F_{\mu\nu}=\partial_\mu A_\nu-\partial_\nu A_\mu+i[A_\mu,A_\nu]$.

We write the gauge field as $A_\mu=A_\mu^a t^a$ where $t^a$ are the $SU(N)$ generators, with $a=1,\dots,N^2-1$. These generators satisfy $[t^a,t^b]=if^{abc}t^c$, and are normalized by ${\rm Tr}\, t^at^b=\delta^{ab}$. The field strength is then $F_{\mu\nu}=F_{\mu\nu}^a t^a$, where $F_{\mu\nu}^a=\partial_\mu A_\nu^a-\partial_\nu A_\mu^a+if^{abc}A_\mu^b A_\nu^c$.

We split the fields into ``fast'' pieces that lie in the region of momentum $\mathbb{S}$ and ``slow'' pieces with momentum in the region $\tilde{\mathbb{P}}$:
\beq
\tilde{\psi}(x)=\int_{\tilde{\mathbb{P}}}\frac{d^{4}p}{(2\pi)^{4}}\psi(p)e^{-ip\cdot x}\!\!\!\!\!&,&\!\!\!\!\! \,\,\varphi(x)
=\int_{\mathbb{S}}\frac{d^{4}p}{(2\pi)^{4}}\psi(p)e^{-ip\cdot x},\;\nonumber\\
\tilde{\bar{\psi}}(x)=\int_{\tilde{\mathbb{P}}}\frac{d^{4}p}{(2\pi)^{4}}\bar{\psi}(p)e^{ip\cdot x}\!\!\!\!\!&,&\!\!\!\!\!\,\,\bar{\varphi}(x)=\int_{\mathbb{S}}\frac{d^{4}p}{(2\pi)^{4}}\bar\psi(p)e^{ip\cdot x}, \nonumber \\
\tilde{A}_{\mu}(x)=
\int_{\tilde{\mathbb{P}}}\frac{d^{4}p}{(2\pi)^{4}}A_{\mu}(p)e^{-ip\cdot x}\!\!\!\!\!&,&\!\!\!\!\!\,\,\,a_{\mu}(x)=\int_{\mathbb{S}}\frac{d^{4}p}{2(\pi)^{4}}A_{\mu}(p)e^{-ip\cdot x}.\nonumber
\eeq

The covariant derivative and the field strength become
\beq
D_\mu=\partial_\mu-i\tilde{A}_\mu-ia_\mu=\tilde{D}_\mu-ia_\mu,\nonumber
\eeq
and
\beq
F_{\mu\nu}=\tilde{F}_{\mu\nu}+[\tilde{D}_\mu,a_\nu]-[\tilde{D}_\nu,a_\mu]-i[a_\mu,a_\nu],\nonumber
\eeq
respectively, where $\tilde{F}_{\mu\nu}=i[\tilde{D}_\mu,\tilde{D}_\nu].$ The pure gauge action is
\beq
S_{\rm gauge}&=&\int d^4x\frac{1}{4g_0^2}\left(\tilde{F}_{\mu\nu}\tilde{F}^{\mu\nu}-4[\tilde{D}_\mu,\tilde{F}^{\mu\nu}]a_\nu\right.\nonumber\\
&&\left.+\left([\tilde{D}_\mu,a_\nu]-[\tilde{D}_\nu, a_\mu]\right)\left([\tilde{D}^\mu, a^\nu]-[\tilde{D}^\nu, a^\mu]\right)-2i\tilde{F}^{\mu\nu}[a_\mu,a_\nu]\right),\nonumber
\eeq
to quadratic order in $a_\mu$.

To do perturbation theory, we add a gauge-fixing term $\frac{1}{2g_0^2}\int d^4x {\rm Tr}\, [\tilde{D}_\mu, a_\mu]^2$ to the action. This reduces the gauge symmetry of the fast fields, and means that we must also introduce Faddeev-Popov ghost fields. The Yang-Mills action becomes
\beq
S_{\rm gauge}=\frac{1}{4g_0^2}\int d^4x \tilde{F}_{\mu\nu}\tilde{F}^{\mu\nu}+\frac{1}{2g_0^2}\int d^4x\left([\tilde{D}_\mu,a_\nu][\tilde{D}^\mu, a^\nu]-2i\tilde{F}^{\mu\nu}[a_\mu, a_\nu]\right)\nonumber.
\eeq
The expansion of the quark-field action into slow and fast components is
\beq
S_{\rm Dirac}=\int d^4x\left(\tilde{\bar{\psi}}\,/\!\!\!\!\partial\,\tilde{\psi}+\bar{\varphi}\,/\!\!\!\!\partial\,\varphi+\bar{\varphi}\,/\!\!\!\!\tilde{A}\,\varphi+\tilde{\bar{\psi}}\,/\!\!\!\!a\,\varphi+\bar{\varphi}\,/\!\!\!\!a\,\tilde{\psi}\right).\nonumber
\eeq
The action is that of free slow and fast fields plus interaction terms
\beq
S=\tilde{S}+S_0+S_{I}+S_{II}+S_{1}+S_{2}+S_{3}+S_{\rm ghost} \nonumber,
\eeq
where
\beq
\tilde{S}&=&\frac{1}{4g_0^4}\int d^4x\tilde{F}_{\mu\nu}\tilde{F}^{\mu\nu}+\int d^4x \tilde{\bar{\psi}}\,/\!\!\!\!\partial\tilde{\psi},\nonumber\\
S_0&=&\frac{1}{2g_0^2}\int_{\mathbb{S}}\frac{d^4q}{(2\pi)^2}q^2a_\mu^b(-q)a^{\mu\, b}(q)-i\int_\mathbb{S}\frac{d^4q}{(2\pi)^4}\bar{\varphi}(-q)/\!\!\!q\varphi(q)\nonumber,\\
S_I&=&\frac{i}{2g_0^2}\int_\mathbb{S}\frac{d^4q}{(2\pi)^4}\int_{\tilde{\mathbb{P}}}\frac{d^4p}{(2\pi)^4}q^\mu f_{bcd}a_\nu^b(q)\tilde{A}_\mu^c(p)a^{\nu\,d}(-q-p)\nonumber\\
&+&\frac{1}{2g_0^2}\int_\mathbb{S}\frac{d^4q}{(2\pi)^4}\int_{\tilde{\mathbb{P}}}\frac{d^4p}{(2\pi)^4}\int_{\tilde{\mathbb{P}}}\frac{d^4l}{(2\pi)^4}f_{bcd}f_{bfg}a_\nu^d(q)\tilde{A}_\mu^c(p)\tilde{A}^{\mu\,f}(l)a^{\nu\,g}(-q-p-l),\nonumber\\
S_{II}&=&\frac{1}{2g_0^2}\int_\mathbb{S}\frac{d^4q}{(2\pi)^4}\int_{\tilde{\mathbb{P}}}\frac{d^4p}{(2\pi)^4}f_{bcd}a_\mu^b(q)\tilde{F}^{\mu\nu\,c}(p)a_\nu^d(-p-q),\nonumber\\
S_1&=&\int_\mathbb{S}\frac{d^4q}{(2\pi)^4}\int_{\tilde{\mathbb{P}}}\frac{d^4p}{(2\pi)^4} \bar{\varphi}(p)/\!\!\!\!A(q)\varphi(-q-p),\nonumber\\
S_2&=&\int_\mathbb{S}\frac{d^4q}{(2\pi)^4}\int_{\tilde{\mathbb{P}}}\frac{d^4p}{(2\pi)^4}\tilde{\bar{\psi}}(p)/\!\!\!\!\,a(q)\varphi(-q-p),
\nonumber\\
S_3&=&S_2^*=\int_\mathbb{S}\frac{d^4q}{(2\pi)^4}\int_{\tilde{\mathbb{P}}}\frac{d^4p}{(2\pi)^4}\bar{\varphi}(-q-p)\,/\!\!\!\!a(q)\varphi(p)\nonumber 
\eeq
and the ghost action, needed to have the correct measure on the fast field, is 
\beq
S_{\rm ghost}&=&\frac{i}{g_0^2}\int_\mathbb{S}\frac{d^4q}{(2\pi)^4}\int_{\tilde{\mathbb{P}}}\frac{d^4p}{(2\pi)^4}q^\mu f_{bcd}G^b(q)\tilde{A}_\mu^c(p)H^d(-q-p)\nonumber\\
&+&\frac{1}{2g_0^2}\int_\mathbb{S}\frac{d^4q}{(2\pi)^4}\int_{\tilde{\mathbb{P}}}\frac{d^4p}{(2\pi)^4}\int_{\tilde{\mathbb{P}}}\frac{d^4l}{(2\pi)^4}f_{bcd}f_{bfg}G^d(q)\tilde{A}_\mu^c(p)\tilde{A}_\mu^f(l)H^g(-q-p)\nonumber.
\eeq
The interaction is therefore $S_{\rm int}=S_{I}+S_{II}+S_{1}+S_{2}+S_{3}+S_{\rm ghost}$.

We start with the functional integral 
\beq
Z=\int_{\tilde{\mathbb{P}}}\mathcal{D}\tilde{\psi}\mathcal{D}\tilde{\bar{\psi}}\mathcal{D}\tilde{A}\,e^{-\tilde{S}}\int_\mathbb{S}\mathcal{D}\varphi\mathcal{D}\bar{\varphi}\mathcal{D}a \,e^{-S_0-S_{\rm int}},\label{functionalintegral}
\eeq
and integrate out the fast fields $\varphi$, $\bar \varphi$ and $a$, to obtain an effective action, $S'$, defined by
\beq
e^{-S'}=
e^{-\tilde{S}}\int_{\mathbb{S}}\mathcal{D}\varphi\mathcal{D}\bar{\varphi}\mathcal{D}a\,e^{-S_0-S_{\rm int}}.\label{effectiveAc}
\eeq
Then the functional integral in terms of slow degrees of freedom only is 
\beq
Z=\int_{\tilde{\mathbb{P}}}\mathcal{D}\tilde{\psi}\mathcal{D}\tilde{\bar{\psi}}\mathcal{D}\tilde{A}e^{-S'}.\nonumber
\eeq
The Green's functions of the slow fields $\psi$, $\bar \psi$ and $\tilde A$, with action $S^{\prime}$ are unchanged from the same Green's functions in the original theory.

The effective action is given explicitly by $S^{\prime}={\tilde S}-\ln\langle e^{-S_{\rm int}}\rangle$, which we evaluate using the connected-graph expansion, as in Section (7.2).  The fast-field propagators are
\beq
\langle a^b_\mu(q)a^c_\nu(p)\rangle=g_0^2\delta^{bc}\delta_{\mu\nu}\delta^4(q+p)q^{-2}(2\pi)^4,\,\,\,\,\,\langle\varphi(p)\bar{\varphi}(q)\rangle=\frac{i\,/\!\!\!\!q}{q^2}\delta^4(p+q)(2\pi)^4.\nonumber
\eeq

We first consider all the contributions quadratic in the slow gauge field, {\em i.e.} vacuum polarization. One contribution comes from the interactions $S_I$ and $S_{\rm ghost}$:
\beq
&\langle S_I\rangle-\frac{1}{2}\left(\langle S_I^2\rangle-\langle S_I\rangle^2\right)+\langle S_{\rm ghost}\rangle-\frac{1}{2}\left(\langle S_{\rm ghost}^2\rangle-\langle S_{\rm ghost}\rangle^2\right)\nonumber\\
&=\frac{C_G}{4}\int_{\tilde{\mathbb{P}}}\frac{d^4p}{(2\pi)^4}\tilde{A}_\mu^b(-p)\tilde{A}_\nu^b(p)P_{\mu\nu}(p),\nonumber
\eeq
where
\beq
P_{\mu\nu}(p)=\int_\mathbb{S}\frac{d^4p}{(2\pi)^4}\left[-\frac{q_\mu(p_\nu+2q_\nu)}{4q^2(q+p)^2}+\frac{\delta_{\mu\nu}}{4q^2}\right],\nonumber
\eeq
and $C_G$ is the quadratic Casimir operator in the adjoint representation, defined by 
$C_G \delta^{bh}=f^{bcd}f^{hcd}$. The tensor $P_{\mu\nu}(p)$ is not symmetric under exchange of indices. We define the integral $I_{\alpha}(p)$ by
\beq
I_{\alpha}(p)=\int_\mathbb{S}\frac{d^4q}{(2\pi)^4}\frac{p_\alpha+2q_\alpha}{q^2(q+p)^2}\nonumber
\eeq
and notice that $I_{\alpha}(p)+I_{\alpha}(-p)=0$ (we can see this by changing the sign of $q$ in the integrand). We can then use this to  replace the tensor $P_{\mu\nu}(p)$ by the manifestly symmetric tensor $\Pi_{\mu\nu}^1(p)$:
\beq
&\langle S_I\rangle-\frac{1}{2}\left(\langle S_I^2\rangle-\langle S_I\rangle^2\right)+\langle S_{\rm ghost}\rangle-\frac{1}{2}\left(\langle S_{\rm ghost}^2\rangle-\langle S_{\rm ghost}\rangle^2\right)\nonumber\\
&=\int_{\tilde{\mathbb{P}}}\frac{d^4p}{(2\pi)^4}\tilde{A}_\mu^b(-p)\tilde{A}_\nu^b(p)\Pi^1_{\mu\nu}(p),\label{poltensorone}
\eeq
where
\beq
\Pi_{\mu\nu}^1(p)=C_G\int_\mathbb{S}\frac{d^4q}{(2\pi)^4}\left[-\frac{\left(p_\mu+2q_\nu\right)\left(p_\nu+2q_\mu\right)}{8q^2(p+q)^2}+\frac{\delta_{\mu\nu}}{4q^2}\right].\nonumber
\eeq
A second contribution to vacuum polarization comes from 
\beq
&-\frac{1}{2}\left(\langle S_{II}^2\rangle-\langle S_{II}\rangle^2\right)=-\frac{C_G}{2}\int_{\tilde{\mathbb{P}}}\frac{d^4p}{(2\pi)^2}\tilde{F}_{\mu\nu}^b(-p)\tilde{F}_{\mu\nu}^b(p)\int_\mathbb{S}\frac{d^4q}{(2\pi)^2}\frac{1}{q^2(p+q)^2}.\label{spinorbit}
\eeq
The third and final contribution to vacuum polarization comes from integration over the fast quark field:
\beq
-\frac{1}{2}\left(\langle S_1^2\rangle-\langle S_1^2\rangle^2\right)=\int_{\tilde{\mathbb{P}}}\frac{d^4p}{(2\pi)^2}\tilde{A}_\mu^b(-p)\tilde{A}_\nu^b(p)\Pi_{\mu\nu}^3(p),\label{vacuumfermion}
\eeq
where
\beq
\Pi_{\mu\nu}^3(p)=\frac{N_{f}}{2}\int_\mathbb{S}\frac{d^4q}{(2\pi)^2}{\rm Tr}\,\left[\frac{/\!\!\!\!q}{q^2}\gamma^\mu\frac{/\!\!\!\!q+/\!\!\!\!p}{(q+p)^2}\gamma^\nu\right],\nonumber
\eeq
and $N_{f}$ is the number of flavors. Combining (\ref{poltensorone}), (\ref{spinorbit}) and (\ref{vacuumfermion}), we find the vacuum-polarizaton contribution:
\beq
\int_{\tilde{\mathbb{P}}}\frac{d^4p}{(2\pi)^2}\tilde{A}^b_\mu(-p)\tilde{A}^b_\nu(p)\Pi_{\mu\nu}(p),\label{polarization}
\eeq
where
\beq
\Pi_{\mu\nu}(p)=\Pi_{\mu\nu}^1(p)+\Pi_{\mu\nu}^2(p)+\Pi_{\mu\nu}^3(p),\nonumber
\eeq
and
\beq
\Pi_{\mu\nu}^2(p)=(p^2\delta_{\mu\nu}-p_\mu p_\nu)C_G\int_\mathbb{S}\frac{-1}{2q^2(p+q)^2}.\nonumber
\eeq

The quark self-energy contribution, which comes from the interactions $S_2$ and $S_3$, is
\beq
-\frac{1}{2}\left(\langle S_2S_3\rangle+\langle S_3S_2\rangle\right)=\int_{\tilde{\mathbb{P}}}\frac{d^4p}{(2\pi)^2}\Sigma(p)\tilde{\bar{\psi}}(p)\tilde{\psi(p)},\label{selfenergy}
\eeq
where
\beq
\Sigma(p)=2g_0^2\int_\mathbb{S}\frac{d^4q}{(2\pi)^4}\left[\frac{i(/\!\!\!\!p+/\!\!\!\!q)}{q^2(p+q)^2}\right].\nonumber
\eeq

The quark-gluon vertex receives a correction from
\beq
&\frac{1}{3!}\left(\langle S_1S_2S_3\rangle-\langle S_2S_3\rangle\langle S_1\rangle-\langle S_3S_2\rangle \langle S_1\rangle\right)\nonumber\\
&+\frac{1}{3!}\left(\langle S_IS_2S_3\rangle-\langle S_2S_3\rangle\langle S_I\rangle-\langle S_3S_2\rangle \langle S_I\rangle\right)\nonumber\\
&\,\,\,\,\,\,\,\,\,\,\,\,\,\,\,\,\,\,\,\,\,\,\,\,\,=\int_{\tilde{\mathbb{P}}}\frac{d^4q}{(2\pi)^4}\int_{\tilde{\mathbb{P}}}\frac{d^4p}{(2\pi)^4}\tilde{\bar{\psi}}(p)\Gamma^{\mu\,a}(p,q)\tilde{A}_\mu^a(q)\tilde{\psi}(-q-p),\label{vertex}
\eeq
where
\beq
\Gamma^{\mu\,a}(p,q)=-2g_0^2 t^a\int_\mathbb{S}\frac{d^4k}{(2\pi)^4}\frac{/\!\!\!\!k \gamma^\mu(/\!\!\!\!k+/\!\!\!\!q)}{(k-p)^2(k+q)^2k^2}.\nonumber
\eeq

\section{QCD with spherical cutoffs}
\setcounter{equation}{0}
\renewcommand{\theequation}{7.7.\arabic{equation}}

As we did for QED, we can recover spherical cutoffs by setting $b=\tilde{b}=1$.
We expand $\Pi_{\mu\nu}(p), \Sigma(p)$ and $\Gamma^{\mu\,a}(p,q)$ in powers of the slow momenta, treating momenta in $\tilde{\mathbb{P}}$ as much smaller than momenta in $\mathbb{S}$. This gives
\beq
&\Pi_{\mu\nu}(p)=\Pi^1_{\mu\nu}(p)+\Pi^2_{\mu\nu}(p)+\Pi^3_{\mu\nu}(p),\nonumber\\
&\Pi^1_{\mu\nu}(p)=C_G\left[\frac{\delta_{\mu\nu}}{4}E-\frac{1}{2}A_{\mu\nu}+\frac{p_\mu p_\alpha}{2}B_{\nu\alpha}+\frac{p_\nu p_\alpha}{2}B_{\nu\alpha}+\frac{p_\nu p_\alpha}{2}B_{\mu\alpha}-\frac{p_\mu p_\nu}{8}D\right.\nonumber\\
&\left.+\frac{p^2}{2}B_{\mu\nu}-2p_\alpha p_\beta C_{\alpha\beta\mu\nu}\right],\nonumber\\
&\Pi^2_{\mu\nu}(p)=-\frac{1}{2}(p^2g_{\mu\nu}-p_\mu p_\nu) C_G D,\nonumber\\
&\Pi^3_{\mu\nu}(p)=\frac{N_{f}}{2}{\rm Tr}\,\gamma^\mu\gamma^\alpha\gamma^\nu\gamma^\beta \left[A_{\alpha\beta}+4C_{\alpha\beta\gamma\delta}p^\gamma p^\delta-p^2B_{\alpha\beta}-2B_{\alpha\gamma}p_\beta p^\gamma\right],\nonumber\\
&\Sigma(p)=2g_0^2N_{f}\gamma^\alpha\left[-2B_{\alpha\beta}p^\beta+p_\alpha D\right],\nonumber\\
&\Gamma^{\mu\,a}=-2g_0^2N_{f}t^a\gamma^\alpha\gamma^\mu\gamma^\beta B_{\alpha\beta},\label{spherical}
\eeq
where
\beq
&A_{\alpha\beta}=\int_\mathbb{S}\frac{d^4q}{(2\pi)^4}\frac{q_\alpha q_\beta}{q^4},\,\,\,\,\,B_{\alpha\beta}=\int_\mathbb{S}\frac{d^4q}{(2\pi)^4}\frac{q_\alpha q_\beta}{q^6},\nonumber\\
&C_{\alpha\beta\gamma\delta}=\int_\mathbb{S}\frac{d^4q}{(2\pi)^4}\frac{q_\alpha q_\beta q_\gamma q_\delta}{q^8},\,\,\,\,\,D=\int_\mathbb{S}\frac{d^4q}{(2\pi)^4}\frac{1}{q^4},\,\,\,\,\,E=\int_\mathbb{S}\frac{d^4q}{(2\pi)^4}\frac{1}{q^2}.\label{defintegrals}
\eeq
The integrals (\ref{defintegrals}) are invariant under $\mathcal{O}(4)$ rotation symmetry. This allows us to write
\beq
\int_\mathbb{S}d^4q\,q_\alpha q_\beta=\frac{\pi^2}{2}\int_{\tilde{\Lambda}}^\Lambda dq\,
\delta_{\alpha\beta}\,q^{2}\;, \label{ofoursymmetryone}
\eeq
and
\beq
\int_{\mathbb{S}}d^4q\,q_\alpha q_\beta q_\gamma q_\delta=\frac{1}{24}\int_\mathbb{S} d^4q\, q^4 (\delta_{\alpha\beta} \delta_{\gamma\delta}+\gamma_{\alpha\delta}\delta_{\gamma\beta}+\delta_{\alpha\gamma}\delta_{\beta\delta}).\label{ofoursymmetrytwo}
\eeq

Using (\ref{ofoursymmetryone}) and (\ref{ofoursymmetrytwo}) we solve (\ref{spherical}):
\beq
&\Pi_{\mu\nu}(p)=-\frac{11C_G}{192\pi^2}(p^2\delta_{\mu\nu}-p_\mu p_\nu)\ln\frac{\Lambda}{\tilde{\Lambda}} +\frac{N_{f}}{12\pi^2}(p^2\delta_{\mu\nu}-p_\mu p_\nu)\ln\frac{\Lambda}{\tilde{\Lambda}}\nonumber\\
&+\frac{C_G}{128\pi^2}(\Lambda^2-\tilde{\Lambda}^2)\delta_{\mu\nu}-\frac{N_{f}}{16\pi^2}(\Lambda^2-\tilde{\Lambda})\delta_{\mu\nu},\nonumber\\
&\Sigma(p)=g_0^2 N_{f} \frac{\gamma^\mu p_\mu}{8\pi^2}\ln\frac{\Lambda}{\tilde{\Lambda}},\nonumber\\
&\Gamma^{a\mu}=g_0^2 N_{f} t^a\frac{\gamma^\mu}{8\pi^2}\ln\frac{\Lambda}{\tilde{\Lambda}}.\label{resultsspherical}
\eeq

The terms in the polarization tensor that are quadratic in the cutoffs produce corrections to the action that break gauge invariance. We can fix this problem by introducing mass counterterms in the action at each scale to cancel these. We keep the gauge invariant part of the polarization tensor, which we call $\hat{\Pi}_{\mu\nu}(p)$, and is defined by
\beq
\hat{\Pi}_{\mu\nu}(p)=\Pi_{\mu\nu}(p)-\Pi_{\mu\nu}(0).\nonumber
\eeq

The resulting action for the slow fields has the coupling $\tilde{g}$, given by
\beq
\frac{1}{4\tilde{g}^2}=\frac{1}{4g_0^2}-\frac{11C_G}{192\pi^2}\ln\frac{\Lambda}{\tilde{\Lambda}}+\frac{N_{f}}{12\pi^2}\ln\frac{\Lambda}{\tilde{\Lambda}}, \label{effectivecoupling}
\eeq
which is the standard result.

\section{QCD with ellipsoidal cutoffs}
\setcounter{equation}{0}
\renewcommand{\theequation}{7.8.\arabic{equation}}

In this section, we generalize to the case where the region 
$\mathbb{S}$ is an ellipsoidal shell. We have already found the results for the integrals $A_{\alpha\beta}, \,B_{\alpha\beta},\,C_{\alpha\beta\gamma\delta},\,$and $D$ in Eq. (\ref{results}). The result for the remaining integral is 
\beq
E=\frac{1}{16\pi^2}\left(\frac{\Lambda^2\ln b}{b-1}-\frac{\tilde{\Lambda}^2\ln\tilde{b}}{\tilde{b}-1}\right).\nonumber
\eeq

We take $b=1$ and $\tilde{b}\approx 1$. We expand $\tilde{b}=1+\ln\tilde{b}+\frac{\ln^2\tilde{b}}{2!}+\cdots$ and $\ln\tilde{b}=\ln\tilde{b}-\frac{\ln^2\tilde{b}}{2}+\cdots$, keeping only the first-order terms in $\ln\tilde{b}$.

The self-energy correction is
\beq
\Sigma(p)=2g_0^2\left[\frac{\gamma^\mu p_\mu}{16\pi^2}\ln\frac{\Lambda}{\tilde{\Lambda}}+\frac{1}{32\pi^2}\ln\tilde{b}\frac{\gamma^Cp_C}{6}-\frac{1}{32\pi^2}\ln\tilde{b}\frac{\gamma^\Omega p_\Omega}{6}\right].\label{selfenergyellipsoid}
\eeq
The vertex correction is
\beq
\Gamma^{\mu\,a}=-2g_0^2 t^a\left[\frac{-\gamma^\mu}{16\pi^2}\ln\frac{\Lambda}{\tilde{\Lambda}}-
\frac{\gamma^\mu}{16\pi^2}\ln\tilde{b}+\frac{g^{C\mu}\gamma_C}{32\pi^2}\frac{5}{6}\ln\tilde{b}+\frac{g^{\Omega\mu}\gamma_\Omega}{32\pi^2}\frac{7}{6}\ln\tilde{b}\right]\!\!.\label{vertexellipsoid}
\eeq

The general form of the quadratic part of the renormalized gauge field action, which is invariant under $\mathcal{O}(2)\times\mathcal{O}(2)$ and gauge symmetry is
\beq
S_{\rm quadratic}=\int_{\tilde{\mathbb{P}}}\frac{d^4p}{(2\pi)^4}A(-p)^T[a_1M_1(p)+a_2M_2(p)+a_3M_3(p)]A(p),\nonumber
\eeq
where the matrices $M_{1,2,3}(p)$ are defined in Eq. (\ref{mmatrices}), and the coefficients $a_1,\,a_2$ and $a_3$ are real numbers. We extract these coefficients from the polarization tensor. Any part that cannot be expressed in terms of (\ref{mmatrices}) (i.e.  $S_{\rm diff}=\int_{\tilde{\mathbb{P}}}\frac{d^4p}{(2\pi)^4}A_\mu(-p)\Pi_{\mu\nu}(p)A_\nu(p)-S_{\rm quadratic})$ must be removed with counterterms in the action. The coefficients $a_i$ are selected such that $S_{\rm diff}$ is maximally non-gauge invariant. From the polarization tensor in Eq. (\ref{spherical}), we find
\beq
a_1&=&-\frac{11C_G}{192\pi^2}\ln\frac{\Lambda}{\tilde{\Lambda}}+\frac{N_{f}}{12\pi^2}\ln\frac{\Lambda}{\tilde{\Lambda}}-\frac{1}{64\pi^2}\frac{31}{9}C_G\ln\tilde{b}\nonumber\\
&&+\left(\frac{5}{48\pi^2}-\frac{1}{128\pi^2}\frac{104}{9}\right)N_{f}\ln\tilde{b},\nonumber\\
a_2&=&-\frac{11C_G}{192\pi^2}\ln\frac{\Lambda}{\tilde{\Lambda}}+\frac{N_{f}}{12\pi^2}\ln\frac{\Lambda}{\tilde{\Lambda}}-\frac{1}{64\pi^2}\frac{67}{9}C_G\ln\tilde{b}+\nonumber\\
&&\left(\frac{5}{48\pi^2}+\frac{1}{128\pi^2}\frac{40}{9}\right)N_{f}\ln\tilde{b},\nonumber\\
a_3&=&-\frac{11C_G}{192\pi^2}\ln\frac{\Lambda}{\tilde{\Lambda}}+\frac{N_{f}}{12\pi^2}\ln\frac{\Lambda}{\tilde{\Lambda}}-\frac{1}{64\pi^2}\frac{59}{9}C_G\ln\tilde{b}+\nonumber\\
&&\left(\frac{5}{48\pi^2}+\frac{1}{128\pi^2}\frac{8}{9}\right)\ln\tilde{b},\label{anumbers}
\eeq
and
\beq
M_{\rm diff} &&\nonumber \\
&=&\frac{C_G\ln\tilde{b}}{64\pi^2}\left(\begin{array}{cc}-\frac{1}{12}p_1^2-\frac{1}{2}p_2^2+\frac{7}{12}p+L^2&0\\0&-\frac{1}{2}p_1^2-\frac{1}{12}p_2^2+\frac{7}{12}p_L^2\\0&0\\0&0\end{array}
\right|\nonumber\\
&&\,\,\,\,\,\,\,\,\,\,\,\,\,\,\,\,\,\,\,\,\,\,\,\,\,\,\,\,\,\,\,\left|
\begin{array}{cc}0&0\\0&0\\ \frac{7}{12}p_\perp^2+\frac{17}{12}p_3^2+\frac{5}{6}p_0^2&0\\0& \frac{7}{12}p_\perp^2+\frac{5}{6}p_3^2+\frac{17}{12}p_0^2\end{array}\right)\nonumber\\
&&\nonumber \\
&&\nonumber \\
&+&\frac{N_{f}\ln\tilde{b}}{128\pi^2}\frac{8}{3}\left(\begin{array}{cccc}\frac{17}{6}p_\perp^2+\frac{4}{3}p_L^2&0&0&0\\0&\frac{17}{6}p_\perp^2+\frac{4}{3}p_L^2&0&0\\0&0&-\frac{7}{6}p_L^2-\frac{14}{3}p_\perp^2&0\\0&0&0&-\frac{7}{6}p_L^2-\frac{14}{3}p_\perp^2\end{array}\right),\nonumber
\eeq
where $M_{\rm diff}$ is defined as
\beq
S_{\rm diff}=\int_{\tilde{\mathbb{P}}}\frac{d^4p}{(2\pi)^4}A(-p)^TM_{\rm diff}A(p).\nonumber
\eeq

\section{The QCD renormalized action}
\setcounter{equation}{0}
\renewcommand{\theequation}{7.9.\arabic{equation}}

We next put together the results of the previous section to obtain the action 
$S'$, defined in (\ref{effectiveAc}). 
This action is 
\beq
S'=\int d^4x\mathcal{L}_{\rm quarks}+\mathcal{L}_{\rm vertex}+\mathcal{L}_{\rm gauge}=\int d^4x\mathcal{L}_{\rm Dirac}+\mathcal{L}_{\rm gauge},\nonumber
\eeq
where to one loop,
\beq
\mathcal{L}_{\rm quarks}&=&\tilde{\bar{\psi}}i(/\!\!\!\partial+\Sigma(\partial))\tilde{\psi},\nonumber\\
\mathcal{L}_{\rm vertex}&=&\tilde{\bar{\psi}}(\gamma^\mu t^a+\Gamma^{\mu\,a})\tilde{A}_\mu^a\tilde{\psi},\nonumber
\eeq
and
\beq
\mathcal{L}_{\rm gauge}=\frac{1}{4g_0^2}\tilde{F}_{\mu\nu}\tilde{F}^{\mu\nu}+\tilde{A}_\mu\left[\sum_{i=1}^3 a_iM_i^{\mu\nu}(\partial)\right]A_\nu.\label{lagrangians}
\eeq
Substituting (\ref{vertexellipsoid}) into (\ref{lagrangians}) yields
\beq
\mathcal{L}_{\rm vertex}&=&R\tilde{\bar{\psi}}\left[\gamma^C\left(1+\frac{N_{f}g_0^2}{8\pi^2}\ln \frac{\Lambda}{\tilde{\Lambda}}+\frac{N_{f}g_0^2}{8\pi^2}\ln\tilde{b}-\frac{N_{f}5g_0^2}{96\pi^2}
\ln\tilde{b}\right)\right.A_C\nonumber\\
&&\left.+\gamma^\Omega\left(1+\frac{N_{f}g_0^2}{8\pi^2}\ln\frac{\Lambda}{\tilde{\Lambda}}+\frac{N_{f}g_0^2}{8\pi^2}\ln\tilde{b}-\frac{N_{f}7g_0^2}{96\pi^2}\ln\tilde{b}\right)A_\Omega\right]\tilde{\psi}\nonumber\\
&=&R\tilde{\bar{\psi}}\left(\gamma^CA_C+\lambda^{\frac{N_{f}g_0^2}{24\pi^2\tilde{R}}}\gamma^\Omega A_\Omega\right)\tilde{\psi},\nonumber
\eeq
where
\beq
&R=\tilde{R}+N_{f}\left(\frac{g_0^2}{8\pi^2}-\frac{5g_0^2}{96\pi^2}\right)\ln\tilde{b}\approx\tilde{R}\tilde{b}^{\frac{7N_{f}g_0^2}{96\pi^2\tilde{R}}}=\tilde{R}\lambda^{-\frac{7N_{f}g_0^2}{48\pi^2\tilde{R}}},\nonumber\\
&\tilde{R}=1+\frac{N_{f}}{8\pi^2}g_0^2\ln\frac{\Lambda}{\tilde{\Lambda}},\label{defr}
\eeq
and where we have identified $\tilde{b}=\lambda^{-2}$ and dropped terms of order 
$(\ln\tilde{b})^{2}$.

We substitute (\ref{selfenergyellipsoid}) into (\ref{lagrangians}) to find
\beq
\mathcal{L}_{\rm quarks}&=&\tilde{\bar{\psi}}\left[\gamma^C\partial_C\left(1+\frac{N_{f}g_0^2}{8\pi^2}\ln\frac{\Lambda}{\tilde{\Lambda}}+\frac{N_{f}g_0^2}{8\pi^2}\ln\tilde{b}-\frac{5N_{f}g_0^2}{96\pi^2}\ln\tilde{b}-\frac{N_{f}g_0^2}{16\pi^2}\ln\tilde{b}\right)\right.\nonumber\\
&&\!\!\!\!\!\left.+\gamma^\Omega\partial_\Omega\left(1+\frac{N_{f}g_0^2}{8\pi^2}\ln\frac{\Lambda}{\tilde{\Lambda}}+\frac{N_{f}g_0^3}{8\pi^2}\ln\tilde{b}-\frac{5N_{f}g_0^2}{96\pi^2}\ln\tilde{b}-\frac{N_{f}g_0^2}{12\pi^2}\ln\tilde{b}\right)\right]\tilde{\psi}\nonumber\\
&\approx &R'\tilde{\bar{\psi}}\,i\left[\gamma^C\partial_C+\lambda^{\frac{N_{f}g_0^2}{24\pi^2\tilde{R}}}\gamma^\Omega\partial_\Omega\right]\tilde{\psi},\nonumber
\eeq
where
\beq
R'=R\lambda^{\frac{N_{f}g_0^2}{8\pi^2\tilde{R}}}.\label{defrprime}
\eeq

To make the effective action manifestly gauge invariant, we need to $\mathcal{L}_{\rm Dirac}=\mathcal{L}_{\rm quarks}+\mathcal{L}_{\rm vertex}$ in terms of covariant derivatives. We accomplish this by redefining
\beq
\lambda^{-\frac{N_{f}g_0^2}{8\pi^2\tilde{R}}}\tilde{A}_\mu\to\tilde{A}_\mu,\,\,\,\,R'\lambda^{-1+\frac{N_{f}g_0^2}{24\pi^2\tilde{R}}}\tilde{\bar{\psi}}\tilde{\psi}\to\bar{\psi}\psi,\label{rescaleda}
\eeq
so that
\beq
\mathcal{L}_{\rm Dirac}=\bar{\psi}\,i\left(\lambda^{1-\frac{N_{f}g_0^2}{24\pi^2\tilde{R}}}\gamma^CD_C+\gamma^\Omega D_\Omega\right)\psi.\nonumber
\eeq
The factor absorbed by the gauge field in the rescaling (\ref{rescaleda}) modifies the pure gauge action. We notice that this factor depends on the effective coupling $\frac{g_0}{\tilde{R}}$ instead of the coupling $\tilde{g}$ from (\ref{effectivecoupling}). This is because the factor from (\ref{rescaleda}) arises from the quark self energy and the vertex corrections, instead of the vacuum polariztion.

Substituting (\ref{anumbers}) into (\ref{lagrangians}) gives us
\beq
\mathcal{L}_{\rm gauge}&=&\frac{1}{4}\left(\frac{1}{g_0^2}-\frac{11}{48\pi^2}C_G\ln\frac{\Lambda}{\tilde{\Lambda}}+\frac{1}{12\pi^2}N_{f}\ln\frac{\Lambda}{\tilde{\Lambda}}\right.\nonumber\\
&&\,\,\,\,\,\,\,\,\,\,\left.-\frac{1}{64\pi^2}\frac{59}{9}C_G\ln\tilde{b}+\frac{1}{9\pi^2}N_{f}\ln\tilde{b}\right)(\tilde{F}_{01}^2+\tilde{F}_{02}^2+\tilde{F}_{13}^2+\tilde{F}_{23}^2)\nonumber\\
&&+\frac{1}{4}\left(\frac{1}{g_0^2}-\frac{11}{48\pi^2}C_G\ln\frac{\Lambda}{\tilde{\Lambda}}+\frac{1}{12\pi^2}N_{f}\ln\frac{\Lambda}{\tilde{\Lambda}}-\frac{1}{64\pi^2}\frac{31}{9}C_G\ln\tilde{b}\right.\nonumber\\
&&\,\,\,\,\,\,\,\,\,\,\,\left.+\frac{1}{9\pi^2}N_{f}\ln\tilde{b}-\frac{7}{72\pi}N_{f}\ln\tilde{b}\right)\tilde{F}_{12}^2\nonumber\\
&&\frac{1}{4}\left(\frac{1}{g_0^2}-\frac{11}{48\pi^2}C_G\ln\frac{\Lambda}{\tilde{\Lambda}}+\frac{1}{12\pi^2}N_{f}\ln\frac{\Lambda}{\tilde{\Lambda}}-\frac{1}{64\pi^2}\frac{67}{9}C_G\ln\tilde{b}\right.\nonumber\\
&&\,\,\,\,\,\,\,\,\,\,\,\,\left.+\frac{1}{9\pi^2}N_{f}\ln\tilde{b}+\frac{1}{36\pi^2}N_{f}\ln\tilde{b}\right)\tilde{F}_{03}^2.\nonumber
\eeq

The pure gauge Lagrangian is then
\beq
\mathcal{L}_{\rm gauge}&=&\frac{1}{4g_{\rm eff}^2}\left(\tilde{F}_{01}^2+\tilde{F}_{02}^2+\tilde{F}_{13}^2+\tilde{F}_{23}^2+\tilde{F}_{03}^2\lambda^{\frac{C_G}{32\pi^2}\frac{8}{9}\tilde{g}^2-\frac{2N_{f}}{9\pi^2}\tilde{g}}\right.\nonumber\\
&&\,\,\,\,\,\,\,\,\,\,\,\,\,\left.+\tilde{F}_{12}^2\lambda^{-\frac{C_G}{32\pi^2}\frac{28}{9}\tilde{g}^2+\frac{7N_{f}}{9\pi^2}\tilde{g}^2}\right),\nonumber
\eeq
where
\beq
g_{\rm eff}^2=\tilde{g}^2\lambda^{-\frac{C_G}{32\pi^2}\frac{59}{9}\tilde{g}^2+\frac{8N_{f}}{9\pi^2}\tilde{g}^2+\frac{N_{f}g_0^2}{4\pi^2\tilde{R}}},\label{geff}
\eeq
and
\beq
\frac{1}{\tilde{g}^2}=\frac{1}{g_0^2}-\frac{11}{48\pi^2}C_G\ln\frac{\Lambda}{\tilde{\Lambda}}+\frac{N_{f}}{3\pi^2}\ln\frac{\Lambda}{\tilde{\Lambda}}.\nonumber
\eeq
The last term in the powers of $\lambda$ in (\ref{geff}) comes from the  redefinitions (\ref{rescaleda}). 

Finally, we longitudinally rescale the slow fields, after removing the tildes, and Wick rotating back to real space. The effective action is given by $S_{\rm eff}=\int d^4x \,\mathcal{L}_{\rm eff}$, where
\beq
\mathcal{L}_{\rm eff}&=&\frac{1}{4g_{\rm eff}^{2}}\left(F_{01}^2+F_{02}^2-F_{13}^2-F_{23}^2+
\lambda^{-2+\frac{C_G}{36\pi^2}\tilde{g}^2-\frac{2N_{f}}{9\pi^2}\tilde{g}^2}F_{03}^{2}
\right.\nonumber\\
&&\,\,\,\,\,\,\,\,\,\,\,\,\,\,\,\,\left.-\lambda^{2-
\frac{7C_G}{72\pi^2}\tilde{g}^2+\frac{7N_{f}}{9\pi^2}\tilde{g}^2}F_{12}^{2}\right) \nonumber \\
&&+ {\bar \psi}_{\alpha}\,i
\left(
\lambda^{1-\frac{N_{f}g_0^2}{12\pi^2\tilde{R}}}\gamma^CD_C+\gamma^\Omega D_\Omega\right)\psi_\alpha , \label{finalaction}
\eeq
where the label $\alpha=1,2,...,N_f$ denotes the flavors of quarks.

\chapter{Discussion}

\setcounter{equation}{0}
\renewcommand{\theequation}{8.\arabic{equation}}

In this final chapter, we present a brief discussion of the new results found in this thesis, and discuss 
possible future research problems. 

Our first original results are the form factors and correlation functions of the principal chiral sigma model. We were able to find all the exact form factors of the Noether current and energy-momentum tensor operators in the 't~Hooft large-$N$ limit.  We then calculated the exact two-point function of these operators, using the new results. This model is the only example of a field theory of propagating particles which has been completely solved (in the sense that correlation functions are known) in the
planar limit. The only other theories solved in this limit are (0+1)-dimensional matrix models. There is hope that N=4 gauge theories will be completely solved in the 't~Hooft limit, but this program is not yet finished \cite{SYMreview}. The planar limit is much harder to solve than the large-N limit of iso-vector theories (such as the O(N) sigma model or the SU(N) Chiral-Gross-Neveu model).

At finite values of $N$, our results are much less ambitious. We are able to find only the two-excitation form factors. In the future it might be possible to calculate form factors with more particles using a more sophisticated method, like the nested off-shell Bethe Ansatz \cite{babujian}.

It would be interesting in the future to study PCSM in a finite volume, with periodic boundary conditions, in the 't~Hooft limit. One can try to find the spectrum of energies as a function of the PCSM mass gap and the length of the $x^1$ direction using the Bethe ansatz, which would be greatly simplified at large $N$.  Some work in this direction has been done for $N=2$ and $N=3$ by Kazakov and Leurent \cite{kazakovleurent}.

The simplest non-integrable model we examined was (1+1)-dimensional massive Yang-Mills theory. 
We saw that in the axial gauge, this model reduces to a principal chiral sigma model perturbed with a current-current interaction. The excitations are confined hadron-like states. Using the exact S-matrix of the sigma model we calculated the meson spectrum in the non-relativistic limit. As we discussed in Chapter 5, in the future we would like to calculate relativistic corrections to this spectrum. We can use the exact Noether current form factors to write a Bethe-Salpeter equation. One can compute corrections to the bound state masses in powers of $1/c$, solving the Bethe-Salpeter equation perturbatively.

In Chapter 6, we use the exact results from the PCSM to compute physical quantities in anisotropic (2+1)-dimensional Yang-Mills theory. These calculations have been done before for $N=2$, using the S-matrix and form factors of the $O(4)$ nonlinear sigma model, in References \cite{glueball},\cite{horizontal},\cite{vertical}. In this thesis, we use our new result for the two-excitation form factor to generalize these results for $N>2$. We computed the first corrections in powers of the rescaling parameter $\lambda$, for the string tension of a static quark-antiquark pair. The theory is not $90$-degrees rotation invariant, so the string tension is different if the particles are separated in the $x^1$ or the $x^2$ direction. We also calculated the low-lying glueball spectrum. This result proved to be very similar to the meson-spectrum of (1+1)-dimensional massive Yang-Mills.

In the future we hope to examine the partition function of (2+1)-Yang-Mills Theory in powers of $\lambda$, away from the integrable limit, in the context of form factor perturbation theory \cite{DMS}. This involves computing matrix elements  $\langle \Psi^\prime \vert H_1 \vert \Psi\rangle$.  This is equivalent to evaluating Noether current correlation functions between the states of the principal chiral model. This is can be done with our form factors from Chapter 4.

 The isotropic theory can be examined through the truncated spectrum approach. This was used by R.M. Konik and Y. Adamov to explore the 3-dimensional Ising model as an array of coupled 2-dimensional Ising chains \cite{konik}. One can discretize the spectrum of the (1+1)-dimensional models by putting them in a box of finite size. The physical states are then ordered by energy,  as $\vert 1\rangle,\,\vert 2\rangle,\dots,\,\vert n\rangle$, with energies $E_1 < E_2< \dots< E_n$, respectively, where $E_n$ is the truncation energy.

We can define a transfer-matrix operator that describes how the system evolves in the $x^2$ direction, as
\begin{eqnarray}
\hat{T}_{x^2-a,\,x^2}=e^{-\frac{1}{2} H_0(x^2-a)-\frac{1}{2}H_0(x^2)-\lambda^2H_1(x^2,\,x^2-a)}.\nonumber
\end{eqnarray}
In the truncated spectrum approach one can build a discrete, $n\times n$ matrix $T_{ij}=\langle i\vert \hat{T}_{x^2-a,\, x^2}\vert j\rangle$, using the set of states with energies $E_i,\,E_j\leq E_n$. The Yang-Mills partition function is 
\begin{eqnarray}
Z={\rm Tr} \,T^{N_2},\nonumber
\end{eqnarray}
where $N_2$ is the total number of sigma models (the size of the $x^2$ direction). The partition function can be computed by diagonalizing the matrix $T_{ij}$, which can be done numerically, or perturbatively in powers of $\lambda$. One can extract the mass spectrum this way, and examine their dependence on the truncation energy $E_n$.

In Chapter 7, we explored the quantum effects of longitudinal rescaling in gauge theories. By rescaling some of the longitudinal coordinates, the ultraviolet momentum cutoffs become ellipsoidal, rather than spherical. We devise an anisotropic version of Wilson's renormalization group, which illustrates how the parameters of the theory flow as the momentum cutoffs become anisotropic. We explicitly computed the quantum longitudinally rescaled actions of QED and QCD with massless fermions in 3+1 dimensions. 

One obvious future project would be study the anisotropic renormalization group in 2+1 dimensions. We also hope in to future to find a gauge-invariant method of calculating the longitudinally-rescaled action. This may be possible with some version of dimensional regularization, where the number of longitudinal and transverse dimensions can vary independently. The background-field method could be used instead of Wilsonian renormalization with sharp momentum cutoffs. This would eliminate the need for counter terms to maintain gauge invariance.


\end{document}